\documentclass[a4paper,fleqn,usenatbib]{mnras}

\usepackage[T1]{fontenc}
\usepackage{ae,aecompl}

\usepackage{graphicx}	
\usepackage{amsmath}	
\usepackage{amssymb}	

\usepackage[usenames,dvipsnames]{color}
\newcommand{\firefly}{\textsf{FIREFLY}~}

\title[Stellar population properties of individual massive high-$z$ ETGs]{Stellar population properties of individual massive early-type galaxies at $1.4 < z < 2$}

\author[I. Lonoce et al.]
{I. Lonoce$^{1,2}$\thanks{E-mail: ilonoce@uchicago.edu},
C. Maraston$^{3}$,
D. Thomas$^{3}$,
M. Longhetti$^{1}$,
T. Parikh$^{3}$,
P. Guarnieri$^{3}$,
\newauthor 
and J. Comparat$^{4}$\\
$^{1}$INAF-Osservatorio Astronomico di Brera, via Brera 28, I-20121 Milano, Italy\\
$^{2}$Department of Astronomy $\&$ Astrophysics, The University of Chicago, 5640 South Ellis Avenue, Chicago, IL 60637, USA \\
$^{3}$Institute of Cosmology and Gravitation, University of Portsmouth, Dennis Sciama Building, Burnaby Road, Portsmouth PO1 3FX, UK\\
$^{4}$Max Planck Institute for extraterrestrial Physics, Giessenbachstrasse 1, 85748 Garching, Bayern, Germany \\
}

\date{Accepted 2019 November 28. Received 2019 November 26; in original form 2018 December 8.}

\pubyear{2019}

\begin{document}
\label{firstpage}
\pagerange{\pageref{firstpage}--\pageref{lastpage}}
\maketitle

\begin{abstract}

We analyse publicly available, individual spectra of four, massive ($M>10^{11}M_{\odot}$) 
early-type galaxies with redshifts in the range $1.4 \leq z \leq 2$ to determine their stellar 
content, extending our previous work up to $z\sim2$. The wide wavelength range of the VLT/X-Shooter 
spectroscopic data in the UV-Optical-NIR arms along with the availability of spectro-photometry 
allows us to explore different techniques to obtain the stellar population properties, namely 
through age/metallicity sensitive spectral indices, full spectral fitting and broad-band photometric 
fitting. Moreover, together with the widely used optical Lick indices we consider further indices 
in the UV rest-frame, and demonstrate that UV indices significantly help the accuracy of the resulting population parameters.

We find galaxy ages ranging from $0.2$ to $4$ Gyr, where the oldest galaxy is found at the lowest 
redshift, with an excellent agreement between ages determined via indices, full spectral fitting or 
broad-band colours. 
These ages are in perfect agreement with ages of local galaxies at the same velocity dispersion 
when we assume pure passive evolution.  Total metallicities derived from indices show some scatter 
(between  less than half-solar to very high values, ($[Z/H]\sim0.6$]). 
We speculate on possible mechanisms explaining these values, but given the sample size and low 
$S/N$ of the spectra no conclusion can be made. 

Indices in the UV-rest frame generally lead to similar conclusions as optical indices. For the 
oldest galaxy (4 Gyr) we show that its UV-indices can only 
be explained by stellar population models including a UV contribution from old stellar populations, 
suggesting that old, UV bright populations start to inhabit mature galaxies of a few Gyr of age.  This is the highest redshift ($z\sim 1.4$) detection of the UV-upturn up to date.

\end{abstract}

\begin{keywords}
galaxies: elliptical and lenticular, cD 
-- galaxies: high-redshift -- galaxies: stellar content.
\end{keywords}
%
%
\section{Introduction}
The stellar population properties of galaxies, such as age, metallicity and chemical abundance ratios, are key to disclose their past formation and evolution at each cosmic epoch, from the local Universe up to high redshift. In this work we focus on early-type galaxies (ETGs). In the local Universe the stellar population properties of early-type galaxies are a clear function of the galaxy stellar mass, with the more massive objects being older, more metal-rich and enhanced in $\alpha$-elements over Iron ($[\alpha/Fe]\sim0.3$, e.g. \citealt{thomas05,thomas10}, \citealt{Clemens06, Allanson09,  parikh19}). At given stellar mass and/or velocity dispersion, a clear homogeneity among the stellar populations of local ETGs is evident, as revealed by the many observed tight scaling relations like the fundamental plane of elliptical galaxies, the age-mass and metallicity-mass relations (e.g.
\citealt{trager2000,bernardi05,gallazzi06,martin-navarro18,cresci19}). It is also well-known that the existence of such tight relations and the properties of ETGs at the massive end challenge predictions from $\lambda$-CDM hierarchical galaxy formation models, where a more gradual and random galaxy build-up with redshift weaken scaling relations and implies a counter dependence between stellar mass, and age/abundance-ratios to what is observed (\citealt{thomas99}; \citealt{spolaor10}).

Current instrumentation is now capable of acquiring galaxy spectra at increasingly larger cosmic distances, hence the question can be addressed directly on how galaxy stellar populations evolve over cosmic time. 

In particular, the chemical content of stellar systems is directly connected to their formation mechanisms from the primordial gas and to the time-scale on which the star formation has taken place (e.g. : \citealt{thomas05}; \citealt{thomas10}; \citealt{spolaor}; \citealt{delarosa11}; \citealt{segers16}; \citealt{mio2}). 
Detailed metallicity information coupled with age estimates allow one to distinguish between natural (passive) ageing processes and other evolutionary paths as drivers of the evolution of ETGs. 

Stellar population properties, in particular the metallicity, are derived from detailed spectral analysis with stellar population models as it is done for galaxies in the local Universe (e.g.: \citealt{gallazzi05}; \citealt{trager09}; \citealt{thomas10}; \citealt{spolaor}). 

Moreover, the younger ages of the high-redshift systems (age $< 2$ Gyr) give the important advantage of reducing the well-known age-metallicity degeneracy (e.g. \citealt{worthey94}), easing the determination of these parameters in spite of the lower signal-to-noise ratios of the data. 

The approach may make use of few selected spectral indices sensitive to the abundance of specific elements or of the whole spectrum. 

Up to now, few pioneering works have been devoted to either the analysis of spectral indices (e.g.: \citealt{ziegler05}; \citealt{onodera}; \citealt{jorgensen13}; \citealt{mio}; \citealt{gallazzi}; \citealt{jorgensen}) or of the full spectrum (e.g.: \citealt{kriek16}; \citealt{mendel15}; \citealt{vandesande}) of high-\emph{z} ETGs, gaining with the confirmation of their passive status and with some indications of $\alpha$-elements enhanced populations.  

In this work, we use a sample of four galaxies in the redshift range $1.4<z<2$ taken in both COSMOS (Cosmological Evolution Survey, \citealt{cracken}) and UKIDSS-UDS (Ultra Deep Survey, \citealt{lawrence}; \citealt{uds}) fields, observed with the VLT/X-Shooter instrument covering a wide spectral range from $3000$ \AA $ $ to $24800$ \AA.

We shall use the classical optical Lick/IDS system (i.e.: H$\gamma$, H$\beta$, Mg$_b$, etc..) with models sensitive to chemical abundance ratios \citep{tmj}, in order to gauge our high-$z$~analysis to the $z=0$~analysis. 

Moreover, and originally with respect to previous work, we also explore spectral indices in the UV region (i.e.: MgI, MgII, 
Mg$_{wide}$, BL3096, BL3580, etc...) using high-resolution models \citep{ma2009} as well as models including various options for the UV upturn phenomenon (from C. Maraston as used in \citet{claire}). 

This paper is organized as follows. In Section \ref{sec:sample} we introduce the sample of ETGs together with the description of the VLT/X-Shooter spectroscopic data. In Section
\ref{analisimetall} we present the spectral analysis including the measurements of Lick and UV indices and the results obtained from their comparison with the models. In the same section we test UV-upturned models with the lowest redshift galaxy of the sample (COSMOS-307881). 
With Section \ref{sub:fsf} and \ref{sec:photometry} we compare the results obtained from the spectral indices analysis with both full spectral fitting and wide photometric fitting. In Section \ref{all:discussion}, we discuss our results and compare them with local scaling relations, discussing possible implications on the redshift evolution of our galaxies. A summary is then found in Section \ref{sec:conclusions}.
 
Throughout this paper, we assume a standard cosmology with H$_0=70$ km s$^{-1}$ Mpc$^{-1}$, $\Omega_m=0.3$ and $\Omega_{\Lambda}=0.7$.

\section{The galaxy sample}
\label{sec:sample}

The galaxy sample we use is composed of $4$ ETGs with spectroscopic redshift in the range $1.4<z<2$, which we selected on the basis of the availability of X-Shooter spectroscopic data. The X-Shooter instrument provides a wide spectral coverage, from the UV to the near-IR, allowing the modelling of several spectral indices (see Section \ref{spectroscopy}). 
Two galaxies, ID-$307881$ and ID-$7447$, lie in the COSMOS field \citep{cracken}, and the other two (identified as UDS-$19627$ and UDS-$29410$ ) have been selected from the UKIDSS-UDS field (\citealt{lawrence}; \citealt{uds}). 

One of the COSMOS object (COSMOS-$307881$) has been already analysed by us in \citet{mio2} using the same spectrum as here, but we include it again as in this paper we perform a wider model analysis including UV indices and upturn models (see Section \ref{sub:upturn}). We shall show that only these up-turn models can match the observed indices.

The spectra of the other $3$ targets, COSMOS-$7447$, UDS-$19627$ and UDS-$29410$ 
have been previously analyzed by \citet{vandesande} (and references therein), who performed the measurement of the velocity dispersion exploiting the high resolution of X-Shooter data, in order to study its evolution to $z\sim0$. They derive stellar population parameters as well as performing spectral fitting on the binned spectra ($\sim10$ \AA) using the FAST code \citep{kriek09}. Differently from this work they used the \cite{bc03} models with an exponentially declining star formation history and a \cite{chabrier03} IMF.

In this work, we perform a detailed spectral and spectro-photometric analysis of the galaxy stellar populations, using a novel approach that combine optical and UV indices and consider several options for the origin of the UV flux besides that of ongoing star formation.

These ETGs are all bright (K $<20$), and according to the analysis of \cite{vandesande} they are all massive ($M>10^{11}M_{\odot}$, from spectral fitting with a Chabrier IMF but also confirmed from our photometric analysis performed with a Salpeter IMF, see below) and dense objects ($\sigma>270$ km/s). 

Their morphology as inferred from available HST (Hubble Space Telescope) \citep{hst} or UKIDSS-UDS \citep{lawrence} deep imaging, is smoothly elliptical, as already discussed in \citet{mio2} and in \citet{vandesande}. Information about the sample and parameters derived from the literature (spectroscopic redshift, stellar mass, effective radius and velocity dispersion) are listed in Table \ref{tab:infoall}. We refer to \citet{vandesande} and \citet{mio2} (and references therein) for more details.

\begin{table*}
\begin{centering}
 \caption{Information and parameters for the sample galaxies: RA (in h, min, sec) , DEC (in deg, arcmin and arcsec), K-band magnitude (AB system), spectroscopic redshift (\emph{z}$_{spec}$), effective radius (R$_e$), observed velocity dispersion ($\sigma$) and dynamical mass (log$\mathcal{M}_{dyn}$) determined with the formula of \citet{cappellari}.}
 \label{tab:infoall}
 \begin{tabular}{lccccccc}
 \hline
 ID & RA & DEC & K    & \emph{z}$_{spec}$ &  R$_e$   & $\sigma$ & log$\mathcal{M}_{dyn}$\\
    &    &     & (AB) &                   &  (kpc)   & (km/s)   & (M$_{\odot}$)         \\
 \hline
 19627  & 02:18:17.06 & -5:21:38.83 & 20.19 & 2.036  &   1.32$\pm$0.17 &  287$\pm$39$^b$ & 11.14 \\
 7447   & 10:00:06.96 & 02:17:33.77 & 19.64 & 1.800  &   1.75$\pm$0.21 &  274$\pm$51$^b$ & 11.21 \\
 29410  & 02:17:51.22 & -5:16:21.84 & 19.36 & 1.456  &   1.83$\pm$0.23 &  355$\pm$98$^b$ & 11.45 \\ 
 307881 & 10:02:35.64 & 02:09:14.36 & 19.44 & 1.426  &   2.68$\pm$0.12 &  385$\pm$85$^a$ & 11.65 \\
 \hline
 
\end{tabular}
\end{centering}

\vspace{3mm}

 $^a$ see \citet{mio2}. \\
 $^b$ see \citet{vandesande}. \\

\end{table*}

\subsection{Spectroscopic data}
\label{spectroscopy}
The spectroscopic observations used in this work have been carried out with the X-Shooter spectrograph on the VLT. The great advantage of X-Shooter is that it obtains high-resolution spectra on a wide spectral range with a single exposure. In particular, it consists of three arms: UVB, VIS and NIR, which together covers spectral window from $3000$ \AA $ $ to $24800$ \AA $ $ with a mean resolution $R\sim6500$. \citet{vandesande} and \citet{mio2} have extensively discussed the X-Shooter data hence here we just summarize their main characteristics (Table \ref{tab:allobs}) . 
\begin{table*}
\begin{centering}
 \caption{Observational details for the $4$ ETGs taken from \citet{vandesande} and \citet{mio2}: observation period, exposure time, slit width, program ID, S/N per dispersion elements before and after the $4000$\AA\space break restframe.}
 \label{tab:allobs}
 \begin{tabular}{lcccccc}
 \hline
 ID & Period & Exposure time & slit & program ID   & S/N [<4000\AA $ $ restframe] & S/N [>4000\AA $ $ restframe]\\
 \hline
 19627  &  October, 2009      & 1h+3h & 0".6+0".9 & 084.A-0303(D) & 4&7\\
 7447   &  January 22th, 2010 & 1.8h & 0".9 & 084.A-1082(A)       & 5&6\\
 29410  &  July, 2010         & 1.7h & 0".9 & 085.A-0962(A)       & 4&6\\ 
 307881 &  February 9th, 2011 & 4.3h & 0".9 & 086.A-0088(A)       & 4&7\\
 \hline
\end{tabular}
\end{centering}
\end{table*}

The observing strategy for all the observations is the standard division in exposure blocks with an alternated ABA'B' dithering of the vertical source position in the slit. 
Close to the sources acquisitions, standard stars have been observed during the same nights. 

We performed data reduction using the public {\small\texttt{ESOREX}} pipeline \citep{goldoni} for the first steps of the process, while the main procedure has been carried out with the standard {\small\texttt{IRAF}} tools. 

In more detail, we relied on the pipeline for the standard calibration process (consisting of correction for bias, dark, flat, etc...), 
checking the output at every single step. Particular care has been given to the identification of the spectral
orders, their straightening and wavelength calibration, where specific hand-made checks have been applied in order to control the pipeline outputs. Furthermore, in each reduction step
where the position of the source was involved, we made use of an ancillary code which identifies the source exact position along the slit with high precision in order to further check the output of the {\small\texttt{ESOREX}} tool. Indeed, the faintness of our high-redshift objects makes their identification over the noisy underlying sky very difficult, and it was suggested that the {\small\texttt{ESOREX}} tools could fail to perform this identification automatically (F. La Barbera, {\it private communication}). We found that the source position of ESOREX was correct in all cases.

During these first phases of the pipeline, we proceeded following the ``stare'' mode (i.e. considering each frame as an independent observation, in opposition to the ``nodding'' mode which pairs subsequent observation frames) in order to obtain at the end a single, order-merged, wavelength calibrated, bi-dimensional spectrum for each observing block. 

We then completed the data reduction by means of {\small\texttt{IRAF}} performing the sky subtraction, the mono-dimensional spectrum extraction and its flux calibration. We found that the sky subtraction was obtained more accurately by means of the {\small\texttt{IRAF}} tools combined with the possibility 
to exploit the A-B observing pattern with respect to the {\small\texttt{ESOREX}} pipeline. 
Mono-dimensional spectra within their effective radius R$_e$ have been extracted (see Table \ref{tab:infoall}).
The construction of the sensitivity function has been performed starting from the reduced standard
stars produced by {\small\texttt{ESOREX}}.

\begin{figure*}
\begin{centering}
\includegraphics[width=17.6cm]{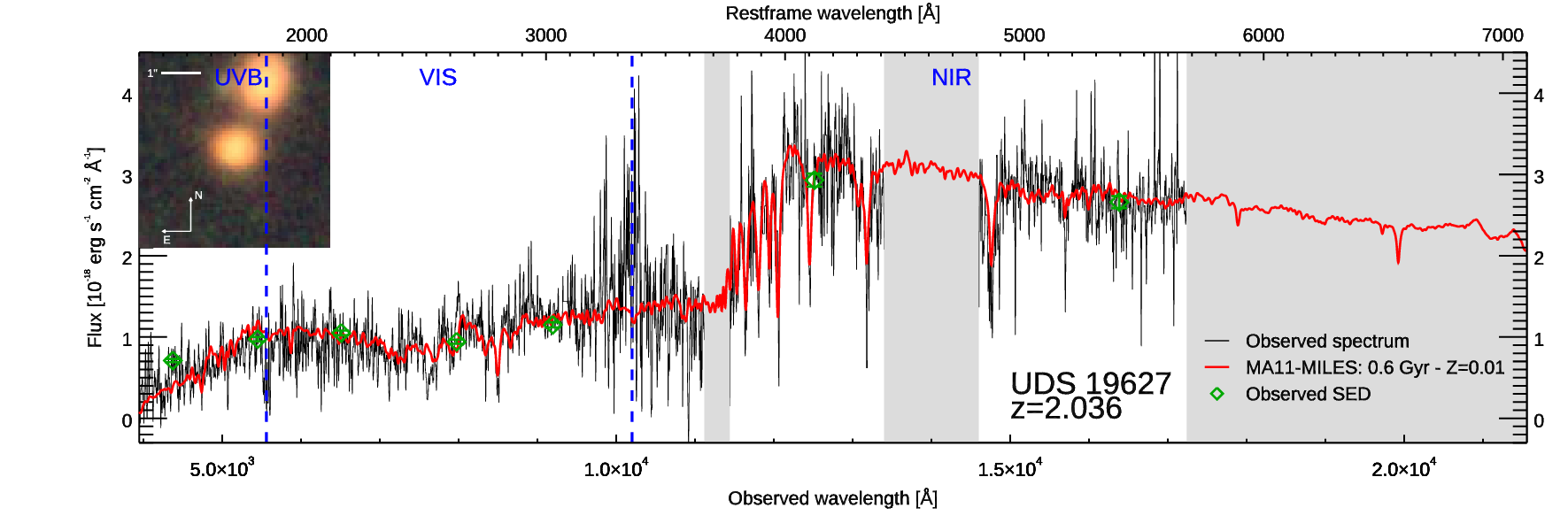}
\includegraphics[width=17.6cm]{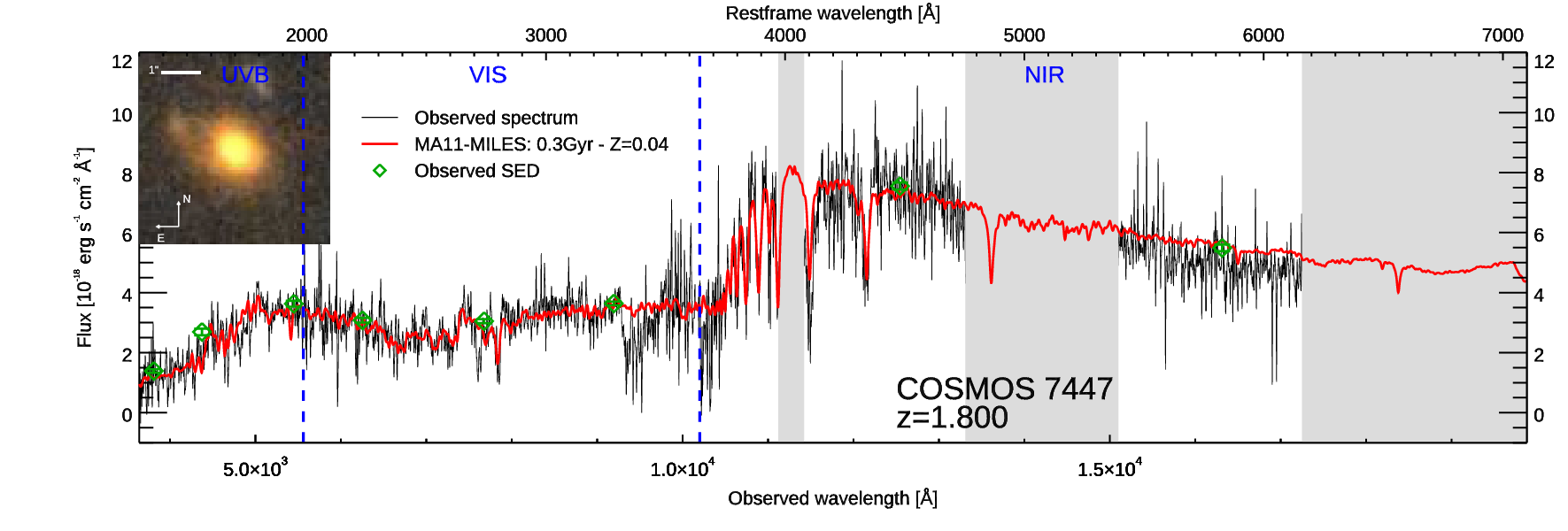}
\includegraphics[width=17.6cm]{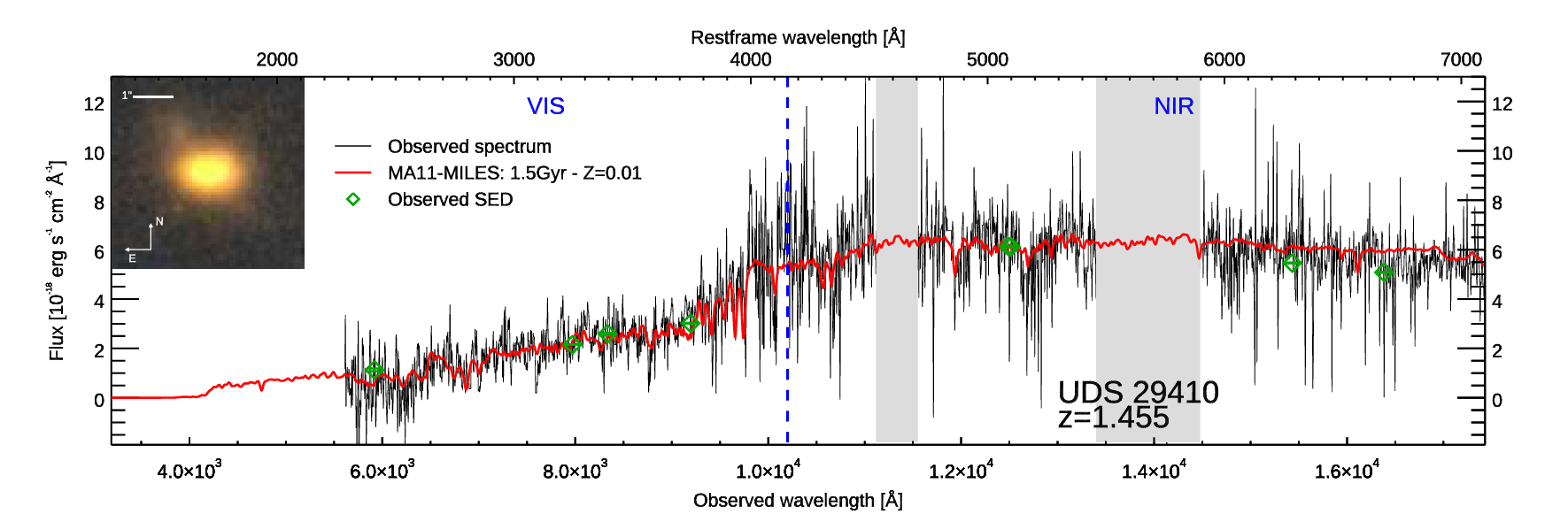}
\includegraphics[width=17.6cm]{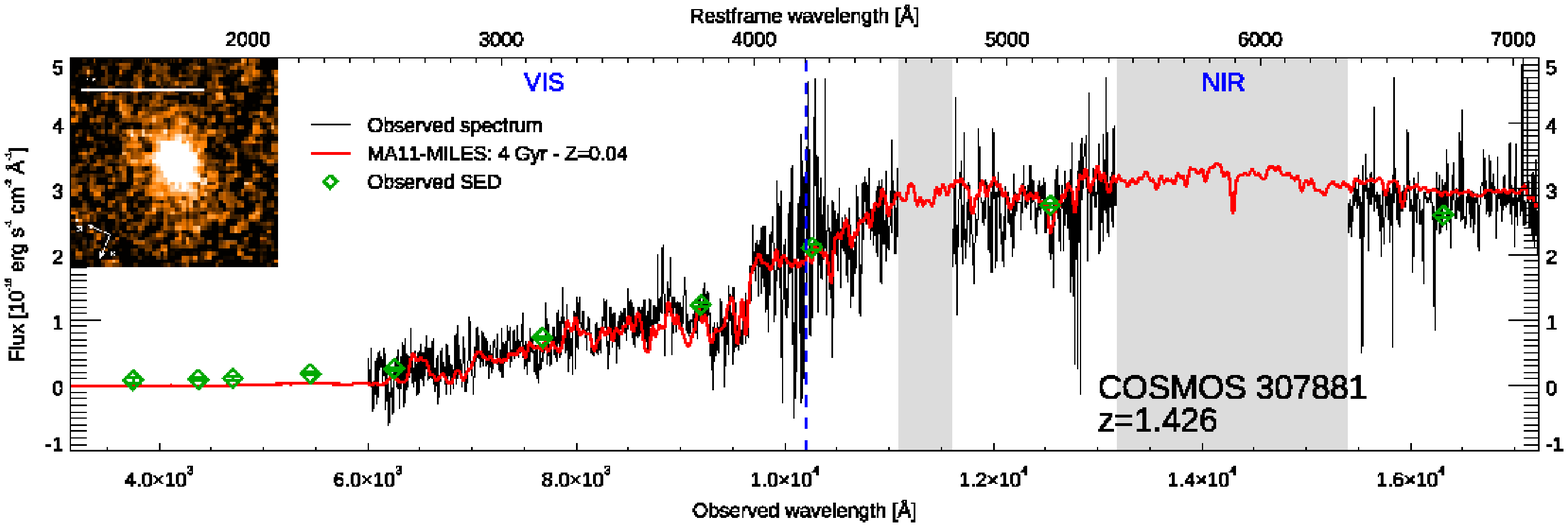}
\caption{Observed spectrum of UDS-19627, COSMOS-7447, UDS-29410 and COSMOS-307881 ordered from the highest to lowest redshift, together with their respective HST images. The observed spectrum (black line) is compared with the available photometric data (green diamonds) and with a representative MS11 model \citep{ma11} (red lines). Vertical dashed blue lines indicate the connection between UVB-VIS and VIS-NIR arms at $\sim5600$\AA $ $ and $\sim10200$\AA. In the NIR band, the strong atmospheric absorptions have been covered. The spectral resolution of the observed spectrum has been downgraded to match the MS11 one, i.e. FWHM $=2.55$ \AA $ $ restframe.}
\label{spettri}
\end{centering}
\end{figure*}

The reduced galaxy spectra can be seen in Fig. \ref{spettri} (black lines). The three spectral windows of the X-Shooter arms (UVB, VIS and NIR) are connected at $\sim5600$\AA $ $ and $\sim10200$\AA $ $ (vertical dashed blue lines) with the help of the available photometric data (see Section \ref{sec:photometry}, green diamonds). Note that objects 
COSMOS-307881 and UDS 29410, due to their lower redshift, did not reveal any non-zero signal in the UVB region, which has not been included in the analysis. 
Nevertheless, for these two objects we were able to measure UV indices such as Mg{\small\texttt{II}} at $2800$\AA\ and FeI(3000) as they fall in the VIS arm at these redshifts  (see Table \ref{tab:allindices}). 
In the NIR region, all the strong atmospheric absorptions have been masked by means of grey shading. Due to the low S/N of the data (typically lower than ten, cfr. Table \ref{tab:allobs}), telluric absorptions cannot be properly removed so we preferred to avoid any index measurement in the affected regions. For each galaxy, a representative model (\citealt{ma11}, hereafter MS11) obtained from the
analysis described in the following sections is shown (red line).

Note that in these figures and in the subsequent analysis, the observed spectra (black lines) have been binned 
to gain a higher S/N ratio, which after downgrading is about $4-7$ per dispersion element (i.e. $1.8$ \AA $ $ in the UV and VIS arms and $2.0$ \AA $ $ in the NIR arm), depending on the spectral range). 
Indeed, the X-Shooter spectra, due to their high resolution, would be totally dominated by noise in case of such faint objects. The decrease in resolution allows to sufficiently characterise the spectral features and make them suitable to be compared to models.

\section{Analysis}
\label{analisimetall}

The wide spectral extension of the X-Shooter data allows the measurement of several spectral indices, which brings a wealth of information about a galaxy stellar content. Moreover, the high-redshift of these galaxies allows us to access their rest-frame UV region. 

Therefore, in this paper, we could explore for the first time the effect of simultaneously modelling the usual optical rest-frame region together with newly-explored UV indices.

\subsection{Measured Indices}
The measured index values are given in Table 3. Note that not all indices could be measured on each spectrum because the redshift and the spectral quality (e.g.: atmospheric ab- sorption regions, telluric residuals, UVB-VIS-NIR arms connections) varies among the targets. This means that a specific analysis tailored to each galaxy needed to be performed, as will be discussed below. In order to understand the uncertainties in stellar population parameters derived with different sets of indices, we performed the same analysis on mock spectra with known population parameters and the same spectral quality as the data. These tests are described in Appendix \ref{appendixA}. We find that the use of different index sets does not affect the derived age and metallicity. A small improvement is noted when using a larger number of indices. We also show that in this young ($< $4 Gyr) age range, the addition of indices in the UV spectral range significantly helps the accuracy of the results.

Errors have been derived by means of a Monte-Carlo approach. We added noise to the observed spectrum by perturbing the flux value at each pixel assuming a Gaussian error distribution. We performed 5,000 Monte Carlo realisations. The flux error was derived from the noise spectrum that we extracted from the 2D spectrum at a location off target.
Hence the latter includes the sky residuals. We then measured the corresponding index values for each Monte Carlo realisation and determined the statistical error from the width of the distribution of index measurements. Following our approach in \citet{mio2} we then divided this value by $\sqrt{2}$ since the observed spectrum already contains the intrinsic error. Systematic errors due for example to flux calibration are neglected since their level is well below the statistic uncertainties.


\begin{table*}
\centering
 \caption{Spectral indices measured on the galaxy spectra and used in the analysis, ordered from the UV to the NIR. Blanks denote those indices that could not be measured due to poor spectral quality.}
 \label{tab:allindices}
 \begin{tabular}{lcccc}
 \hline
 Index                         &   COSMOS-307881      &  COSMOS-7447        &  UDS-19627          &   UDS-29410         \\
 \hline 
 BL(1617)   		       &   -		      &  -     &  4.86$\pm$1.78      &  -			\\
 BL(1664)   		       &   -		      &  5.11$\pm$1.33      &  4.11$\pm$1.88      &  -			\\
 BL(1719)		       &   -  		      &  -     &  -1.07$\pm$1.76     &  -		        \\
 Fe{\small\texttt{II}}(2402)   &   -		      &  5.86$\pm$1.25      &  4.51$\pm$2.13      &  -			\\
 BL(2538)   		       &   -		      &  4.35$\pm$1.30      &  10.54$\pm$1.66     &  -			\\
 Fe{\small\texttt{II}}(2609)   &   -		      &  5.46$\pm$0.87      &  6.92$\pm$1.26      &  -	\\
 Mg{\small\texttt{II}}	       &   14.00$\pm$2.79     &  12.40$\pm$0.79     &  13.02$\pm$1.28     &  13.62$\pm$2.27     \\
 Mg{\small\texttt{I}}	       &   13.18$\pm$2.06     &  2.50$\pm$0.82      &  3.43$\pm$1.28      &  10.41$\pm$1.83     \\
 Mg$_{wide}$   		           &   51.41$\pm$7.51     &  19.63$\pm$14.52    &  9.43$\pm$3.83      &  8.79$\pm$6.66      \\
 Fe{\small\texttt{I}}(3000)    &   9.60$\pm$3.72      &  -                  &  3.71$\pm$2.76      &  -                  \\
 BL(3580)   		       &   8.99$\pm$1.80      &  -		    &  -		  &  -	\\
 CN(3883)   		       &   0.16$\pm$0.07      &  0.07$\pm$0.03      &  -		  &  0.12$\pm$0.05	\\
 D4000   		       &   2.44$\pm$0.12      &  1.450$\pm$0.06     &  1.69$\pm$0.04      &  1.75$\pm$0.04      \\
 CN(4170)   		       &   -		      &  -     &  -0.04$\pm$0.03     &  -                  \\
 H$\delta_A$   		       &   -		      &  -		    &  4.93$\pm$1.26      &  -     \\
 H$\delta_F$   		       &   -		      &  8.99$\pm$0.69      &  4.87$\pm$0.87      &  -                  \\
 H$\gamma_A$   		       &   -		      &  10.68$\pm$0.86     &  8.37$\pm$1.24      &  6.30$\pm$1.43   	\\
 H$\gamma_F$   		       &   -1.56$\pm$0.92     &  7.91$\pm$0.54      &  8.00$\pm$0.78      &  5.22$\pm$0.94   	\\
 G4300   		       &   6.52$\pm$1.12      &  -     &  0.70$\pm$1.31      &  3.44$\pm$1.38   	\\
 Fe(4383)   		       &   7.40$\pm$1.74      &  -     &  -		  &  -  		\\
 Ca(4455)   		       &   1.06$\pm$0.83      &  -     &  -		  &  -  		\\
 Fe(4531)   		       &   3.20$\pm$1.40      &  -      &  -		  &  -  		\\
 H$\beta$   		       &   2.52$\pm$0.93      &  -		    &  -		  &  2.78$\pm$1.18	\\
 Fe(5015)   		       &   3.89$\pm$1.91      &  -		    &  -     &  -     \\
 Mg$_b$   		       &   5.75$\pm$0.81      &  -		    &  -      &  -    \\
 Fe(5270)   		       &   -		      &  -		    &  1.61$\pm$0.76      &  -      \\

 \hline
\end{tabular}
\end{table*}

Within the optical rest-frame region, we could measure most of the well-known Lick/IDS indices \citep{worthey}, following the band-passes definitions proposed by \citet{trager}, together with the H$\delta$~and H$\gamma$~index definitions by \citet{wortheyott}. 
These indices, coupled with stellar population models sensitive to element abundance ratios \citep{tmj} allow the derivation of $\alpha$-element enhancement (and other specific abundance ratios) which are key to derive star formation timescales \citep{thomas05}.

In addition we measured UV-rest-frame spectral indices, which we could analyse using the models by \citet{ma2009}, which have been tested on young star clusters in the Magellanic Clouds. 
The UV indices follow the definition by \citet{fanelli} and \citet{davidge}, as used in \cite{ma2009}. 
\subsection{Stellar population models}
\label{sec:models}
For interpreting the observed Lick optical indices we adopt the models by \citet{tmj} (hereafter TMJ, for a Salpeter IMF), whose stellar index values are based on the MILES library \citep{miles} as from the index calibration by \citet{johansson10}. The underlying evolutionary population synthesis model is from \cite{ma2005}, which is consistently used also in the UV indices, the high-resolution model spectra and the spectro-photometric fits (see below). 

For the UV indices we used the theoretical high-resolution models of \cite{ma2009}, which is connected around 3700\AA $ $ with the models of \citet[][MS11\footnote{\url{www.maraston.eu}}]{ma11}, in their version based on the MILES library, herein MS11-MILES UV-extended (all for a Salpeter IMF). MS11 offer a suite of models with the same energetics (i.e. the same mass-luminosity relation on the Main Sequence and fuel consumption on the post-Main Sequence) and different stellar libraries. 
The reason for opting for the MILES version here is primarily to ensure consistency with the model Lick indices. Besides, the MS11-MILES allow the widest metallicity range to be explored among the other available options.

We used the TMJ models for the Lick indices and the MS11 for the UV indices, homogenising the model age grids as to cover from $0.1$ Gyr to $12$ Gyr at steps of $0.1$ Gyr, and the metallicity grid to cover from $[Z/H]=-0.30$ to $0.30$ at step of $0.01$. These ranges are those in common to both models so that no extrapolation is needed.
The TMJ models provide index values at the MILES spectral resolution (i.e. $2.5$ \AA $ $ rest-frame), while MS11 models provide the whole SED (we used the set at the MILES resolution here) on which any index can be measured. We measured the mid-UV indices (located below $4000$ \AA $ $) from the MS11 SEDs for all available ages and metallicities and add this new mid-UV index set to the TMJ-based Lick indices at the correspondent age and metallicity. The finer grid described above was obtained in interpolation when necessary.

UV-indices are indicators of the age and/or metallicity (see \citealt{ma2009} for a description). The effect of element abundance ratios on these indices is not yet known. In this paper we shall simulate the qualitative effect of an enhancement on MgII(2800) and MgI(2852) by applying the same differential index variation of the optical Mg$_b$(5175) as modelled in TMJ models for a [$\alpha$/Fe]=0.3 (see Section \ref{sub:upturn} and Fig. \ref{upturn}).

In addition, we employ models including a UV-upturn (by C. Maraston, based on the early modelling by Maraston \& Thomas (2000)) as used in \citet{claire}. The UV-upturn is an enhanced flux in the ultraviolet observed in local massive galaxies and massive spiral bulges (see \citealt{maraston00} and references therein). It is thought to be due to old, hot stellar components, which manage to expose high effective temperatures and become UV emitters. The astrophysical origin of such an evolution is still mysterious, as widely discussed by Greggio \& Renzini (1990; see also Maraston \& Thomas 2000), but the association to old (at least a few Gyr), metal-rich populations is an empirical fact.  
In order to simulate the variety of upturns observed in local galaxies and to allow for an exploration into the mysterious nature of the UV-upturn, several models are calculated as a functions of two parameters, namely the temperature (or temperature distribution) of the old UV components and the fuel consumption at those temperatures (as in \citealt{maraston00}). Table~1 in \citet{claire} provides full details of the models.
In \citet{claire} the UV-upturn models were applied to a large sample of massive galaxies at $z\sim0.7-0.9$ from SDSS-III, finding that a large fraction of them has spectra consistent with the presence of an up-turn component. We also find as expected that the strength of such a component decreases with increasing redshift, following the rejuvenation of the stars. The $z\sim1$~galaxies have ages and stellar masses (see \citealt{ma2013} and \citealt{claire}) which are similar to our lowest redshift galaxy, COSMOS-307881. Therefore we shall probe whether the UV indices of this galaxy are consistent with the presence of an upturn. 

\subsection{Data-model comparison procedure}
The analysis of spectral indices of high-\emph{z} ETGs must be carefully treated. Indeed, $z>1.4$ ETGs contain rather young stellar populations (age $\sim1-2$ Gyr), and at
these ages spectral indices change quickly for small changes of age and metallicity. 
This is not the case of local ETGs, which are old (age $>>5$ Gyr) and for which indices are stable in their dependence with stellar parameters (and also more degenerate, as well known).

\begin{figure}
\begin{centering}
\includegraphics[width=\columnwidth]{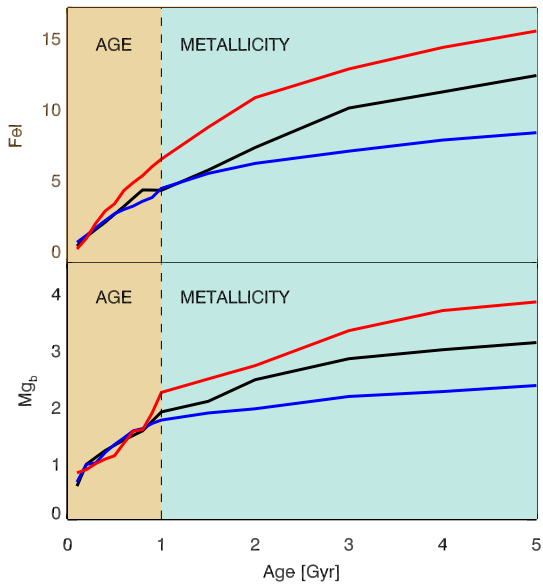}  
\caption{Examples of the different trends with age of two representative spectral indices, i.e. Fe{\small\texttt{I}}(3000) and Mg$_b$, for MS11 and TMJ models respectively, for different stellar
metallicities: subsolar (blue lines), solar (black lines) and supersolar (red lines). For ages $<1$ Gyr both indices are solid \emph{age}-indicators, while for older ages they become better
\emph{metallicity}-indicators.}
\label{exampleindices}
\end{centering}
\end{figure}

Thus, we attempted a separation between more \emph{age}-dependent and \emph{metallicity}-dependent indices
even if - it should be noted - it is not possible to perfectly disentangle the two parameters and the dominance of age and/or metallicity sensitivity depends on the explored ages range. 
As an example, in Fig. \ref{exampleindices}, the trends of two indices, one UV index,
Fe{\small\texttt{I}}(3000), and one Lick index, Mg$_b$(5175), are shown as a function of the age of the stellar population for three different stellar metallicity values, subsolar (blue lines), solar (black lines) and
supersolar (red lines). We can see that for ages $<1$ Gyr both indices are solid \emph{age}-indicators showing a monotonic increase with age, with negligible metallicity effects, while for older ages the age-dependence weakens (due to the milder evolution of the Main sequence and Red Giant Branch, e.g. \citealt{ma2005}) and become better  \emph{metallicity}-indicators.

In addition (see Section \ref{analisimetall}), it has not been possible to measure all indices for all objects due to their different redshift and spectrum quality. This means that the sets of indices used in the analysis are different for each object. 

Therefore, in order to avoid meaningless comparisons, the most efficient way to extract information from the many spectral indices was to perform a single-case analysis for each index and for each object. For each object we classified each index as (mostly) \emph{age}-indicator or (mostly) \emph{metallicity}-indicator. 
Note that we have checked that if all indices were analysed simultaneously, the results would be less constrained (i.e. with larger uncertainties) since residuals of the 
age-metallicity degeneracy are left. A similar effect is also found if a separation between blue indices and Lick indices is adopted: the results point to the same parameters values but the degeneracy is more marked.

\begin{figure*}
\begin{centering}
\includegraphics[width=8.8cm]{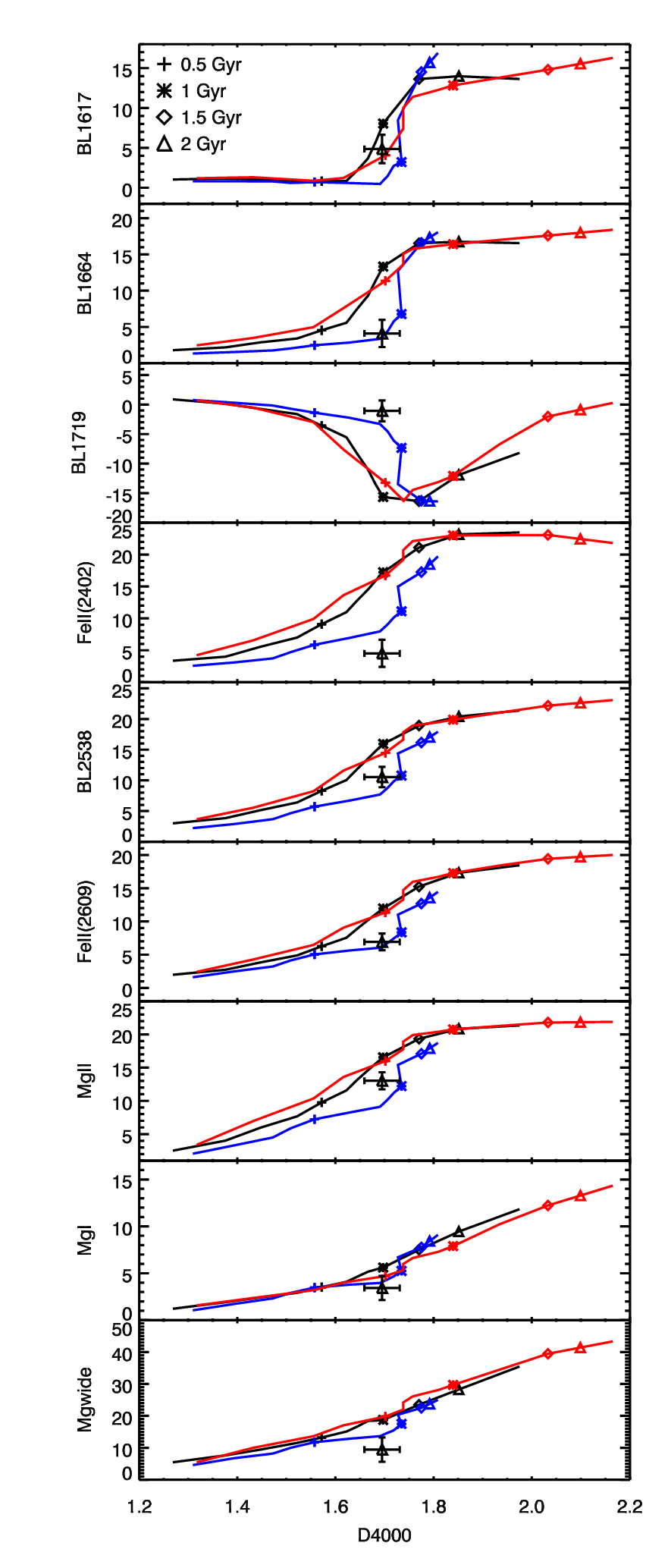}
\includegraphics[width=8.8cm]{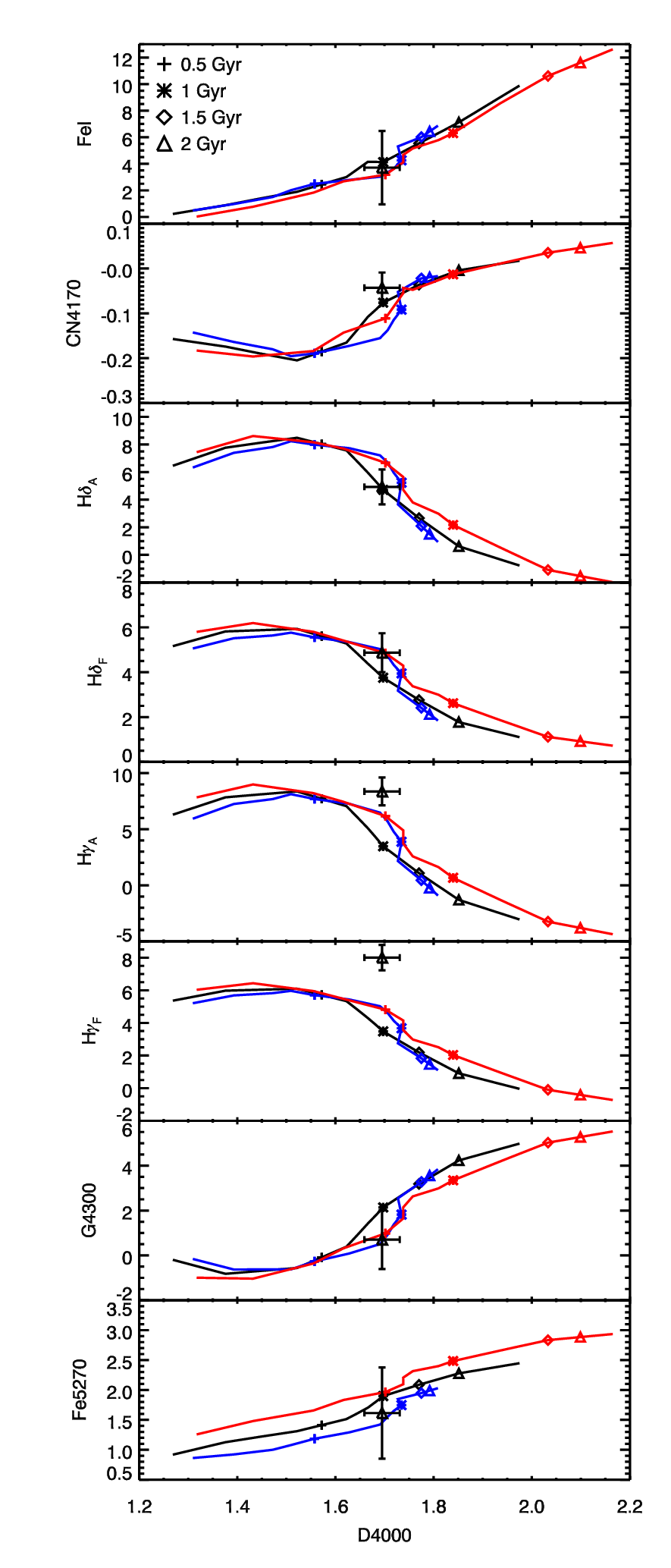}
\caption{UDS-19627. Spectral indices as a function of the D4000 break, with measured data showed as black diamonds with error bars, are compared to stellar population models. Blue, black and red lines refer to sub-solar, solar and super-solar metallicity. MS11 models are adopted for blue-UV indices and TMJ models for Lick indices. Ages run from $0.1$ to the limit of the Universe age at this redshift, intermediate age values are highlighted with different symbols. Index values are corrected for the measured velocity dispersion
(see Table \ref{tab:infoall}).}
\label{19627usati}
\end{centering}
\end{figure*}

\begin{figure*}
\begin{centering}
\includegraphics[width=8.8cm]{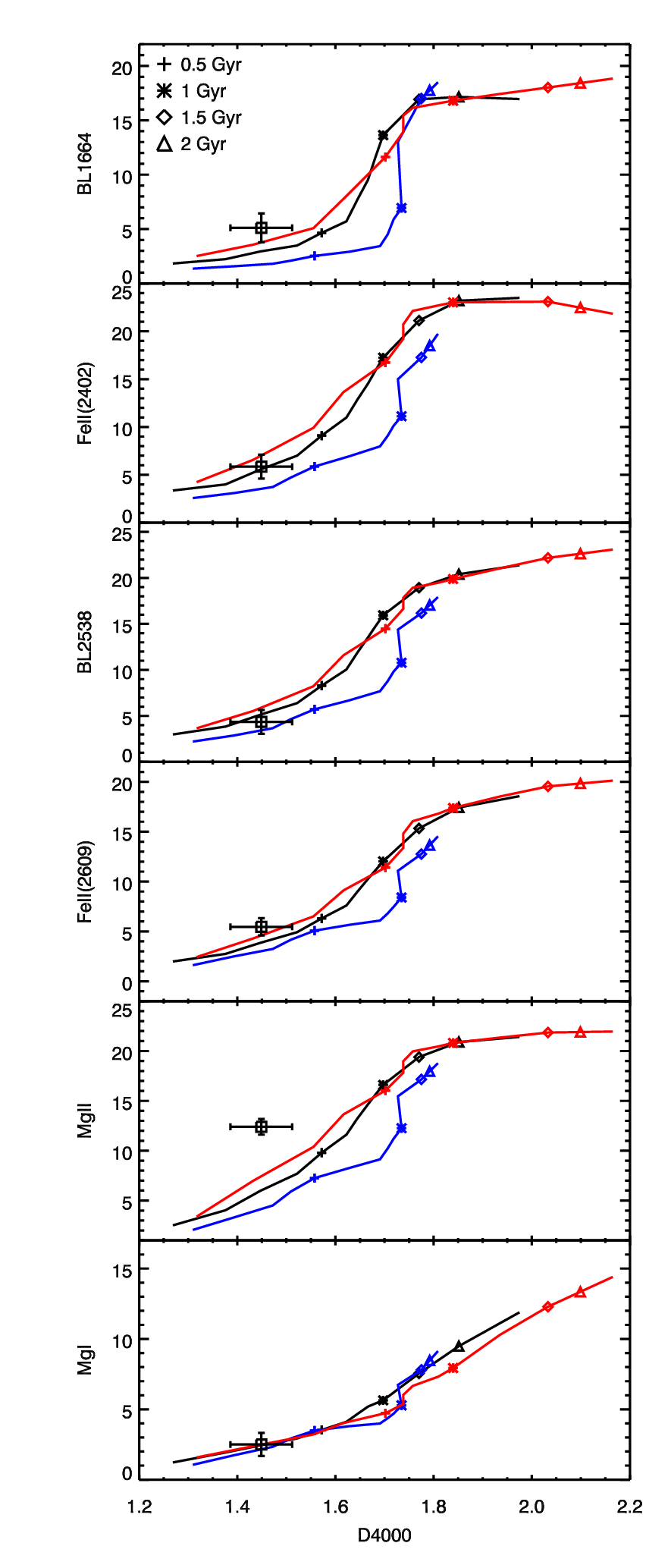}
\includegraphics[width=8.8cm]{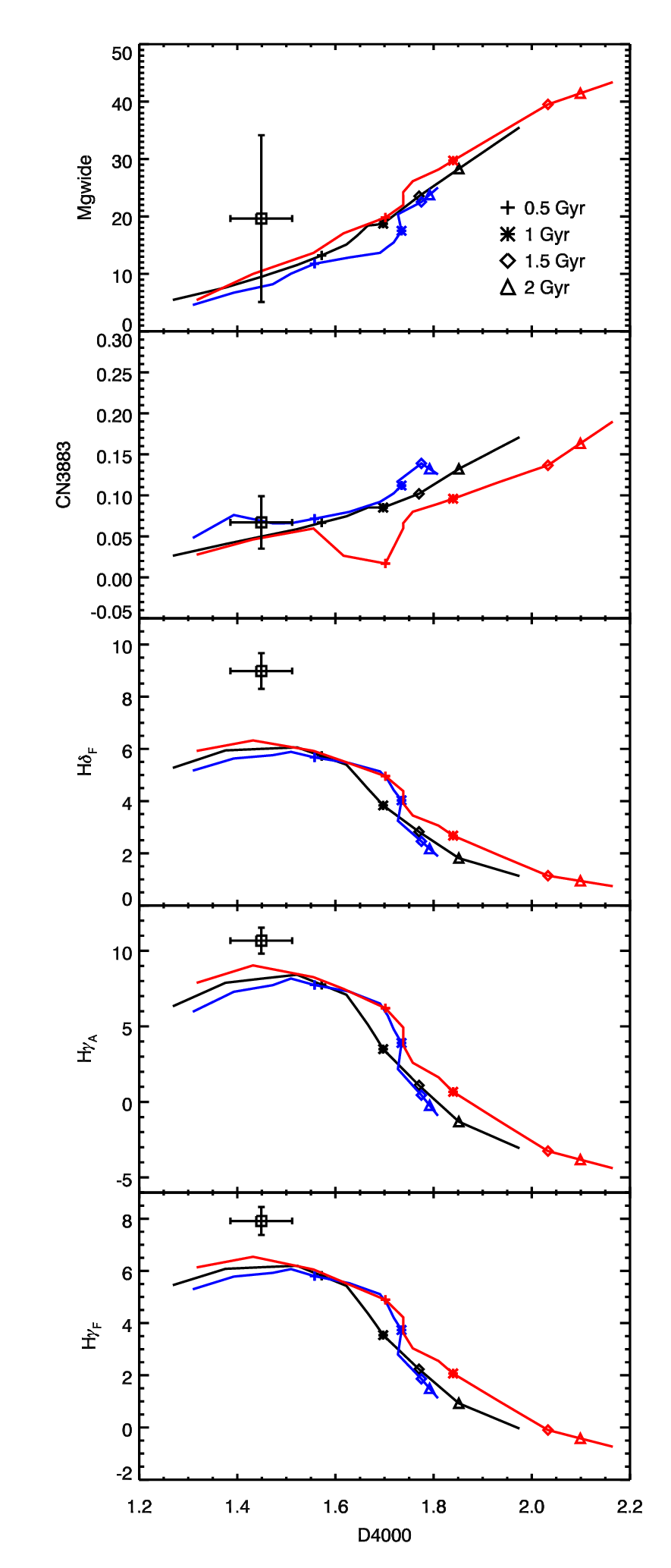}
\caption{As in Fig. \ref{19627usati} for COSMOS-7447.}
\label{7447usati}
\end{centering}
\end{figure*}

\begin{figure}
\begin{centering}
\includegraphics[width=8.8cm]{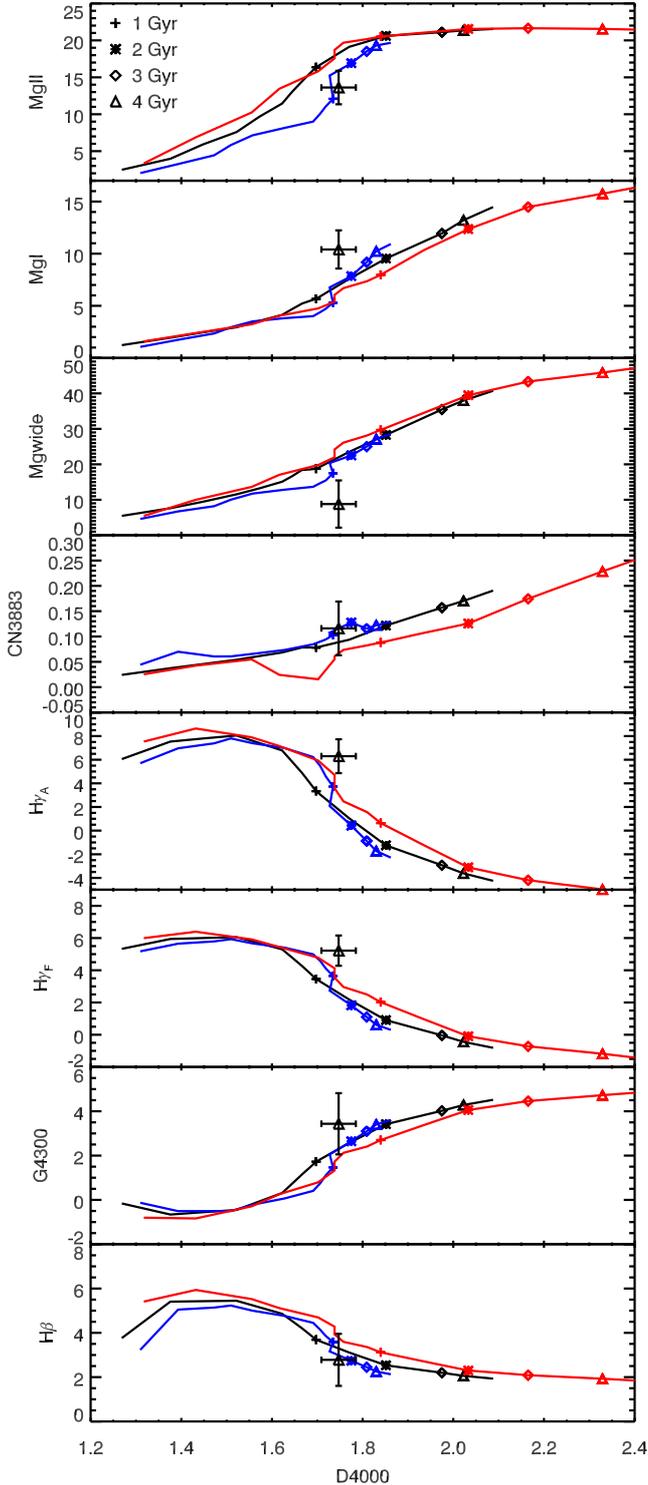}
\caption{As in Fig. \ref{19627usati} for UDS-29410.}
\label{29410usati}
\end{centering}
\end{figure}

\begin{figure*}
\begin{centering}
\includegraphics[width=8.8cm]{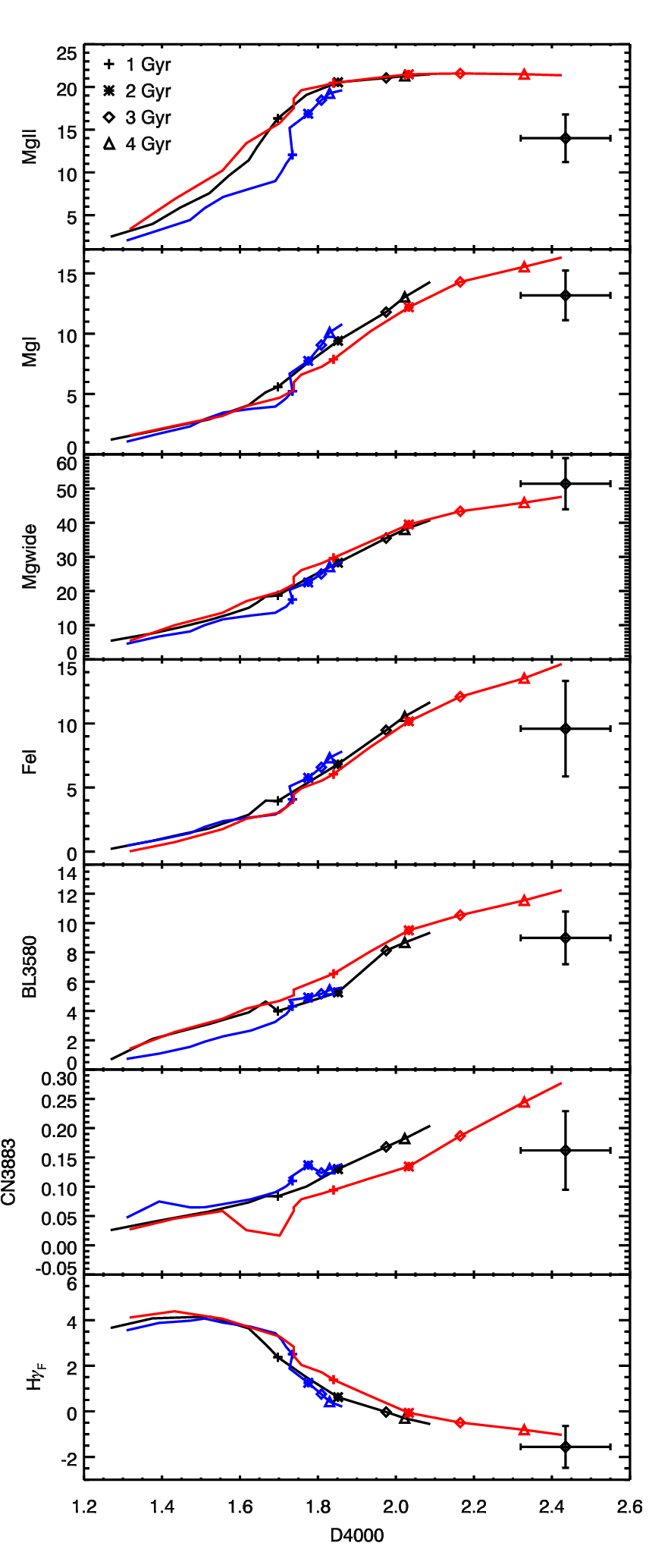}
\includegraphics[width=8.8cm]{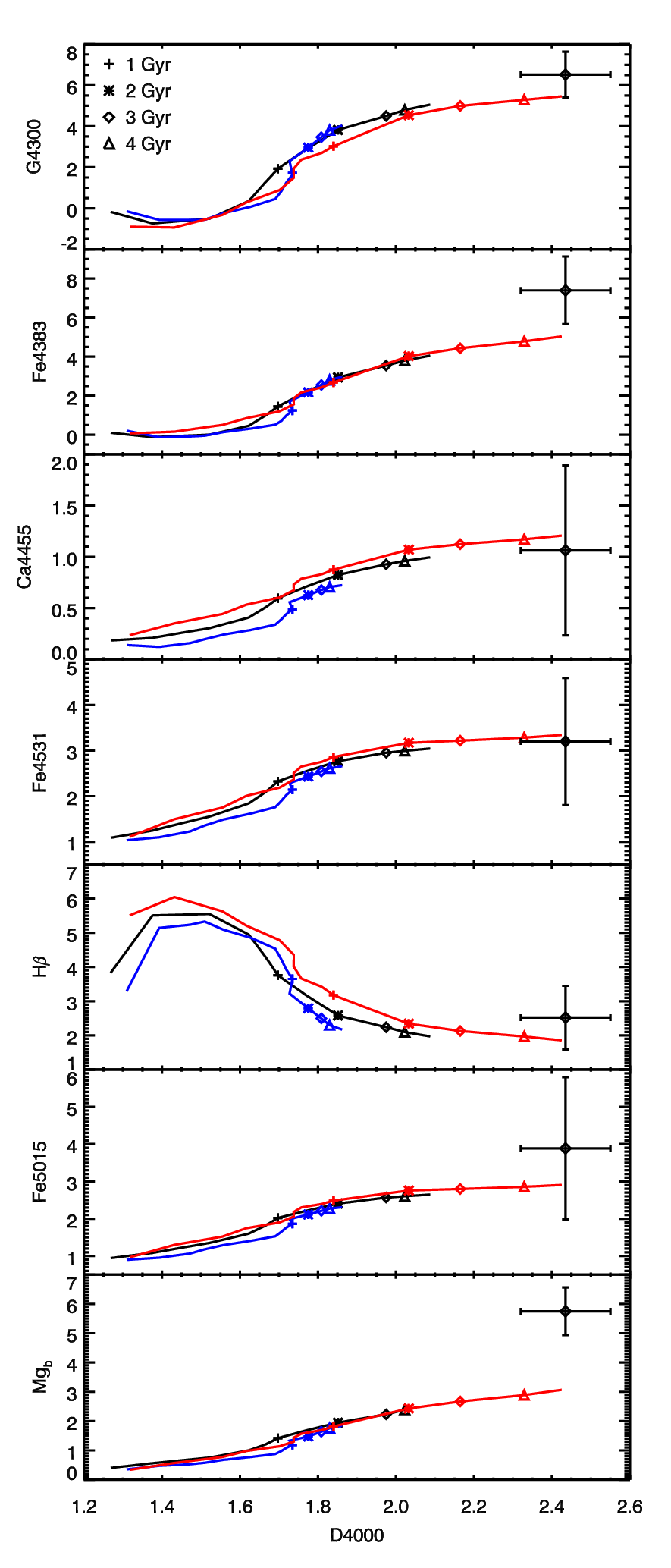}
\caption{As in Fig. \ref{19627usati} for COSMOS-307881.}
\label{307881usati}
\end{centering}
\end{figure*}

In Fig. \ref{19627usati}-\ref{307881usati} for each galaxy the measured indices (black symbols with error bars) used in the final analysis are shown as a function of the most solid
age-dependent broad index D4000, in comparison with the expectations of stellar population models (coloured lines for three different stellar metallicity values), MS11 models for blue-UV indices and TMJ models for Lick indices. 

For the analysis of each single object we proceeded as follows. We firstly compared by means of a $\chi^2$ minimization the values of the set of \emph{age}-dependent
indices with the merged MS11-TMJ models (described above) assuming ages from $0.1$ Gyr to the limit of the age of the Universe determined by each galaxy's redshift, and 
metallicity from $[Z/H]=-0.30$ to $0.30$. During these comparisons, index values have been corrected to remove the intrinsic broadening of the galaxy velocity dispersion which affects absorption lines. Index values were multiplied by correction factors, which were obtained by measuring indices on models broadened to the relevant velocity dispersions.

The best-fit solution and the distribution of all fitting-solutions with similar associated probabilities provided an estimate of the stellar population age with its error \citep{mio2}. The same procedure was repeated with the set of \emph{metallicity}-dependent indices, but now including the constraint from the age analysis, i.e. ages are allowed to run only in the range indicated by the \emph{age}-dependent indices, and the best fitting metallicity value is obtained within this range. 
Results are summarized in Table \ref{tab:allresults}.  First of all, we notice that our spectral ages are in excellent agreement with the results obtained from the photometric fitting shown in Table \ref{tab:infoall}, and discussed in Section \ref{sec:photometry}. 

In a second step, we repeated the whole procedure excluding the contribution of Balmer lines indices (e.g. H$\beta$, H$\gamma$, H$\delta$) in order to verify that possible residuals of star formation,
which could affect these indices, or any younger minor stellar component would not alter the results. As can be seen in Table \ref{tab:allresults}, the results
obtained in the two cases, with and without the Balmer lines indices, are completely consistent. This result may also be suggestive that Balmer lines are not crucial to constrain ages when all other spectral indices are available. 


\begin{table*}
\centering
 \caption{Best-fit values of age and metallicity with their reduced $\chi^2$ values obtained from the independent analysis of the set of \emph{age}-dependent (first lines) and \emph{metallicity}-dependent indices (second lines). The ``NO Balmer'' columns refer to the analysis performed excluding the Balmer line indices (see text).}
 \label{tab:allresults}
 \begin{tabular}{llcccccc}
 \hline
 ID & Set of used indices & Age   & [Z/H]  & $\chi^2$  & Age (Gyr)  & [Z/H]            & $\chi^2$      \\
    &                     & (Gyr) &        &           & NO Balmer  &  NO Balmer       &               \\
 \hline
 19627  &  [Mg{\small\texttt{I}}, Mg$_{wide}$, Fe{\small\texttt{I}}(3000), & 0.6$\pm0.1$ & & 2.40 &  0.7$\pm0.1$ & & 0.30 \\
        &  D$4000$, H$\delta_A$, H$\delta_F$,&&&&&&\\
     	&  H$\gamma_A$, H$\gamma_F$, Gband] &&&&&&\\
        &  [BL($1617$), BL($1664$), BL($1719$), & & -0.30$_{-0.0}^{+0.1}$ & 2.69 &&&\\
        &  Fe{\small\texttt{II}}($2402$), BL($2538$), Fe{\small\texttt{II}}($2609$),&&&&&&\\
	&  Mg{\small\texttt{II}}, CN($4170$), Fe($5270$)]  &&&&&&\\
 \hline 
 7447   &  [Mg{\small\texttt{I}}, CN($3883$), D$4000$,  & 0.2$_{-0.05}^{+0.2}$ & & 4.30 & 0.3$\pm0.1$ & & 0.04 \\ 
        &  H$\delta_F$, H$\gamma_A$, H$\gamma_F$] &&&&&&\\
        &  [BL($1664$), Fe{\small\texttt{II}}($2402$), BL($2538$), & & 0.26$_{-0.4}^{+0.04}$ &  4.44  &&&\\
        &  Fe{\small\texttt{II}}($2609$), Mg{\small\texttt{II}}, Mg$_{wide}$] &&&&&&\\
 \hline 
 29410  &  [D$4000$, Gband, H$\beta$] & 1.8$\pm$0.3 & & 0.21  & 1.8$\pm$0.4 && 0.30  \\ 
        &  [Mg{\small\texttt{II}}, Mg{\small\texttt{I}}, Mg$_{wide}$, & & -0.30$_{-0.0}^{+0.2}$ & 2.94 &  & -0.30$_{-0.0}^{+0.1}$ & 1.87  \\ 
        &  CN($3883$), H$\gamma_A$, H$\gamma_F$] &&&&&&\\ 
 \hline
 307881 &  [CN($3883$), D$4000$, H$\beta$]  &  4.1$\pm0.4$ &      &   0.90    & 4.2$\pm0.3$           &           & 1.17 \\
        &  [Mg{\small\texttt{II}}, Mg{\small\texttt{I}}, Mg$_{wide}$, & &  0.30$_{-0.14}^{+0.0}$ & 2.40 & & 0.24$\pm0.08$ & 2.56 \\
        &  Fe{\small\texttt{I}}(3000), BL($3580$), H$\gamma_F$, &&&&&&\\
	&  Gband, Fe($4383$), Ca($4455$), &&&&&&\\
	&  Fe($4531$), Fe($5015$), Mg$_b$] &&&&&&\\
 \hline
\end{tabular}
\end{table*}

An estimate of $\alpha$-enhancement can be derived from suitable Lick indices using TMJ models which are calculated for various values of this parameter. However, this measurement is challenging for distant galaxies due to the low S/N ratio of the spectra. In order to understand the feasibility of such measurement given our data quality, we determine the stellar population parameters on mock spectra simulated to have the same S/N as the data, using the exact same set of indices as in the real data analysis (see Appendix \ref{appendixA}). We find that we are able to constrain the [$\alpha$/Fe] values for 3 out of 4 objects, albeit with rather large uncertainties. 
We estimated the $\alpha$-enhancement on our data by comparing the observed Lick indices with TMJ models again through a $\chi^2$ minimization process, with age running again from $0.1$ Gyr to the limit of the age of the Universe at step of $0.1$ Gyr, covering a wider range
of metallicities from $[Z/H]=-2.25$ to $0.67$ at step of $0.01$, and with [$\alpha$/Fe] values in the range from $-0.30$ to $+0.50$ at step of $0.01$. 

In Appendix \ref{appendixB} we compare the best-fit model indices (both from the combined MS11-TMJ and from the sole TMJ models) with the observed ones. We find that most of them agree well within $1\sigma$ and more generally they are consistent within $2\sigma$.

The resulting stella population parameters are shown in Table \ref{tab:allresultsalpha}. As expected, and also confirmed by the simulations described in Appendix \ref{appendixA}, it was not possible to constrain the stellar population parameters for object 7447 with only Lick indices as too few could be measured on the observed spectrum (see Table \ref{tab:allindices}). 
From Table \ref{tab:allresultsalpha} it can be immediately seen that galaxy ages are well consistent between the two different analysis, which is an important result as galaxy ages are key to galaxy formation and evolution and also to use galaxies as cosmological probes (e.g. as cosmic chronometers, \citealt{moresco16}).

All results from Table \ref{tab:allresults} and \ref{tab:allresultsalpha} will be discussed in more detail in Section \ref{all:discussion}. 

\begin{table*}
\centering
 \caption{Results from the analysis of the sole Lick indices using TMJ models, which allow us to also determine the [$\alpha$/Fe] parameter together with age and metallicity. For object 7447 it is not possible to retrieve the stellar parameters due to the too small number of available Lick indices (see also Appendix \ref{appendixA}). }
 \label{tab:allresultsalpha}
\begin{tabular}{llccc}
 \hline
 ID & Set of used indices & Age   & [Z/H]  & [$\alpha$/Fe]       \\
    &                     & (Gyr) &        &                      \\
 \hline
 19627  &  [D$4000$, H$\delta_A$, H$\delta_F$, & 0.7$_{-0.2}^{+0.3}$ & -0.82$_{-0.6}^{+0.3}$ & -0.18$_{-0.12}^{+0.2}$ \\
	&  H$\gamma_A$, H$\gamma_F$,Gband, &&& \\
	&  Fe($5270$)]  &&& \\
 \hline 
 7447   &  [D$4000$, H$\delta_F$, & - & - & - \\ 
        &  H$\gamma_A$, H$\gamma_F$] &&& \\
 \hline 
 29410  &  [D$4000$,Gband,H$\beta$, & 1.2$_{-0.7}^{+0.6}$ & -0.70$_{-0.3}^{+0.2}$ & 0.42$_{-0.1}^{+0.08}$  \\ 
        &  H$\gamma_A$,H$\gamma_F$] &&& \\ 
 \hline
 307881 &  [D$4000$, H$\gamma_F$, Gband, &  4.1$_{-0.8}^{+0.5}$ &  0.61$_{-0.05}^{+0.06}$    &   0.45$_{-0.19}^{+0.05}$  \\
        &  H$\beta$, Fe($4383$), Ca($4455$), &&& \\
        &  Fe($4531$), Fe($5015$), Mg$_b$] &&& \\
 \hline
\end{tabular}

\end{table*}

A special case is that of COSMOS-307881 as already discussed in \citet{mio2}. The previous paper by analyzing only the Lick indices concluded for a very metal rich and $\alpha$-enhanced stellar population, as recalled in Table \ref{tab:allresultsalpha}. The extended analysis presented here includes the information brought by bluer indices and it confirms the galaxy old age and high metallicity at the limits of the merged modelling (hereafter for this galaxy we will consider the metallicity value found from Lick indices with TMJ models 
[Z/H]=0.61$_{-0.05}^{+0.06}$ as in \citealt{mio2}). At the same time, we noticed that some UV indices are not consistent with any of the adopted models (Fig. \ref{307881usati}), showing systematically lower values that a metallicity cannot explain. 

In order to find an explanation for these values we further compare the UV indices with models including a UV-upturn, in the next section (\ref{sub:upturn}).

\subsection{A possible UV up-turn in COSMOS-307881}
\label{sub:upturn}

The emission in the UV region is dominated by the hotter component of a galaxy stellar populations. Generally, the hottest component originates from the youngest stars (age $<1$ Gyr). However, also old stellar populations (age $>1$ Gyr) can become UV-bright after a post main-sequence phase of sufficient mass loss (\citealt{greggio}, \citealt{maraston00}). This phenomenon causes the so-called UV up-turn (see \citet{yi} for a review). The effect of the UV up-turn on the spectrum shape is clearly visible in the far UV (around $1000-2500$ \AA, see \citealt{burstein88}). \citet{claire} show that also spectral indices in the $\sim2500$ \AA $ $ region are affected by this phenomenon (e.g. Mg{\small\texttt{II}}(2800), Mg{\small\texttt{I}}(2852) and Fe{\small\texttt{I}}(3000)).

The origin of the UV-upturn is still under debate. One of the most accredited possibilities, proposed by \citet{greggio}, is that post main-sequence 
stars with an enough high-metallicity content do increase significantly the opacity in their stellar atmospheres thus causing mass loss during their horizontal branch phase. Consequently, the 
internal hotter shells become exposed and are responsible for a UV emission. 

COSMOS-307881 is the best candidate in our sample where to search for the presence of an UV up-turn because it is the oldest galaxy and stellar populations need to age in order to possess low-mass stars which can become hot emitters. Moreover, this galaxy has got the highest metallicity which should also be a favouring factor for the onset of an upturn.

\begin{figure}
\begin{centering}
\includegraphics[width=\columnwidth]{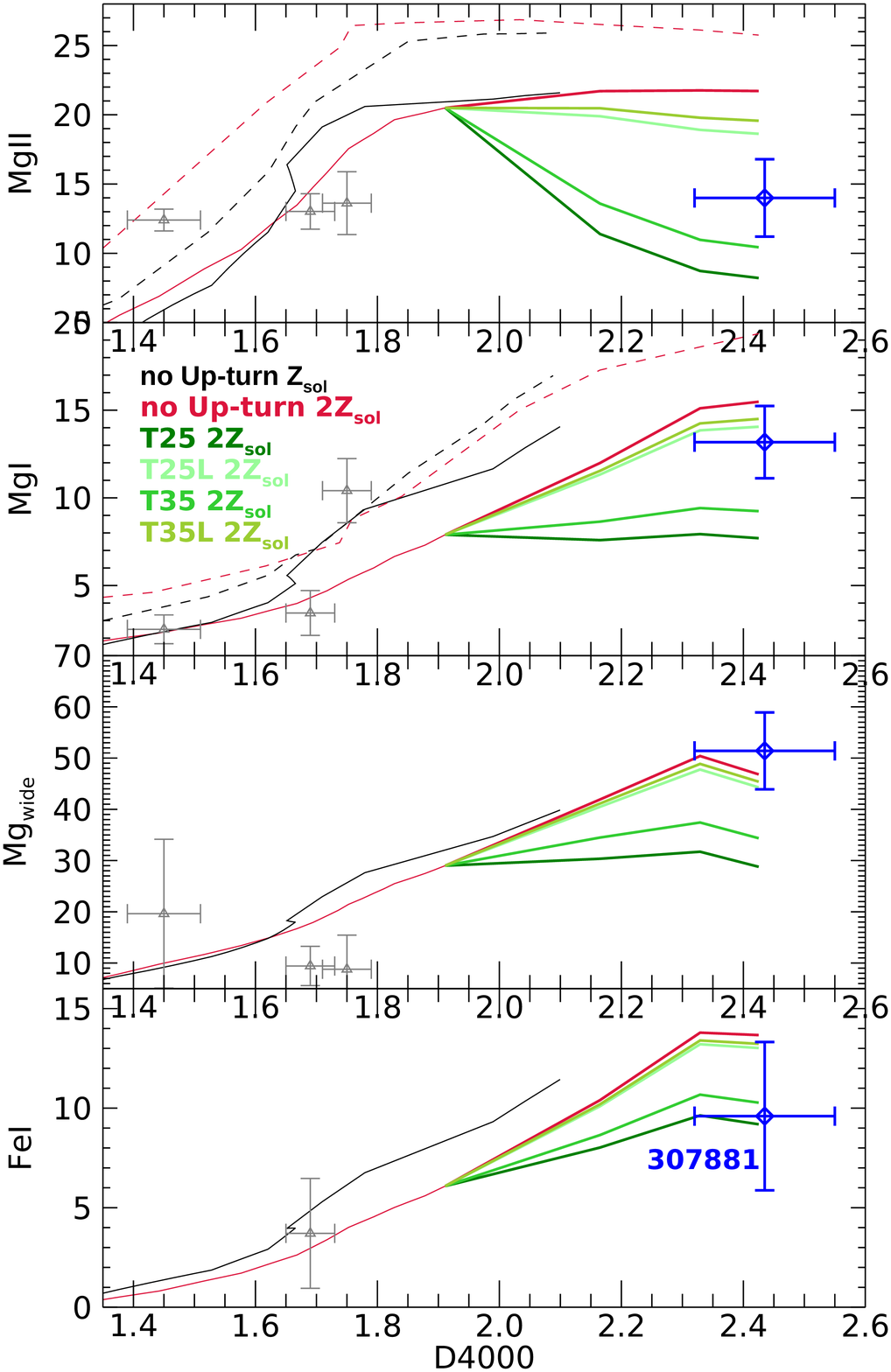} 
\caption{Mg{\small\texttt{II}}, Mg{\small\texttt{I}}, Mg$_{wide}$ and Fe{\small\texttt{I}}(3000) vs D4000 for COSMOS-307881 (blue symbols with error bars) compared to single-burst stellar population models with metallicities Z$_{_\odot}$ (black) and 2Z$_{\odot}$ (red). Solid and dashed lines refer to solar-scaled and super-solar values of [$\alpha$/Fe]. Green lines show models with UV-upturns for various combinations of 
temperature (T$_{eff}$/K) and fuel ($f/$M$_{\odot}$). Details on up-turn models can be found in \citet{claire}. Grey points show the index values for the other galaxies analyzed in this work. No UV upturn is required to model the indices of these other galaxies.}
\label{upturn}
\end{centering}
\end{figure}

Most importantly, UV indices (see Fig. \ref{upturn}): 
a discrepancy between the measured values of some UV indices (e.g. Mg{\small\texttt{II}}(2800), Mg{\small\texttt{I}}(2852) and Fe{\small\texttt{I}}(3000)) 
and the models previsions which cannot be explained even assuming higher metallicity values has been found. In particular, Mg{\small\texttt{II}}, Mg{\small\texttt{I}} and Fe{\small\texttt{I}} values 
are lower than expected (black and red solid lines in Fig. \ref{upturn}), pointing toward younger ages which are not consistent with the indication of the D4000 and with the age resulted from the all indices analysis.

Recently, \citet{claire} has explored the occurrence of the UV up-turn in a wide sample of massive, red and passive galaxies from \emph{z}$\sim0.6$ up to \emph{z}$\sim1$ from SDSS-III/BOSS \citep{dawson}, finding that a 
significant fraction of the sample ($\sim40\%$) shows the signatures of this phenomenon. In order to reveal the UV up-turn effect, the authors investigated UV spectral indices as well as full spectral fitting, using stellar population models calculated with different assumptions for the up-turn component, as we recall in Section \ref{sec:models}. The effect of the UV-upturn starts at ages older than $1$ Gyr (see \citet{claire} for further details).
Here we compare these models with the Mg$_{wide}$, Mg{\small\texttt{II}}(2800), Mg{\small\texttt{I}}(2852) and Fe{\small\texttt{I}}(3000), which \citet{claire} find to be sensitive to the UV up-turn effect and whose values for COSMOS-307881 cannot be reproduced by any standard model at whatever metallicity. The comparison is shown in Fig. \ref{upturn}. 

It is clear that standard models (black and red lines) are not able to explain the measured values, independently of metallicity. For Mg{\small\texttt{II}}(2800) and Mg{\small\texttt{I}}(2852), we also verified that an enhanced
$\alpha$/Fe ratio - as derived from the optical absorptions (Table \ref{tab:allresultsalpha}) - does not help explaining the detected discrepancy. This is show by models (dashed lines) in which we simulated the effect of [$\alpha$/Fe]$=0.3$
 on the UV indices by applying to these indices the differential index variation with [$\alpha$/Fe] obtained from the TMJ models.
 
Only models including a UV up-turn (green lines) are able to reach the observed values. With different tones of green we display a range of calculations with ages $> 1$ Gyr, a high 2Z$_{_\odot}$ metallicity and various assumptions for the temperature (T$_{eff}=25000,35000$ K) and fuel consumption (f$=6.5 \times 10^{-3},6.5 \times 10^{-2}$ /M$_{\odot}$) in the upturn. The best match would probably be achieved by a tuned mix of these two parameters.

The presence of a bright-UV old stellar population also helps explaining the detected [O{\small\texttt{II}}](3727\AA) weak emission, as already hinted in \citet{mio2}. Indeed, by locating COSMOS-307881 on a BPT diagram involving [O{\small\texttt{II}}](3727\AA), [O{\small\texttt{III}}](5007\AA) (null flux) and H$\beta$
(in absorption) as shown in \citet{thomas13}, we could classify this object as a LINER (Low-ionization nuclear emission-line region) galaxy. From optical spectroscopic studies it is known that a significant fraction of old elliptical galaxies belong to the LINERs population \citep{yan}, although 
the ionization mechanism in LINERs is still under debate (\citealt{annibali}, \citealt{belfiore16}). One of the most accredited scenario is that photo-ionization by old post-asymptotic giant
branch stars is the responsible of the atomic excitations since their radiation is able to reproduce the observed emission-line ratios in LINERs
\citep{trinchieri}. Our hypothesis is that the same old population which gives origin to the UV up-turn also
contributes to the excitation of some residual gas in the interstellar medium of this ETG, thus causing the detected [O{\small\texttt{II}}] weak emission.

Fig. \ref{upturn} also shows as grey points the index values for the other galaxies in our sample. It is clear that no UV upturn is required to model the indices of these other galaxies, which are all younger than COSMOS-307881. This result is then supportive of the idea that evolved, hot stellar components giving origin to the UV-upturn pertain to old galaxies, where the typical evolutionary mass is small enough such that mass-loss can remove stellar envelope effectively and shift the effective temperature towards high values.


\section{Full-spectrum fitting}
\label{sub:fsf}

An additional analysis we performed was a full-spectrum fitting using the code 
\firefly \citep{wilkinson_et_al_2017}. Briefly, \firefly\footnote{\url{www.icg.port.ac.uk/firefly}} 
(\textsf{F}itting \textsf{I}te\textsf{R}ativ\textsf{E}ly \textsf{F}or \textsf{L}ikelihood 
anal\textsf{Y}sis) is a full-spectrum fitting code based on $\chi^2$ minimization for comparing 
combinations of single-burst models (SSPs) to a given observed or mock spectrum (of a galaxy, a 
star cluster, etc.). \firefly iteratively fits these combinations controlling them with a Bayesian 
information criterion until convergence in order to avoid overfitting. This code does not require 
any other prior than the assumed grid of population models. Dust attenuation is derived 
directly from the data before the fitting such that the continuum is corrected before the stellar 
population fitting is performed. This step helps fixing problems related to dust but also poor 
flux calibration of the data. The output of \firefly consists of age and metallicity (light- and 
mass-weighted), stellar masses and their partition into remnants, and E(B-V) values, for the 
combined model fit as well as the individual SSP components. For a full description and details of 
the algorithm, and testing of the code performance on mocks of different S/N ratios, we refer the 
reader to \citet[][]{wilkinson_et_al_2017}, and for verification of its performance for a wide 
variety of galaxies and signal-to-noise ratios (down to S/N $\sim 5$) to 
\citet[][]{comparat_et_al_2017}. We shall return on this when interpreting the results.

The models that \firefly used to perform the full-spectrum fitting for this work, were the MS11-MILES 
UV-extended version of the MS11 models with a Salpeter IMF. Ages are constrained to be younger than the age of the 
Universe at each galaxy redshift, as we did in the spectral index and photometric analysis. Metallicity values run from [Z/H]$=-0.3$ to $0.3$. 
With the data sample already at the 
same resolution as the models, the code 1. obtained the linear combination of the models that best 
fits the spectrum, 2. computed the contributions of the stellar populations, 3. converted the 
$\chi^{2}$ into probabilities, and 4. calculated the average properties and errors. 

\begin{figure*}
\begin{centering}
\includegraphics[width=16cm]{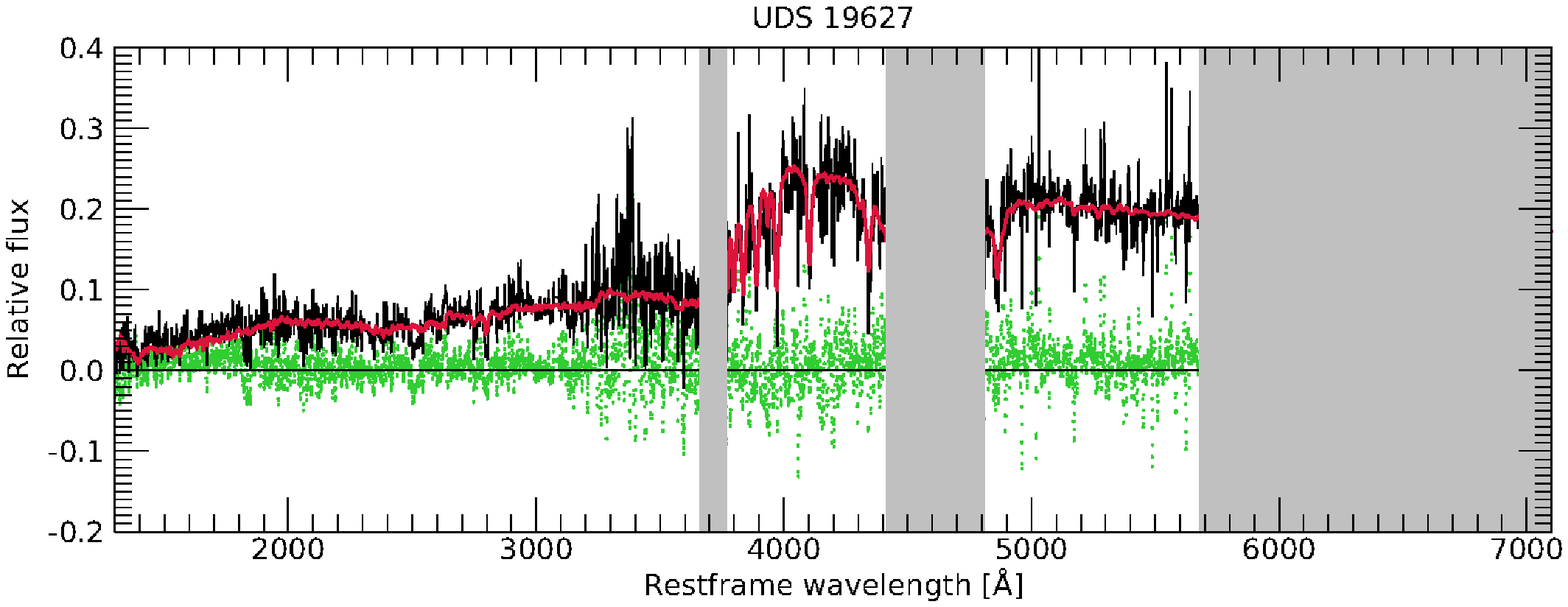}
\includegraphics[width=16cm]{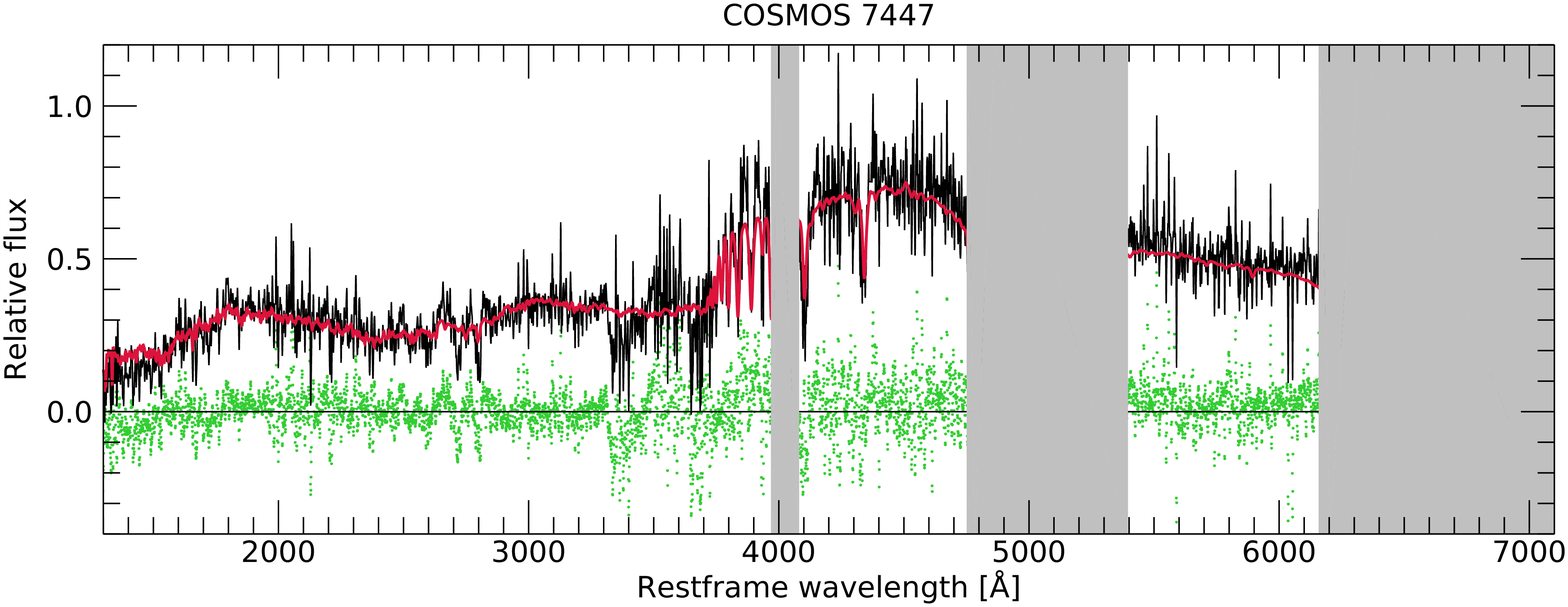}
\includegraphics[width=16cm]{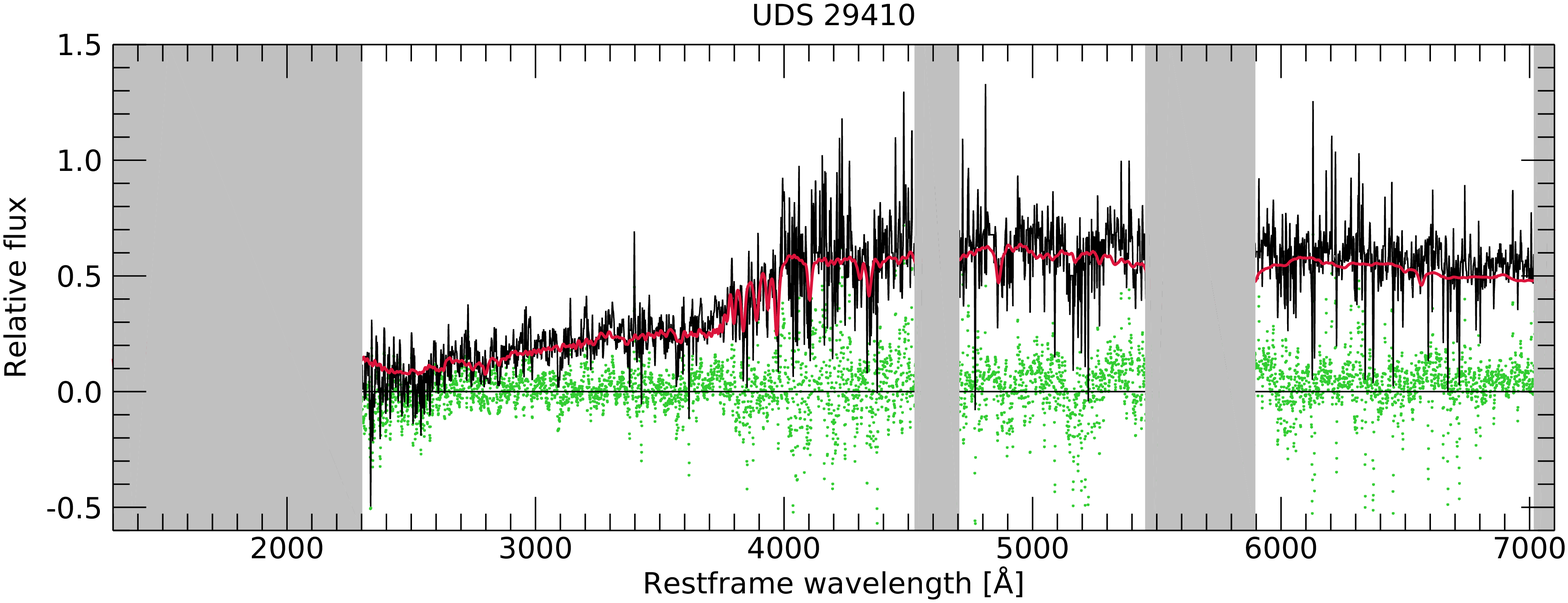}
\includegraphics[width=16cm]{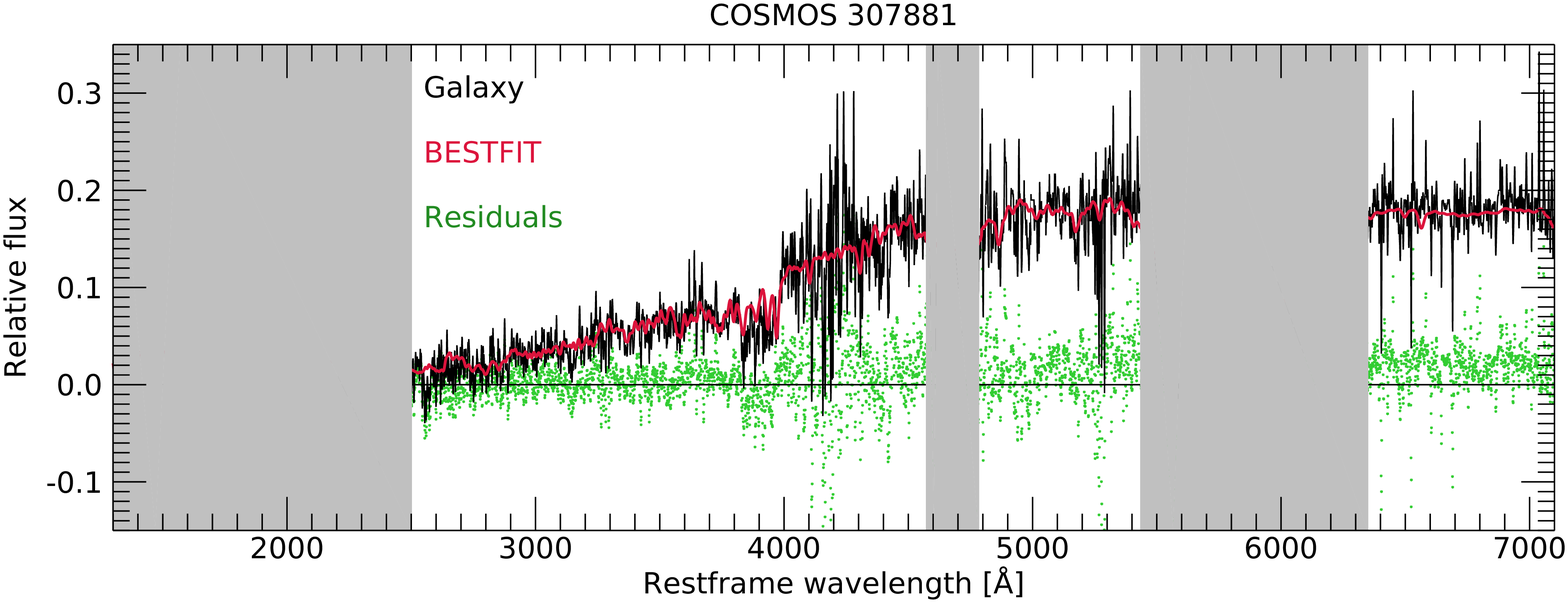}
\caption{\firefly fits of the sample galaxies using UV-extended MS11-MILES models. 
The black spectrum represents the data, the red one is the best-fit model and green residuals. Grey areas are sections of the spectrum that 
were not considered in the fit.}
\label{bestfits}
\end{centering}
\end{figure*}

The mass- and light- weighted ages and metallicities obtained with \firefly for our ETG sample are given in
Table \ref{tab:FSF}, together with the reddening $E(B-V)$. The fits are shown in Fig. \ref{bestfits}. 
The full spectral fitting describe the ETGs as young, except for COSMOS-307881, which results to be $\sim 4$~Gyr old. Metallicities are slightly sub-solar with the oldest galaxy having essentially a solar-metallicity. There is little difference between light-weighted and mass-weighted quantities, which signifies that the duration of star formation is short and galaxies are essentially passive or on the verge of becoming passive. The reddening is larger for the youngest objects and becomes small as the age increases, which is consistent with the absence of extended star formation. As we shall comment more extensively in the Discussion session, these results are consistent with the output of the index analysis described in Section \ref{analisimetall}.

We should stress that performing full spectral fitting of galaxy spectra with $S/N < 10$~is challenging. \cite{wilkinson_et_al_2017} show the results of extensive simulations on mock galaxy spectra with different S/N, star formation histories, reddening, etc. By glancing at their Figures 8, 9 and 11, we learn that when fitting the spectra of galaxies with $S/N=5$ and short star formation histories (single bursts and $\tau$-models with $\tau=0.1-1$~Gyr) in the age regime relevant to our galaxies, metallicity can indeed be underestimated, while age is overestimated. Indeed, ages between a few hundreds of millions to a few billion years are most prone to the age/metallicity degeneracy. While we should not push these comparisons too quantitatively, the low S/N has an effect on the degradation of results from full spectral fitting. And yet, as we shall discuss in Section \ref{all:discussion}, the overall agreement between results from different techniques is encouraging.

\begin{table*}
\begin{centering}
 \caption{\firefly full spectral fitting results: light and mass weighted age and metallicity and reddening E(B-V). Note that for 19627 the lower bound of the error on the light-weighted metallicity is formally zero because the determination coincides with the bottom of the model grid.}
 \label{tab:FSF}
 \begin{tabular}{lccccc}
 \hline
 ID & Light weighted age & Mass weighted age & Light weighted metallicity & Mass weighted metallicity & E(B-V) \\
    & (Gyr)              & (Gyr)             &                            &                           &         \\
 \hline
 19627  & 0.19$^{+0.12}_{-0.12}$  & 0.23$^{+3.08}_{-0.04}$  & -0.30$^{+0.17}_{-0.00}$ & -0.20$^{+0.10}_{-0.10}$    & 0.29  \\
 7447   & 0.15$^{+0.10}_{-0.03}$  & 0.21$^{+0.98}_{-0.05}$  & -0.12$^{+0.12}_{-0.08}$ & -0.18$^{+0.18}_{-0.07}$    & 0.27  \\
 29410  & 1.10$^{+0.93}_{-0.32}$  & 1.61$^{+2.03}_{-0.55}$  & -0.28$^{+0.15}_{-0.02}$ & -0.29$^{+0.25}_{-0.01}$    & 0.08  \\ 
 307881 & 3.93$^{+0.25}_{-0.43}$  & 4.89$^{+0.11}_{-0.40}$  & -0.05$^{+0.04}_{-0.02}$ &  0.00$^{+0.03}_{-0.02}$    & 0.14  \\
 \hline
 
\end{tabular}
\end{centering}
\end{table*}

\section{Photometric fitting}
\label{sec:photometry}

It is a general question whether broad-band photometry which is what is usually available at high-redshift is able to allow the determination of robust stellar ages of galaxies, and in general how ages determined via photometry or spectroscopy compare.

For this reason we have performed spectrophotometry template fitting to the photometry of our four galaxies and determined their photometric ages, chemical composition and stellar masses.

As in our previous works \citep{ma2006,ma2010} we use the \texttt{Hyperzspec} code \citep{bolzonella} combined with ancillary scripts for the calculation of the stellar mass (\citealt{daddi05}; \citealt{ma2006}), 
and a selection of $32$ sets of stellar population models based on the \citet{ma2005} evolutionary population synthesis with a Salpeter IMF, spanning a variety of star formation histories including single-burst simple stellar populations, exponentially declining star formation histories ($\tau$-models), truncated star formation and constant star formation. Each model star formation history spans a grid of 221 ages, and is available for four metallicities values between $\frac{1}{5} Z_{\odot}$ to $2Z_{\odot}$. 

We performed the fitting for three options for the description of dust reddening, from no reddening to dust reddening following the Calzetti's law \citep{calzetti00} or the so-called SMC law \citep{prevot84} as available in  \texttt{Hyperzspec}. The extinction parameter $A_{V}$ was allowed to vary between 0 and 3, in steps of 0.5. Furthermore, we used an age cut to retain only solutions older than 0.1 Gyr, which is commonly used in order to avoid age-dust degeneracy pushing the fits towards too low ages. It should be said that for these galaxies the setting or not of such limit has no influence on the derived properties. 

In performing the spectrophotometric fitting, we tried to use the same, or similar bands for all the objects, depending on their availability. Simulations of photometric fitting varying the number of filters and their wavelength extension (\citealt{pforr12}, \citealt{capozzi16}) demonstrate that the fitting results depend on these parameters. We also excluded bands that at the putative redshift of the object would sample wavelengths long ward the K-band rest-frame (following \citealt{ma2006}). This is because standard population models such as M05, BC03 do not model PAH or other effects long ward K hence the model fitting to those data is meaningless.
 
COSMOS 7447 \citep{vandesande11} was fitted using bands $u$ from CFHT/Megacam, $B$, $V$, $r$, $i$, and $z$ from Subaru/Suprime-Cam, $J$ and $K$ from UKIDSS, and [3.6], [4.5], [5.8], and [8.0] $\mu$m from Spitzer/IRAC.
For UDS 19627 \citep{toft12} we used the same bands as COSMOS 7447, with the exclusion of channels [5.8] and [8.0].
On the other hand, compared to COSMOS 7447, UDS 29410 \citep{galametz13} has band $R_{c}$ instead of $r$ among the Subaru/Suprime-Cam bands, and also has HST\/ACS F606W and HST\/WFC3 F125W.
Lastly, for COSMOS 307881 the same bands as for UDS 29410 were available with the exception of F606W and F125W.

\begin{figure*}
\begin{centering}
\includegraphics[width=14cm]{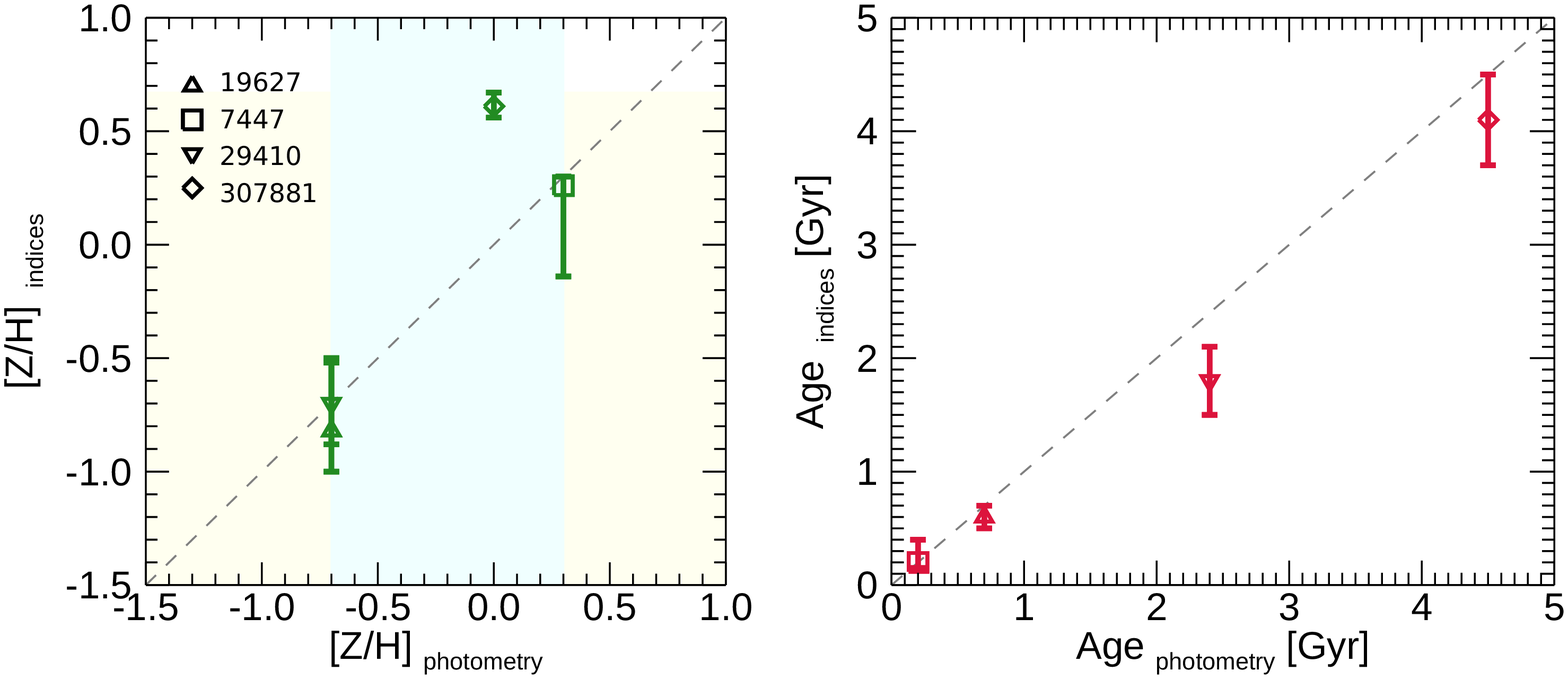} 
\includegraphics[width=14cm]{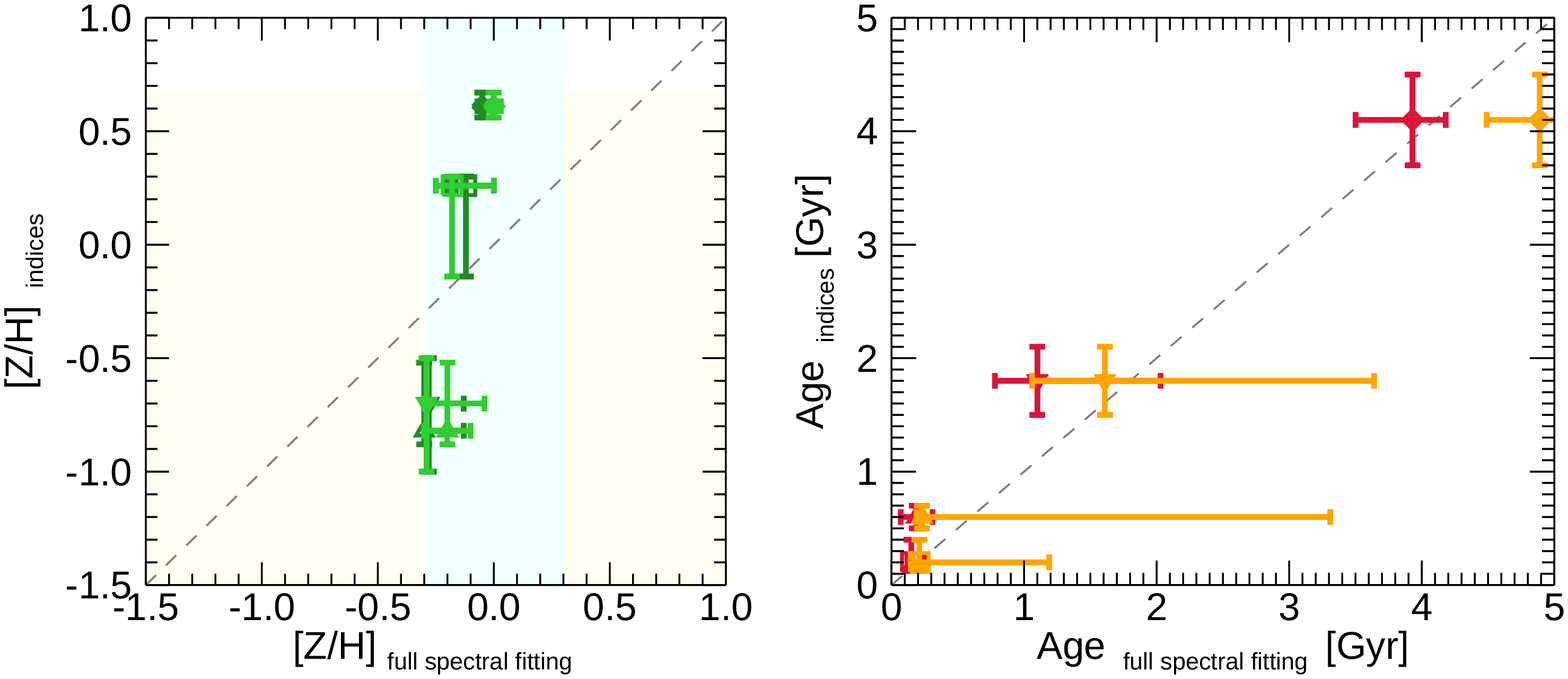} 
\includegraphics[width=14cm]{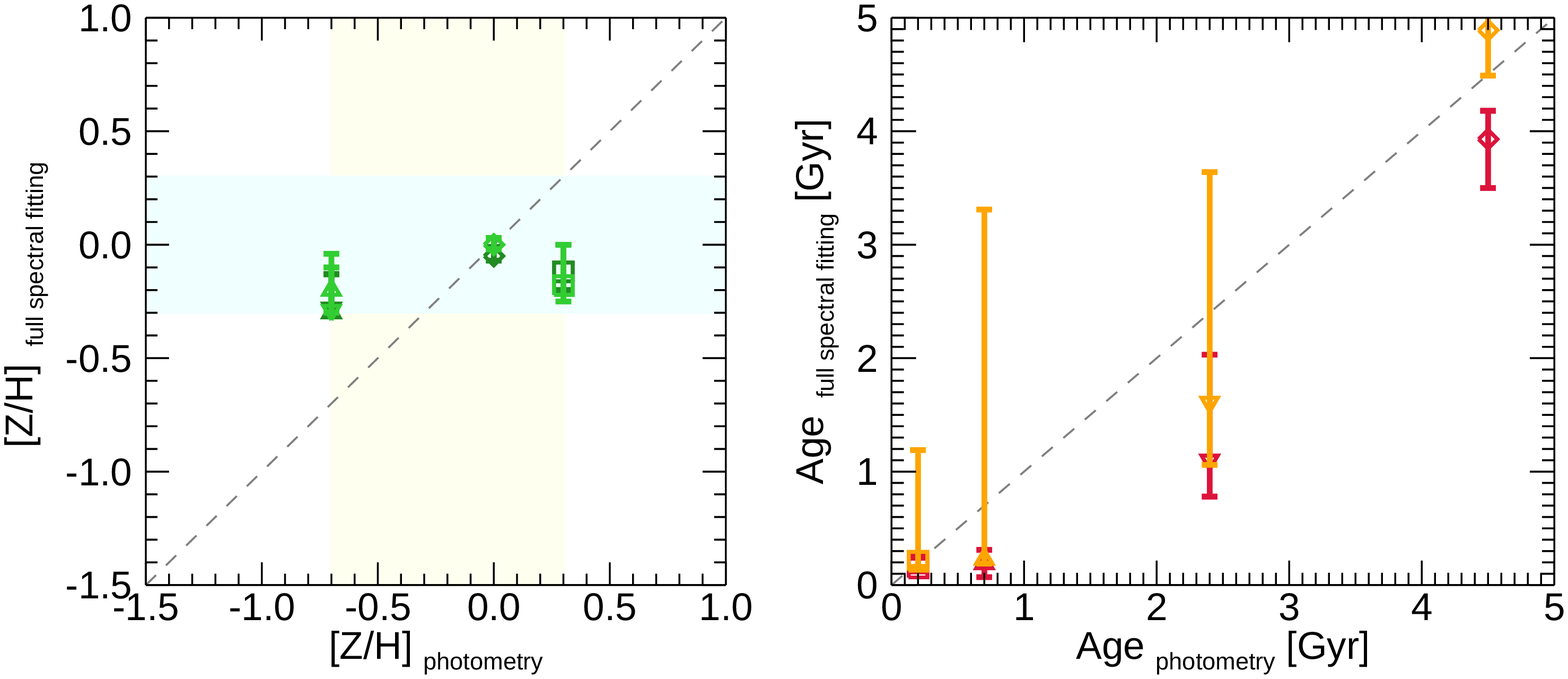} 
\caption{\small{Age and metallicity values derived from indices analysis (Table \ref{tab:allsummary}), full spectral fitting analysis (Table \ref{tab:FSF}) and photometric analysis (Table \ref{tab:photometry}).
Upper panels: indices analysis vs photometry; middle panels: indices analysis vs full spectral fitting; lower panels: full spectral fitting analysis vs photometry. For full spectral
fitting analysis we report both light-weighted quantities (age red and metallicity dark green) and mass-weighted (age orange and metallicity light green).
Shaded areas show the metallicity ranges of the models used in each method.}}
\label{phot_vs_index}
\end{centering}
\end{figure*}

\begin{figure*}
\begin{centering}
\includegraphics[width=8.5cm]{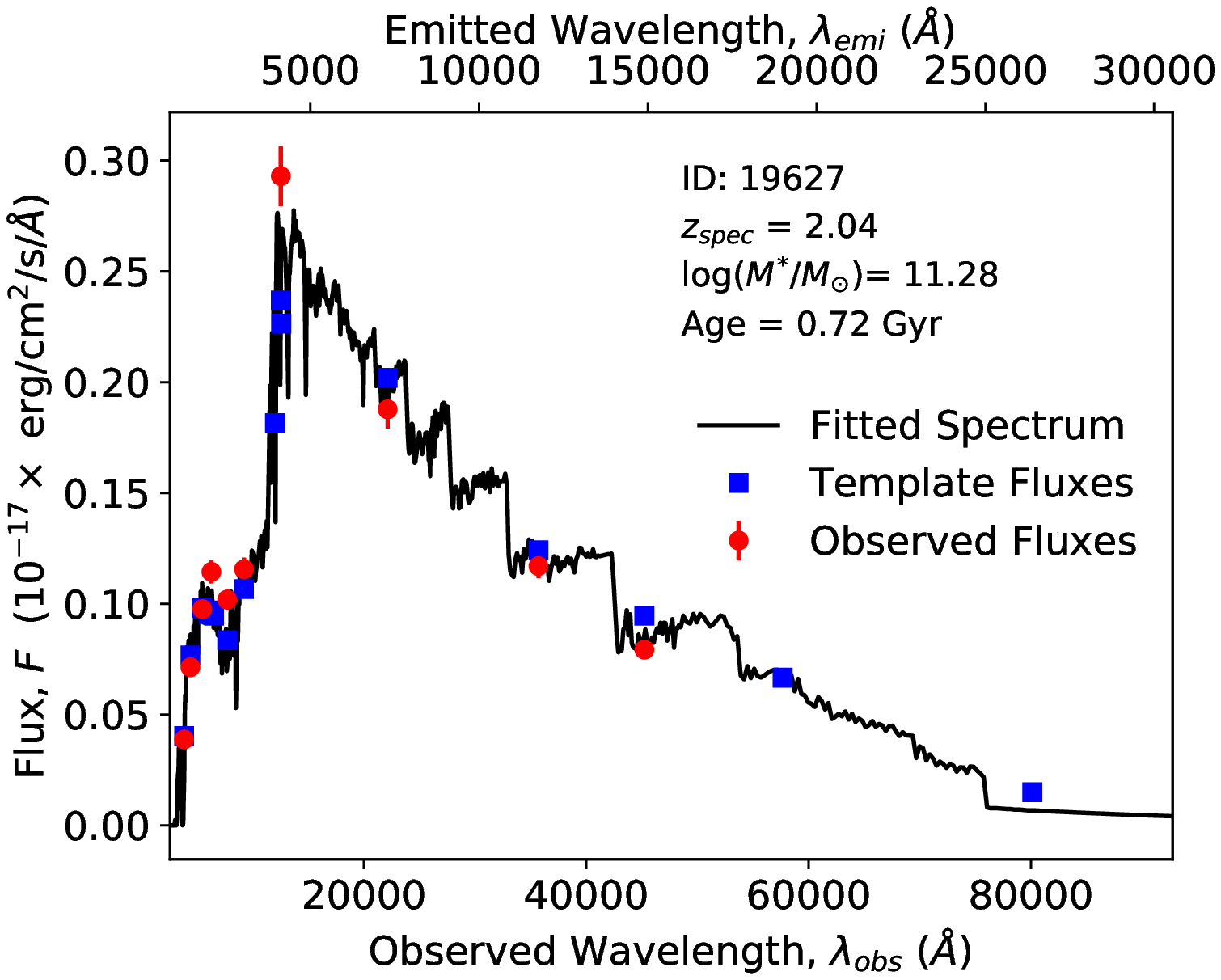}
\includegraphics[width=8.5cm]{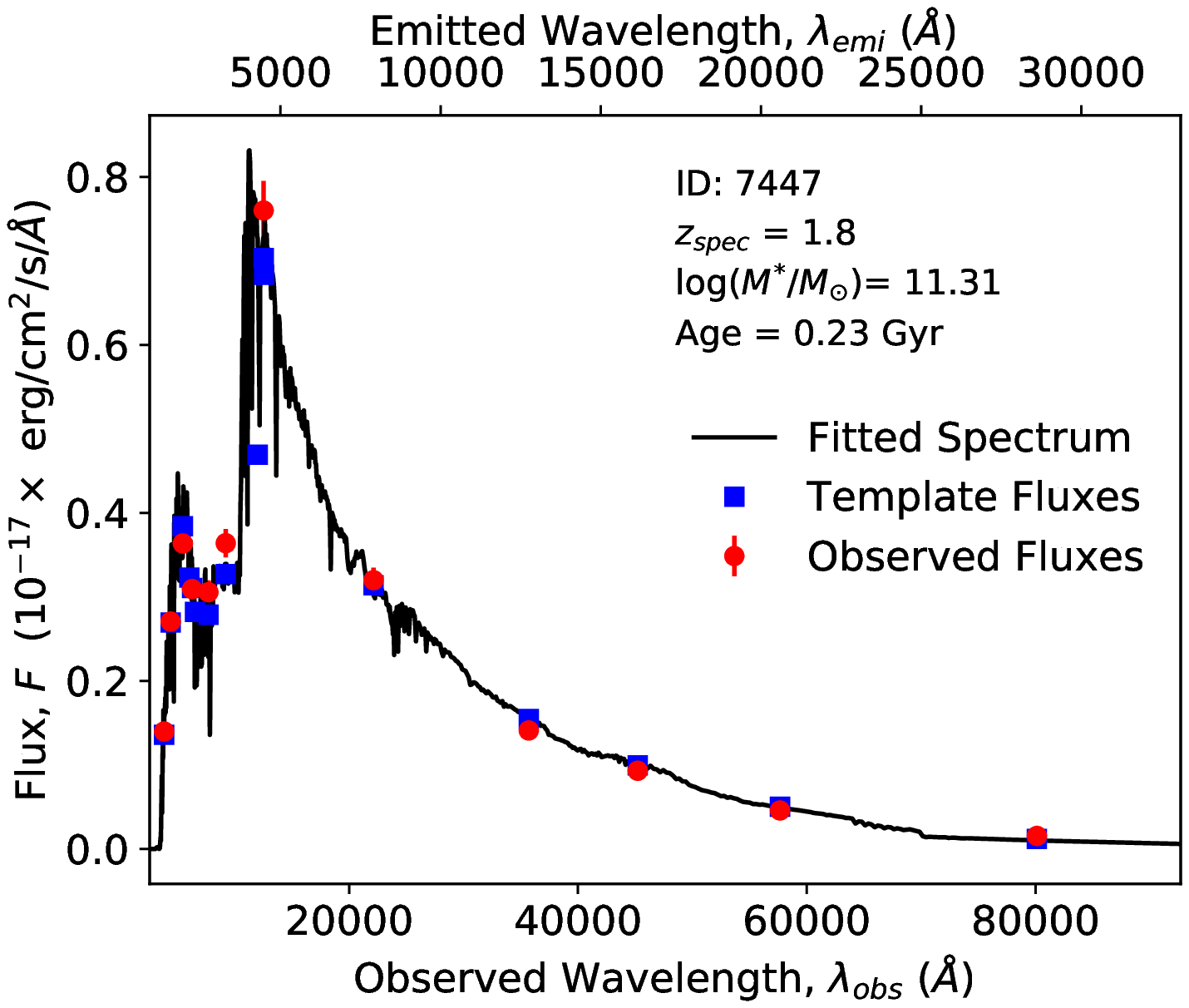}
\includegraphics[width=8.5cm]{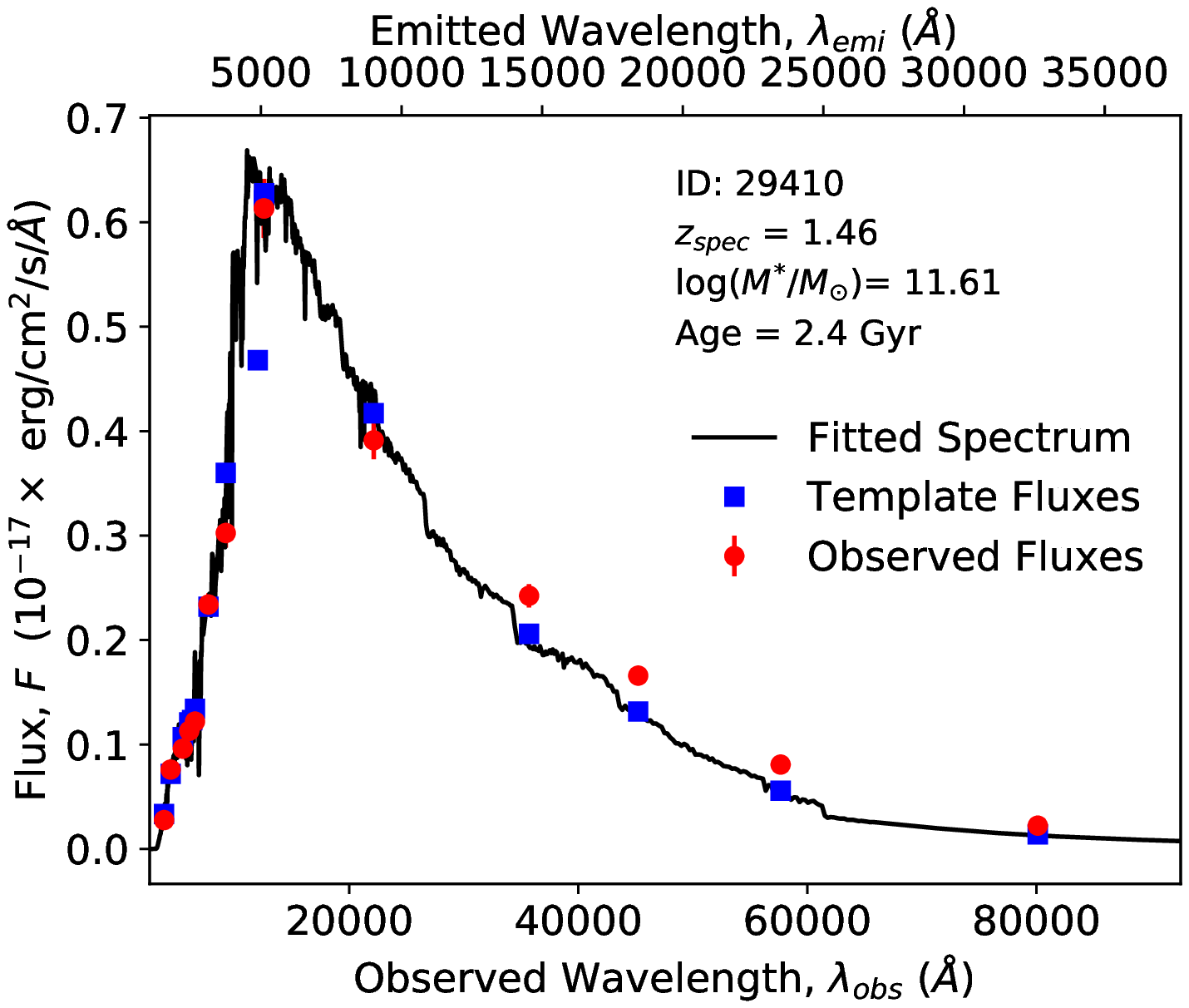}
\includegraphics[width=8.5cm]{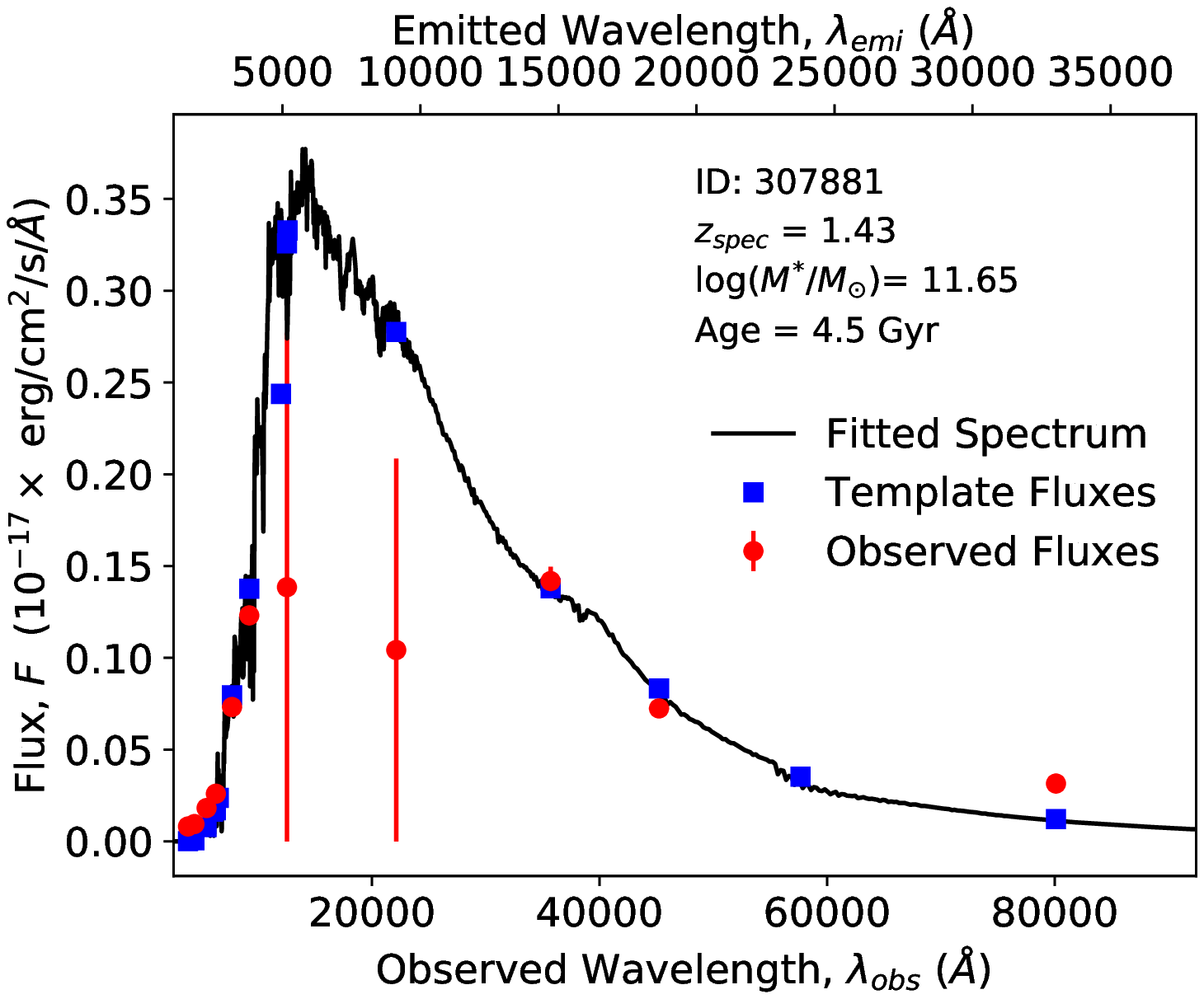}
\caption{Results of photometric model fitting. Galaxies are ordered as decreasing spectroscopic redshift from top left to bottom right. In each panel, the observed data (red circles with error-bars) are shown together with the template fluxes (blue squares) to which the data has been fitted to (in the same filters) and also the best fitting model spectrum (solid black line). The derived properties age (Gyr) and stellar mass (in log and solar units) are labelled, along with the galaxy ID and the spectroscopic redshift.}
\label{fig:sedfit}
\end{centering}
\end{figure*}

Results are shown in Table \ref{tab:photometry}, where we quote the best fit among all templates. From Fig. \ref{phot_vs_index}
we see that the photometric broad-band fitting release galaxy ages that are in excellent agreement with what obtained from the spectral 
analysis (compare also with Table \ref{tab:allresults}). This is an important result considering that most future large galaxy surveys 
will essentially be based on photometry (e.g. EUCLID). Also note that the photometric analysis suggest these galaxies to be passive, which is also 
consistent with what derived from the spectra. Stellar masses, quoted for a Salpeter IMF, are in agreement with those determined by 
\citet{vandesande} using \citet{bc03} models for a Chabrier IMF (we quote their masses augmented by 0.17 which is the 
conversion between Chabrier and Salpeter's IMFs, see \citealt{pforr12}). Fig. \ref{fig:sedfit} shows the photometric fittings, the 
observed data being the red points with error bars. 
We note that the $\chi_{r}^{2}$ of COSMOS-307881 is rather high, probably due to the small errors of some of the photometric data. Nonetheless, the other photometric points are still able to constrain the fits to the spectral age of $\sim ~4$ Gyr.

\begin{table*}
\begin{centering}
 \caption{Results from photometric fitting (for a Salpeter IMF): stellar mass, age, metallicity, SFH, reduced $\chi^2$. The values of stellar mass in brackets are from  \citet{vandesande} using \citet{bc03} models with a Chabrier IMF. The quoted constant ($+0.17$) is the conversion between Chabrier and Salpeter's IMFs from \citealt{pforr12}). }
 \label{tab:photometry}
 \begin{tabular}{lccccc}
 \hline
 ID & log$\mathcal{M}_*$  & Age$_{phot}$ & Z$_{phot}$ & SFH$_{phot}$& $\chi_{r}^{2}$   \\
    & (M$_{\odot}$)       & (Gyr)        &            &             &     \\
 \hline
 19627  &  11.34  (11.24+0.17) & 0.7   & 0.004 & SSP        &  7.971  \\
 7447   &  11.36  (11.27+0.17) & 0.2   & 0.04  & SSP        &  2.649  \\
 29410  &  11.69  (11.29+0.17) & 2.4   & 0.004 & SSP        &  13.418 \\ 
 307881 &  11.65  (11.50+0.17) & 4.5   & 0.02  & $\tau=$0.3 &  24.990 \\
 \hline
 
\end{tabular}
\end{centering}
\end{table*}

\section{Discussion}
\label{all:discussion}

\subsection{Comparison of results from fitting techniques}
In spite of the low S/N of the spectra, we could perform several types of model fitting finding interesting 
results for the stellar population content of these distant galaxies. The use of different techniques on the same objects also informs us on the most efficient approach and on the consistency of results, which will be useful in view of the advent of large telescopes and the JWST which will acquire a multitude of high-$z$~galaxy spectra.

Let us first compare the results obtained with different techniques and assess their consistency.

\begin{table*}
\centering
 \caption{Summary of fitting results for all adopted techniques. From left to right, age (Gyr), [Z/H]  and [$\alpha$/Fe] from spectral  
 indices, full spectral fitting (using light-weighted quantities) and photometry fitting. 
 The [$\alpha$/Fe] for COSMOS-7447 is not constrained since only Balmer lines are usable for this galaxy.}
 \vspace{5mm}
 \begin{tabular}{lccccccc}
 \hline
 ID   &  Age$_{IND}$ (Gyr) & [Z/H]$_{IND}$ & [$\alpha$/Fe]$_{IND}$ &  Age$_{FSF}$ (Gyr) & [Z/H]$_{FSF}$ &  Age$_{PHOT}$ (Gyr) & [Z/H]$_{PHOT}$  \\
 \hline
 \\
 19627   & $0.6\pm0.1$          & $-0.82_{-0.06}^{+0.3}   $ & $-0.18_{-0.12}^{+0.20} $ & 0.19$^{+0.12}_{-0.12}$  &  -0.30$^{+0.17}_{-0.00}$ & 0.7   & -0.7 \\
 7447    & $0.2^{+0.2}_{-0.05}$ & $+0.26_{-0.40}^{+0.04}  $ &         -                & 0.15$^{+0.10}_{-0.03}$  &  -0.12$^{+0.12}_{-0.08}$ & 0.2   & +0.30  \\
 29410   & $1.8\pm0.3$          & $-0.70_{-0.3}^{+0.2}    $ & $+0.42_{-0.10}^{+0.08} $ & 1.10$^{+0.93}_{-0.32}$  &  -0.28$^{+0.15}_{-0.02}$ & 2.4   & -0.7 \\ 
 307881  & $4.1\pm0.4$          & $+0.61^{+0.06}_{-0.05}  $ & $+0.45^{+0.05}_{-0.19} $ & 3.93$^{+0.25}_{-0.43}$  &  -0.05$^{+0.04}_{-0.02}$ & 4.5   & +0.0  \\
 \\
 \hline
\end{tabular}
\label{tab:allsummary}
\end{table*}

Table \ref{tab:allsummary} provides a summary of the final derived stellar population parameters according to the different techniques, from spectral indices through full spectral fitting to photometric fitting. It should be noticed that the parameters from indices given in Table \ref{tab:allsummary} are drawn from both Table \ref{tab:allresults} and \ref{tab:allresultsalpha}:
 ages are better constrained using all indices (UV+Lick), whereas metallicities are better derived from only Lick indices and the TMJ models. The $\alpha$-enhancement is derived from Lick indices plus TMJ models only. The comparison between the same parameters as derived with the three methods is visualised in Figure \ref{phot_vs_index}.

The parameter we could determine most robustly turned out to be the galaxy age, which appears to be very stable independently of the fitting technique. Ages from spectral indices, full spectral fitting and broad-band photometry fitting are all very similar. This is an important result for performing galaxy evolution with large photometric surveys.

The metallicities $[Z/H]$~derived from indices show quite a scatter from very sub-solar 
([Z/H]$=-0.82$) to the very high value of COSMOS-307881 ([Z/H]$=0.61$]). Two galaxies have subsolar metallicities, two galaxies super-solar metallicity. Interestingly, 
the metallicities derived with the other two methods, although being different in absolute values (but consider the errors) do display the same trend, namely the same 
galaxy is found to be solar(sub-solar) independently of the technique. An exception is COSMOS-7447 for which the result from the full spectral fitting value is in disagreement with those from 
the other two methods. COSMOS-307881 is found to be very metal-rich according to spectral indices while the other techniques agree towards solar metallicity.

Regarding the $\alpha$-enhancement, the two oldest galaxies have a $[\alpha/Fe]\sim+0.4$, consistent with massive ellipticals in the local Universe (e.g. \citealt{thomas10}) and at high redshift (\citealt{onodera}; \citealt{kriek16}), one galaxy is solar-scaled and for object 7447 this parameter could not be determined as we only had Balmer line measurements for it.

In summary, ages are robust and consistent and metallicities show a similar trend. If we consider the cosmic distance and the low S/N of our data, the general agreement between results from various methods is encouraging.

\subsection{Formation epochs and evolution to $z\sim0$.}

As noted before, the parameter we could determine most robustly turned out to be the galaxy age, which resulted to be very stable independently of the adopted fitting. 

\begin{figure}
\begin{centering}
\includegraphics[width=\columnwidth]{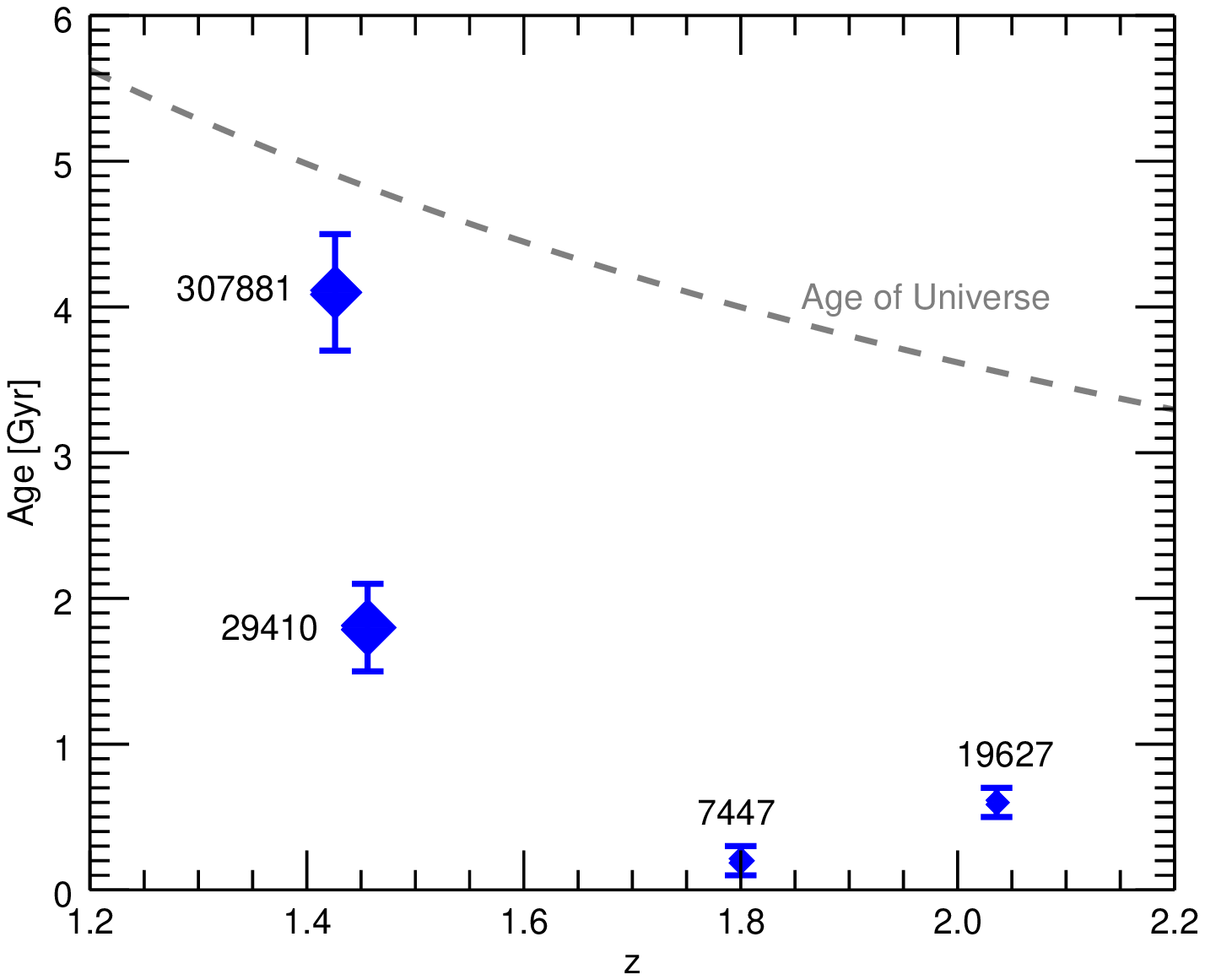} 
\caption{Galaxy ages (in Gyr) vs spectroscopic redshift, with the age of the Universe for a Planck (2013) cosmology indicated as a dashed line. The age of the oldest galaxy lies close to the age of the Universe at the given redshift. Younger galaxies lie at progressively higher redshift.}
\label{fig:agez}
\end{centering}
\end{figure}

Galaxy ages show an approximately monotonic behaviour with redshift, with the oldest galaxy being found at the lowest redshift (Fig. \ref{fig:agez}). 

From the age estimates we can derive the epochs of stellar formation. With the exception of the object COSMOS-307881 that has a redshift of formation $z_{form}\sim5$, the other three galaxies have formed the
bulk of their stars around $z_{form}\sim2-2.5$, i.e. around the peak of the cosmic star formation (\cite{madau14} and references therein). The high formation redshift for COSMOS-307881 suggests this galaxy is a descendant of massive, $z\sim3-5$~ objects recently found by various authors (e.g. \citealt{glazebrook17}, \citealt{guarnieri19}, \citealt{santini19}).

\begin{figure*}
\begin{centering}
\includegraphics[width=8cm]{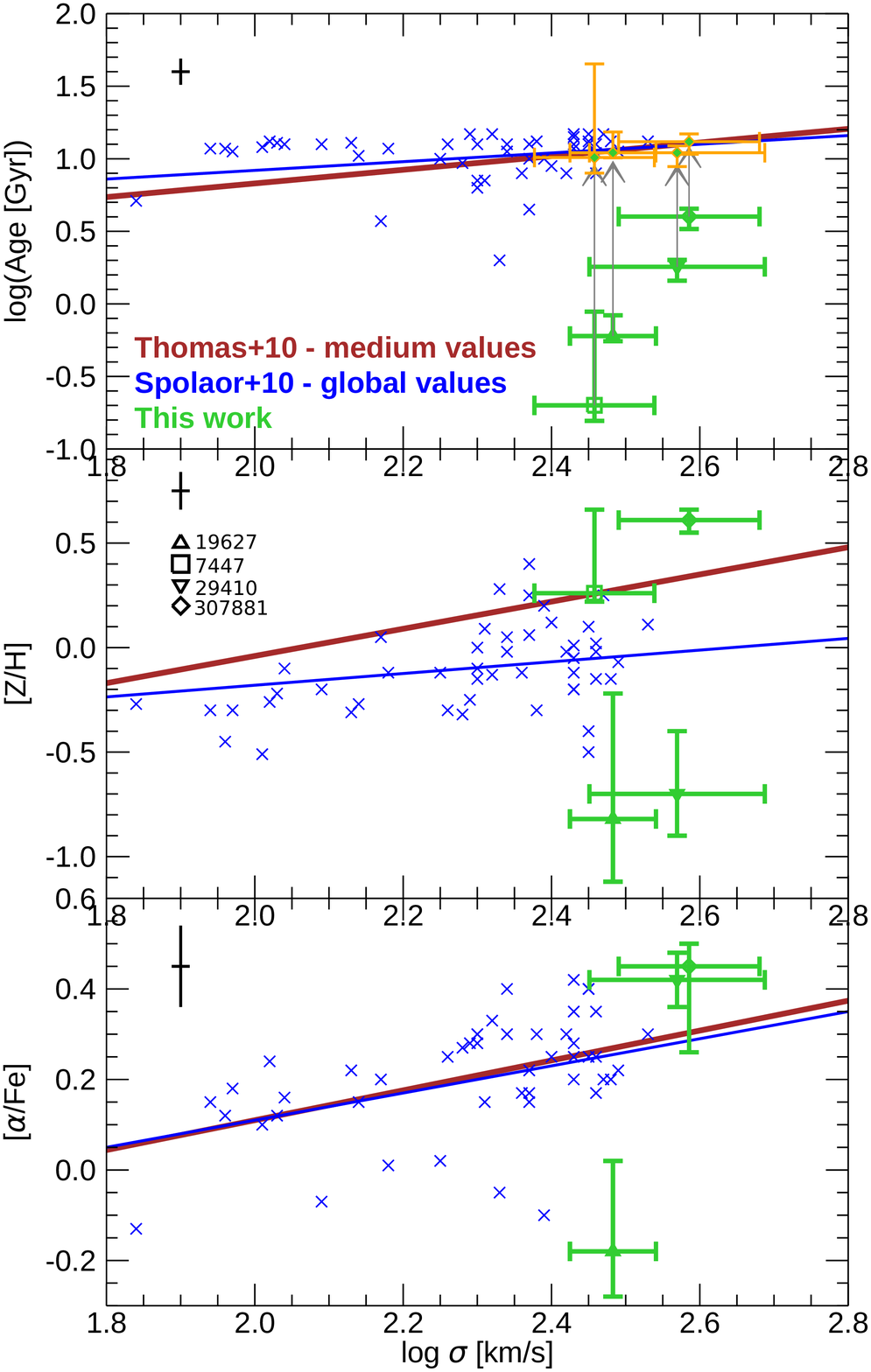} 
\includegraphics[width=8cm]{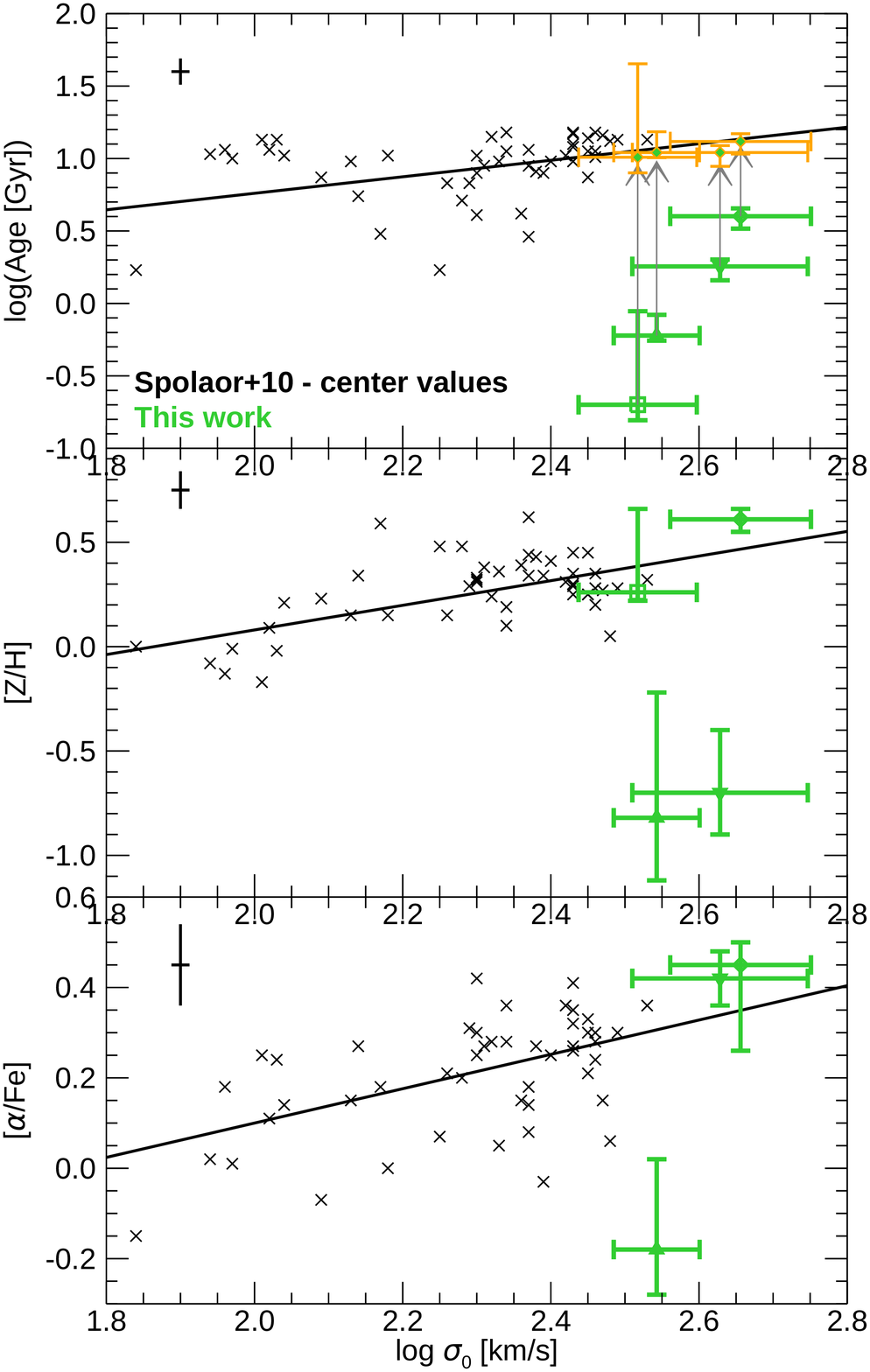} 
\caption{Comparison with local scaling relations of age (top panels), metallicity (middle panels) and $\alpha$-enhancement (bottom panels) as a function of the velocity dispersion at $1$r$_e$ (left panels) and at r$_e/8$ (right panels). Brown lines refer to the local scaling relations from \citet{thomas10}. Black and blue crosses are the central (r$_e/8$) and mean values for the local sample of ETGs by \citet{spolaor10} with the correspondent scaling relation shown as black and blue lines, respectively. Typical error bars are shown in the upper left corner of each panel. Our galaxies are shown as green symbols. Ages after assuming passive evolution are shown in orange, and the symbols are connected with grey arrows.}
\label{scalingrelation}
\end{centering}
\end{figure*}

The expected evolution of these galaxies from the epochs of observation to \emph{z} $\sim0$, can be explored by comparing the stellar population properties derived from this analysis with the results from the local Universe, i.e. by using local scaling relations between stellar parameters and the velocity dispersion. As comparison sets we adopt the results by \citet{thomas10} based on the Sloan Digital 
Sky Survey data and those for the sample of $\sim50$ local ETGs by \citet{spolaor10}. In both cases the data were analyzed using spectral indices and $\alpha$-sensitive models as we do here (see Fig. \ref{scalingrelation}). In particular, in Thomas et al. one set of parameters is given, which corresponds to an intermediate value between the center and one effective radius, while in Spolaor et al. the stellar parameters are measured both in the inner core (within a radius of r$_e/8$) and at the effective radius (r$_e$, considered as the mean values of the global stellar population, see \citealt{spolaor10}). 
This double information is precious because it not only provides an indication of the 
parameters gradients within the galaxies, but also gives us a tool to understand if the compact galaxies of our sample will resemble the centres 
of local elliptical (\citealt{naab2009}; 
\citealt{huang2013}) or will they manage to enlarge their radius without changes in their stellar content.

In Fig. \ref{scalingrelation}, the stellar properties age (top panels), metallicity (middle panels) and $\alpha$-enhancement (bottom panels) are plotted as a function of
the velocity dispersion at $1$r$_e$ (left panels) and at r$_e/8$ (right panels). Our objects are labelled with green symbols. The central (r$_e/8$) values of $\sigma$ for the galaxies of our sample have been derived from the global ones assuming the formula by \citet{cappellari}. Black
crosses in the right panels refer to the central values (r$_e/8$) of the local sample of \citet{spolaor10}, and solid black lines show the scaling relations. The blue crosses and lines in the left panels analogously refer to the mean values. Typical error bars are shown in the upper left corner of each panel. 

The values for the high-$z$ galaxies (green diamonds in Fig. \ref{scalingrelation}) lie into the right region of all plots, which corresponds to the largest velocity dispersions. This is not
surprising because our sample is composed by compact ETGs which are typically found to be denser than local objects \citep{Trujillo09}. 
Indeed, as shown in Table \ref{tab:infoall}, the dynamical masses of our objects are consistent with those of local ellipticals.

In order to compare the ages obtained at \emph{z} $>1.4$ with those in the local Universe we have aged the stellar populations of our objects 
assuming passive evolution (Fig.~\ref{scalingrelation}, top panel). Ageing is indicated by vertical grey arrows pointing towards \emph{z} $\sim0$ (orange symbols). 
It can be seen that our high-$z$~objects will evolve onto the local scaling relation. We can then conclude that the star formation histories of the galaxies of the distant sample are consistent with a star formation which started at
\emph{z} $\sim2$ ($\sim5$ for COSMOS-307881)  followed by passive evolution. We can also conclude that the four distant objects are likely precursors of local early-type galaxies. 

In the middle panel of Fig. \ref{scalingrelation} we compare our derived metallicities with the local ones. A larger
scatter with respect to the age diagram is seen in the [Z/H] vs $\sigma$ plane, even if it is of the order of the local sample one. 
The four galaxies point to two distinct behaviours: COSMOS-307881 and COSMOS-7447 seem to lie on the local scaling
relation related to the central values, while UDS-19627 and UDS-29410 on the one built with the local mean values. 

\begin{figure}
\begin{centering}
\includegraphics[width=\columnwidth]{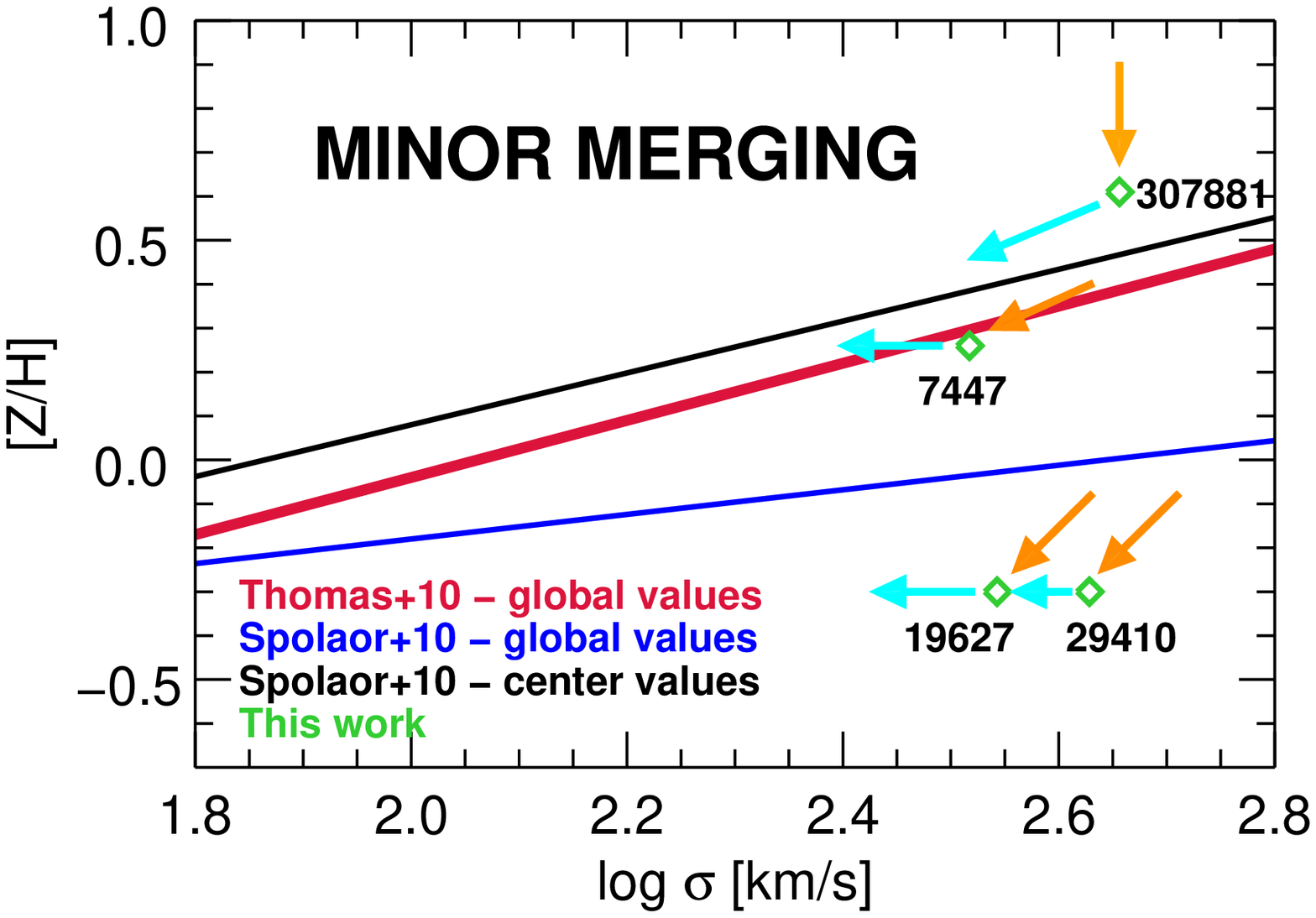}
\includegraphics[width=\columnwidth]{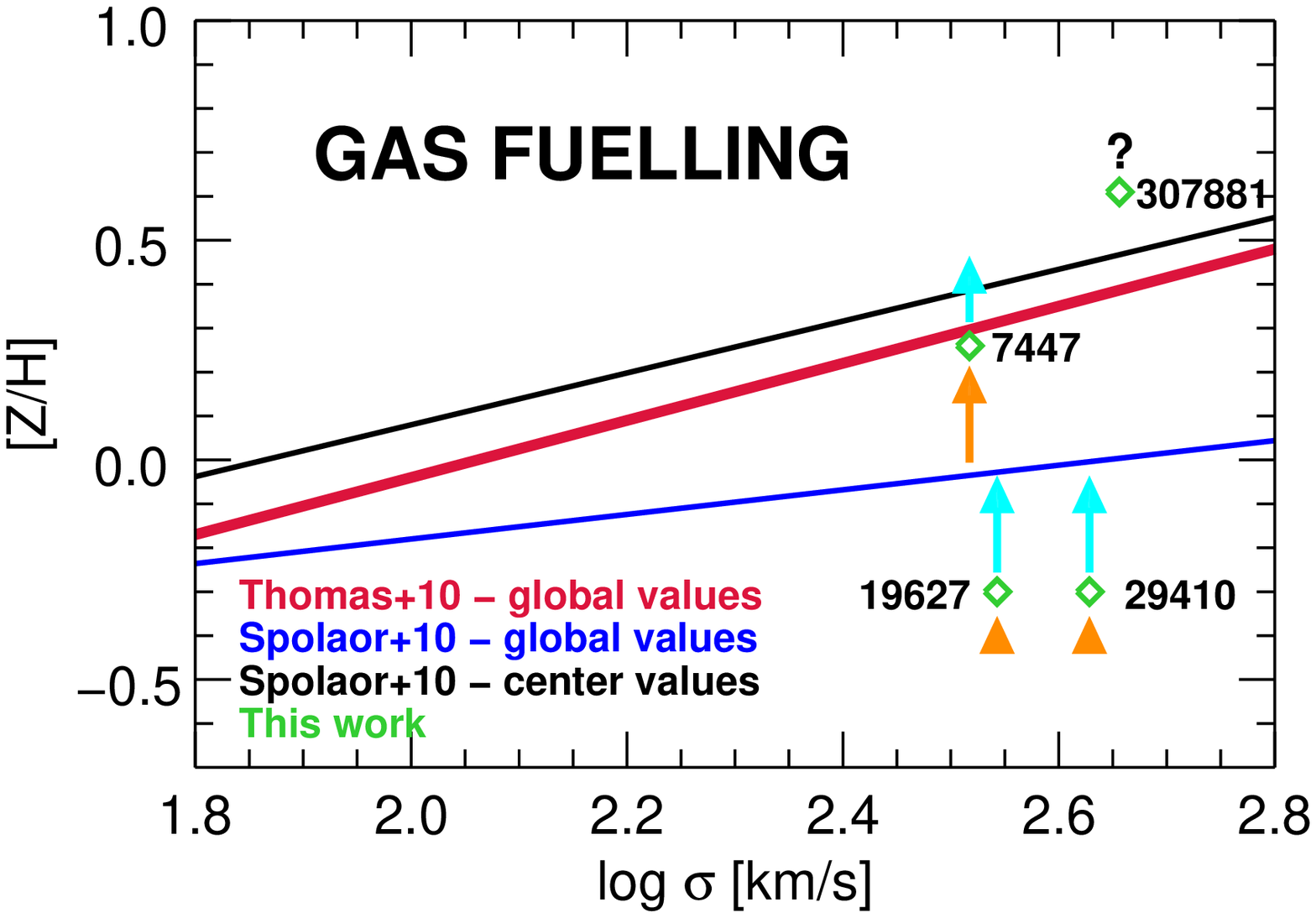}
\caption{Mass-metallicity relation: two possible scenarios for the evolution of the sample galaxies (green diamonds) in comparison with the local relation of \citet{spolaor10} 
(see Fig. \ref{scalingrelation}). Upper panel: minor merging scenario. Bottom panel: gas fuelling scenario. Orange arrows refer to the formation and cyan arrows to the evolution of these galaxies.}
\label{2evolution}
\end{centering}
\end{figure}

Different scenarios may be envisaged for the formation and evolution of these galaxies. As far as COSMOS-307881 is concerned, its high metallicity 
value is expected from the local scaling relation at high velocity dispersion. A possible explanation proposed by the model of \citet{pipino}, is that high mass objects
formed in a rapid star formation burst with the consequent increasing of the star formation efficiency and thus of the metals production. 
This scenario is supported also by the high value of the $\alpha$/Fe ratio of this object (as shown in Fig. \ref{scalingrelation}, bottom panel), which is tightly
connected with the star formation time-scale as modelled by \citet{thomas05}, which in this case results to be short ($\Delta t \sim 0.1$ Gyr). 
Comparable super-solar gaseous metallicities are for example observed in quasars at $z>4$ \citep{juarez}. Such metallicity values in the local Universe are not
found in the global values of local ellipticals, but only in their centers as observed in many works focused on metallicity gradients of local ETGs (\citealt{trager}; \citealt{spolaor}; 
\citealt{martin}). This means that probably COSMOS-307881 will experience minor merging events with
lower metallicity systems during its evolution toward $z=0$, which are able to dilute both the velocity dispersion and the chemical content \citep{naab2009}.
Indeed, low-mass systems (as dwarf galaxies) are expected to contain low metallicity stars as it is observed in the local mass-metallicity relation \citep{thomas05}. 
This interpretation is shown in Fig. \ref{2evolution}, upper panel: COSMOS-307881 probably comes from an even more metal rich position (orange arrow) and will
probably evolve in the direction (cyan arrow) of the cores local scaling relation (black line), decreasing both its metal content and its velocity dispersion by means of mass accretion.

Different scenarios must be invoked for the other three objects, which, although as massive and dense as COSMOS-307881, seem to be featured by a lower metallicity.
COSMOS-7447 shows a super-solar metallicity, but its large error bar makes it consistent also with sub-solar values. A possibility similar to the previous one is that, 
given their high mass and the arguments previously exposed, also these galaxies have formed in a short burst at $z\sim2-2.5$ 
from high-metallicity gas, but their metal content have been already diluted by merging events happened soon after the formation and that involved the whole galaxy, not only
its external regions. This scenario is outlined in Fig. \ref{2evolution}, upper panel: orange arrows show the direction from which the sample galaxies (green diamonds) are
formed in the [Z/H] vs $\sigma$ plane, and cyan arrows show their possible future evolution.

Still adopting the merging scenario, we can also suppose that these three ETGs have formed from a metal poor nebular gas as that of Lyman-break 
galaxies (LBG), observed at $z\sim3$ as less metal enriched than local star forming galaxies \citep{mannucci}. Indeed, LBGs could be
the progenitors of $z\sim1$ compact ETGs \citep{williams14}. In this case the sample ETGs are formed in the position where they are observed in the [Z/H] vs
$\sigma$ plane (Fig. \ref{2evolution}, upper panel, not considering the orange arrows), and their evolution will push them towards the region 
occupied by lower density systems thanks to minor merging events with sub-solar to solar metallicity systems (cyan arrows).

Another scenario is that of gas fuelling (Fig. \ref{2evolution}, bottom panel): supposing again that these galaxies started their evolution from a metal poor stellar
composition (and e.g. being the descendant of LBGs), residual cold gas streams reaching the inner part of ETGs would activate new star
formation events prolonged in time, thus increasing the metal content in the galaxy centres. Indeed, it seems that field ETGs are able to retain gas coming from evolving stars
which is able to form new generation of metal rich stars \citep{peebles}. This is consistent with the results obtained in \citet{mio}, where
the constant presence of small amounts of young stars in $z\sim1$ ETGs is observed. This process, as shown in Fig. \ref{2evolution} (bottom panel) with the same notation as before, 
would leave the galaxies with metallicity (and age) gradients, as it is observed in the local Universe (\citealt{spolaor10}; \citealt{labarbera12}; \citealt{goddard17}). 
Cold gas streams with prolonged star formation would lower the [$\alpha$/Fe] value, as we observe in UDS-19627, but not in COSMOS-307881 and UDS-29410 (see Figure \ref{scalingrelation} lower panels).
Moreover, this scenario cannot explain the evolution of COSMOS-307881, as it is unlikely that its metallicity will further increase.

The minor-merging scenario is expected to trigger central star formation by funnelling cold gas towards galaxy centres. 
The gas is consumed just after falling in the potential well of the galaxy, and the star formation triggered by this kind of event is expected to be 
bursty.  In case of dry mergers, no star formation is expected. 
On the other hand, the gas-funneling scenario foresees the presence of neutral cold gas filaments between galaxies, feeding a low but continuous 
star formation activity. 
 Discriminating between the two scenarios would be possible by means of observations searching for molecular gas around high redshift ETGs with ALMA. Deep searchings for molecular gas in high redshift galaxies have been recently carried out 
\citep{decarli16}, but present surveys usually aim at studying star forming galaxies for which the amount of molecular gas is 
much higher than the one expected around quiescent galaxies. Indeed, the deepest ALMA observation able to quantify the molecular gas content of a 
non-starforming galaxy at $z > 1$ \citep{bezanson19} did not reach a reliable detection, providing an upper limit to the molecular H2 gas mass of $1.1 \cdot 10^{10}$ M$_{\odot}$ in a post star burst galaxy at $z=1.5$. The more promising way to distinguish between the two scenarios seems to be spatially 
resolved studies of the stellar content within quiescent high redshift galaxies, which could be obtained by the next generation of large telescopes (e.g.: JWST, E-ELT, GMT).

\vspace{0.5cm}
Finally, from the bottom panels of Fig. \ref{scalingrelation}, we obtain indications on the star formation time-scales of those galaxies for which 
the $\alpha$/Fe ratio could be derived, i.e.: COSMOS-307881, UDS-19627 and UDS-29410. As already mentioned, high values of [$\alpha$/Fe] imply short star
formation time-scales since the lack of Fe-peak elements is a sign that supernovae type Ia did not had enough time to explode before the star formation quenching. 
Massive systems are expected to show high values of $\alpha$/Fe (i.e., short star formation time scale), as indicated by the local mass-[$\alpha$/Fe] scaling relation. This is the case 
for COSMOS-307881 and UDS-29410, which are
consistent with the local scaling relation (see Fig. \ref{scalingrelation}, bottom panel), confirming the expectations of \citet{thomas05} and \citet{pipino}. UDS-19627 instead shows a 
sub-solar value of [$\alpha$/Fe] in contrast with the expectations. However its value is consistent within the
observed scatter of the local scaling relation, i.e. similar massive local ETGs with low [$\alpha$/Fe] do exist.

\section{Summary and conclusions}
\label{sec:conclusions}

In this paper we present new age, metallicity and $[\alpha$/Fe] estimates for four individual ETGs at $z>1.4$ in the COSMOS and UDS fields. These measures have been performed thanks to the unique possibility of analyzing many spectral indices from the UV to the near-IR given by X-Shooter spectra, that cover a wide spectral range with high spectral resolution. In particular, besides the optical Lick indices, we have exploited UV spectral indices (down to $\sim1500$
\AA $ $ restframe). The large number of indices we could measure significantly helped in disentangling the residual age-metallicity degeneracy in the comparison with stellar population models. The addition of the UV indices helps with age determinations.

For obtaining the galaxy stellar population parameters, we have employed three popular techniques, namely fitting of individual absorptions, full spectral fitting and photometric fitting, and we have compared the results. 

In case of spectral indices, galaxy-tailored analyses have been performed. Galaxy ages have been constrained with a mean error of $\sigma_{Age}\sim0.3$ Gyr, stellar metallicities with a mean $\sigma_{[Z/H]}\sim0.24$ and we obtained the $\alpha$/Fe abundance ratio for $3$ out of $4$ objects.

By using population models including UV-bright, evolved stellar components \citep{claire}, we infer the presence of a UV-upturn in the oldest, most massive and most metal-rich galaxy of the sample, COSMOS-307881. Only UV-upturn models were able to explain the strong UV indices of this galaxy. This is the highest redshift detection of the UV upturn with spectroscopic measurements.

With the same spectroscopic data and public broad-band photometry we have derived ages and metallicities from full spectral fitting using the
 \firefly code and from broadband SED fit using the HyperZ code. Noticeably, we found an excellent agreement between galaxy ages determined via indices, full spectral fitting or broad-band colours. Even if for just a few objects, 
 this is an important result that will be useful to exploit data from future large photometric surveys, as Euclid or LSST, for the vast majority of which spectra will not be available.

We compared the stellar population properties of the high-\emph{z}
galaxies with those measured in the local Universe. We find that our galaxies will evolve on the local scaling relations following pure passive evolution.

Additionally we speculate on possible formation and evolution scenarios for our objects, being aware that higher S/N spectra will be needed to confirm 
metallicity and element ratios.

In the future it will be very interesting to repeat our analysis on a larger sample of high redshift ETGs. 

\section*{Acknowledgements}

We are grateful to the anonymous referee for his/her constructive comments and suggestions which have improved the manuscript.
IL acknowledges the Institute of Cosmology and Gravitation of the University of Portsmouth for the hospitality during which this project and collaboration started.




\bibliographystyle{mnras}
\bibliography{bibliography} 

\appendix

\section{Analysis on Mock spectra}
\label{appendixA}

In this section we discuss the derivation of stellar population parameters using different sets of spectral indices, in order to secure ourselves on the robustness of the results.  Recall that - due to the variable data quality and different redshift of the data - we were forced to use different combinations of indices in our analysis. 

We constructed mock spectra having the same spectral characteristics of our data. 
We used stellar population models from MS11, with similar (known) stellar parameters as those obtained for our real galaxies. We downgraded the models to the velocity dispersion of each object and added a Gaussian random noise whose standard deviation is derived from the real S/N values in different spectral regions. Systematic errors are considered well below the statistical ones so they were neglected.
We repeated the addition of noise 1000 times in order to obtain a statistical sample of measured indices to be compared to models in the same way as we did for the real data.

As a first test, we compared the 1000 index realizations with the MS11-TMJ merged models (see Section \ref{sec:models}) and retrieved ages and metallicities. We then selected the best-fit solutions from the $\chi^2$ values. 
This procedure of parameter derivation was repeated for each mock using different sets of indices as it follows:

\begin{itemize}
    \item  All indices listed in Table \ref{tab:allindices}.
    \item  The set of indices actually used for the given galaxy.
    \item  The sets of indices used for the other three objects.
    \item  Only UV indices, i.e. all the non-Lick indices located below 4000\AA, plus the D4000 break.
    \item  Only Lick indices plus D4000.
\end{itemize}

\begin{figure*}
\begin{centering}
\includegraphics[width=10.cm]{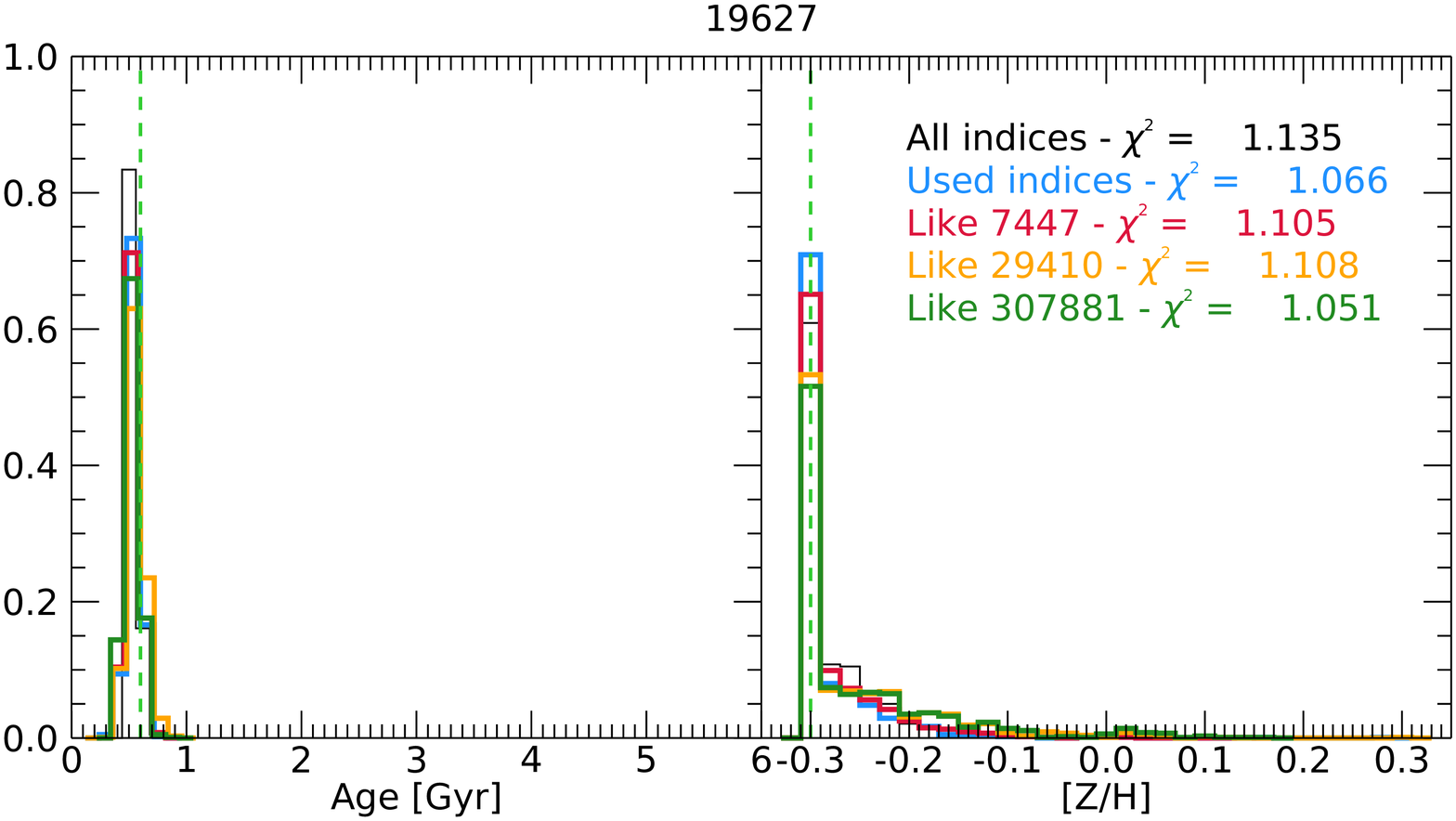}
\includegraphics[width=10.cm]{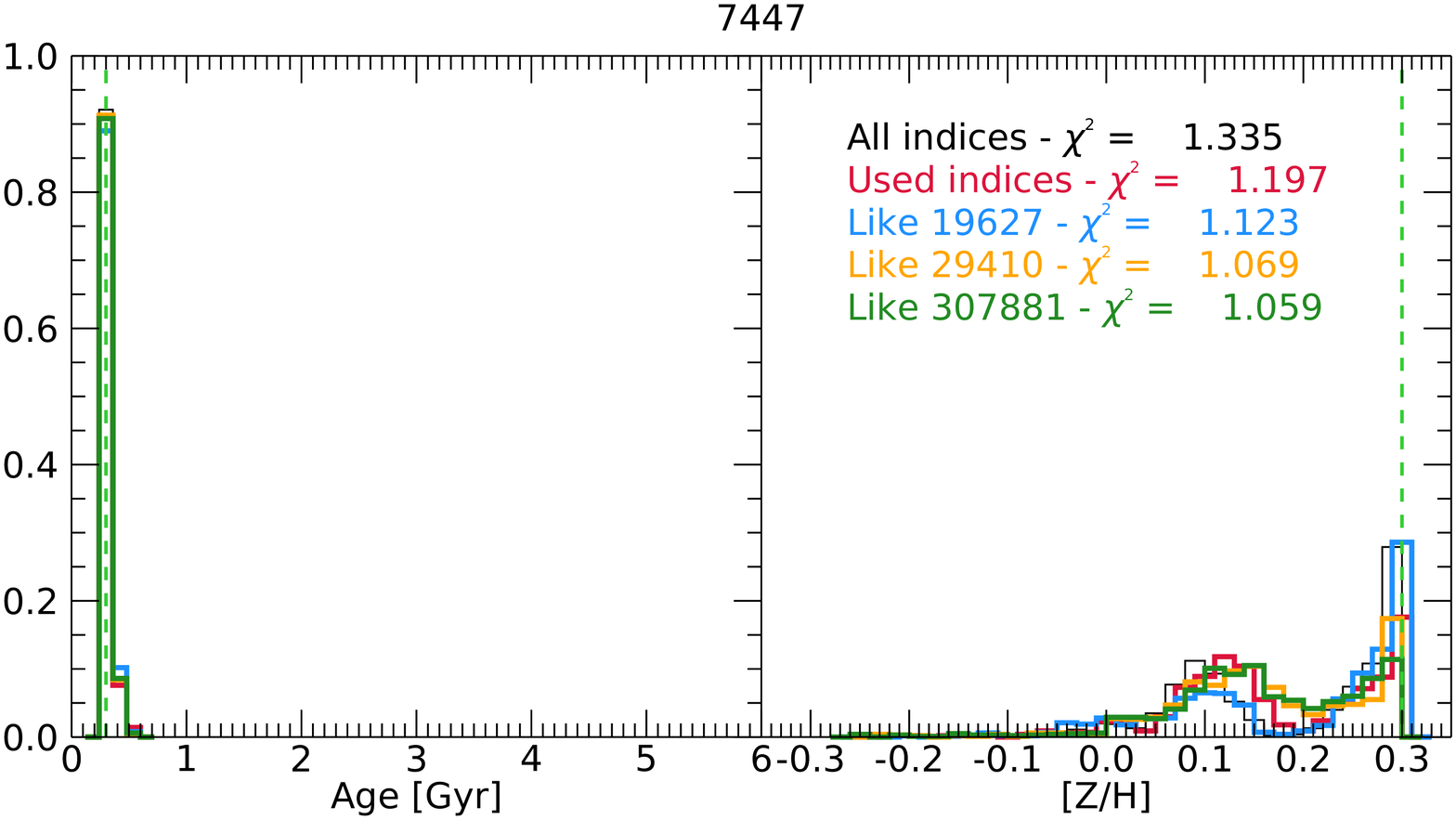}
\includegraphics[width=10.cm]{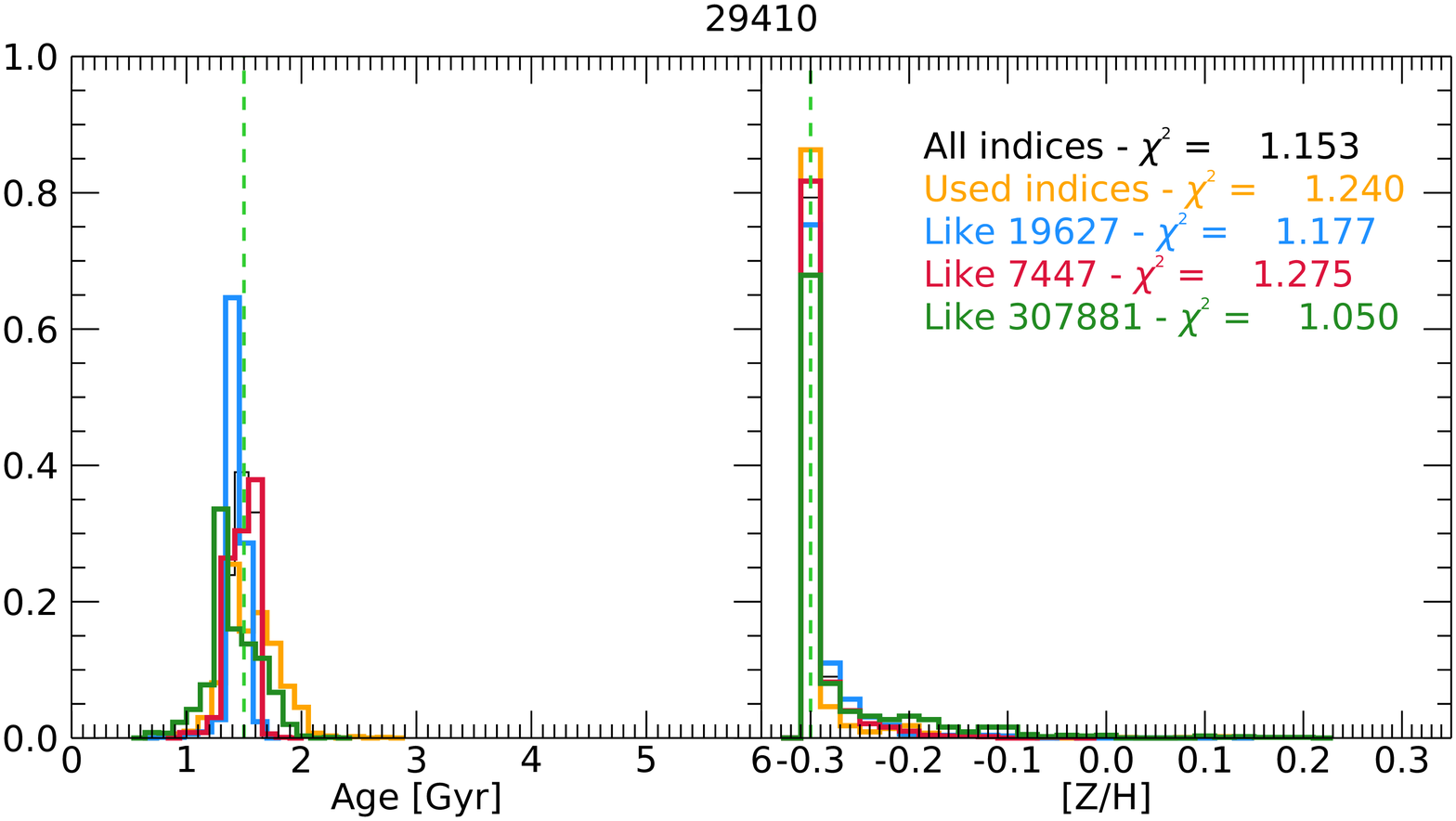}
\includegraphics[width=10.cm]{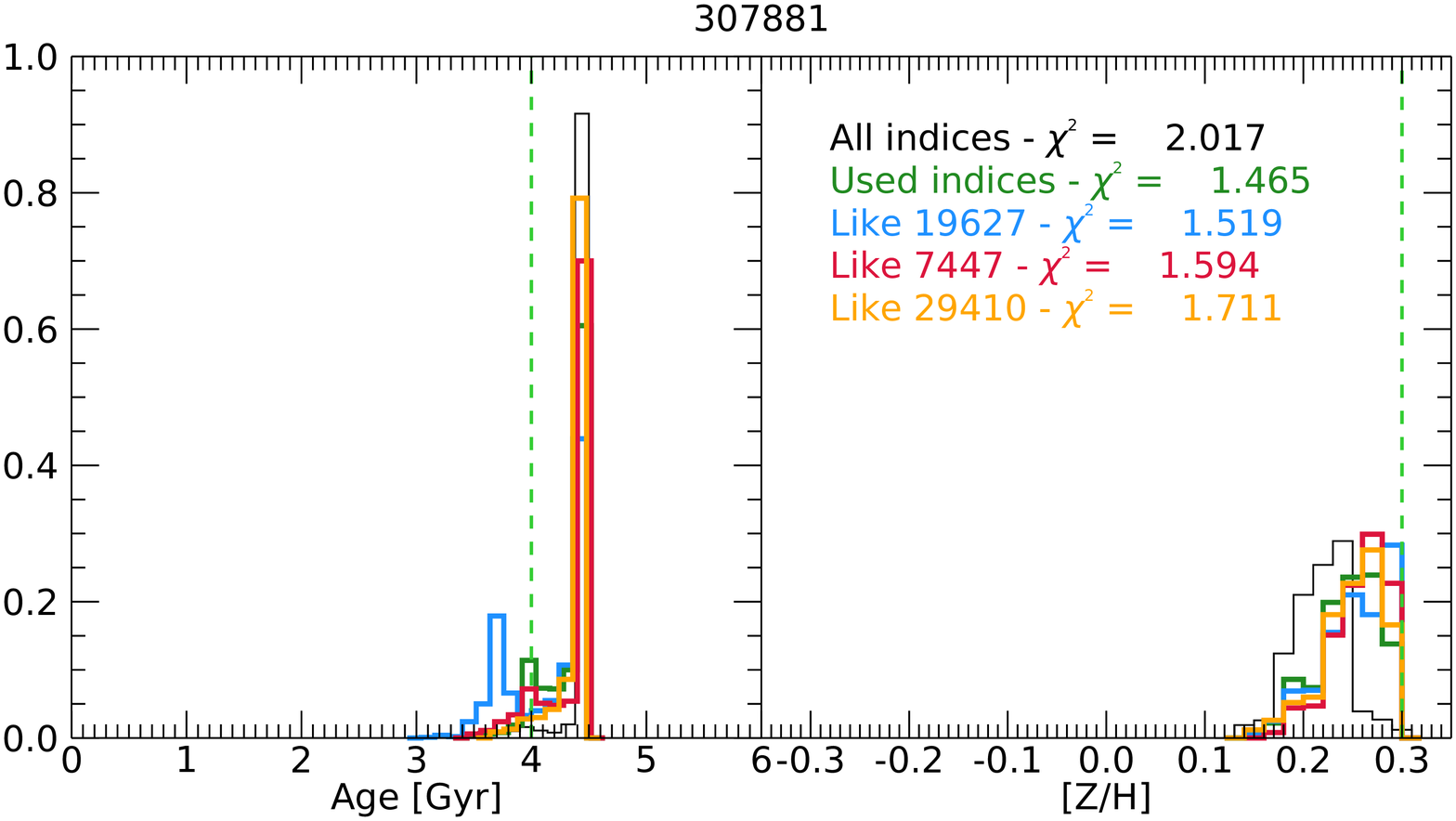}
\caption{\small{Mock spectra analysis: bestfit age and metallicity distributions for different sets of indices. Mocks correspondent to the different galaxies are colour coded as: 19627-blue, 7447-red, 29410-orange, 307881-green. All index distributions are shown in black. Vertical dashed green lines indicate the true input age and metallicity values. Reduced $\chi^2$s for each index set are also displayed.}}
\label{fig_appA:set1}
\end{centering}
\end{figure*}

\begin{figure*}
\begin{centering}
\includegraphics[width=10cm]{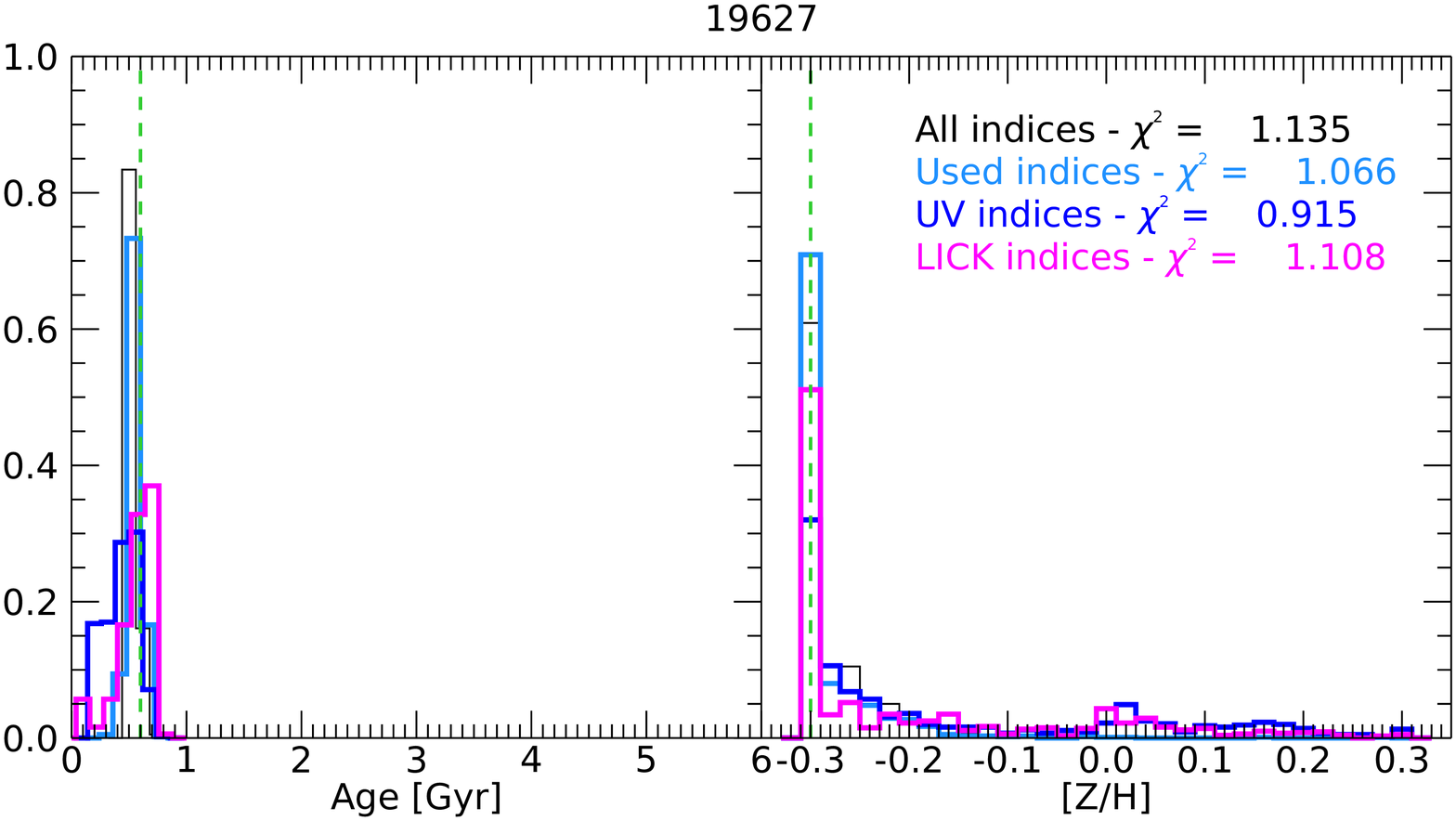}
\includegraphics[width=10cm]{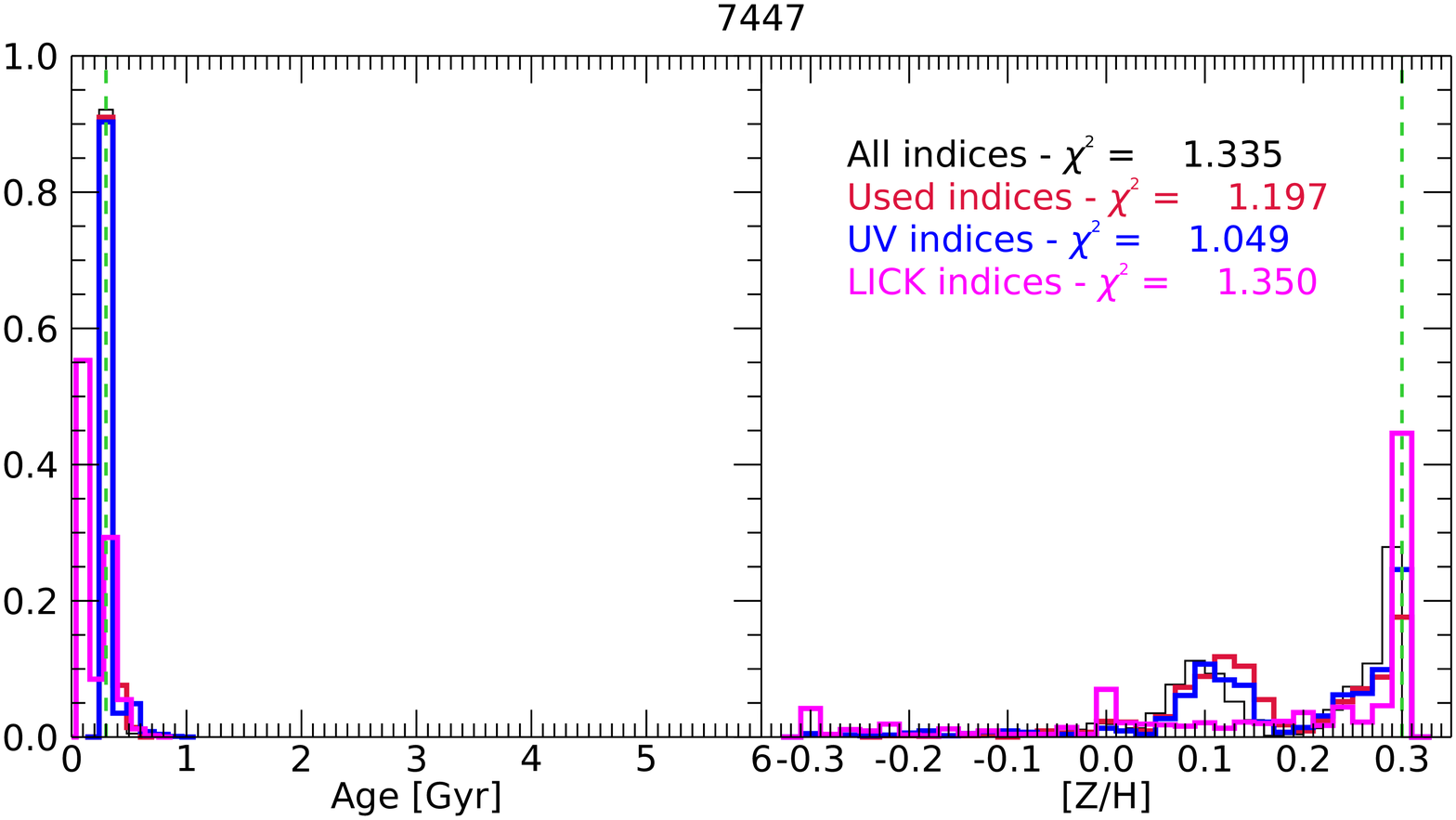}
\includegraphics[width=10cm]{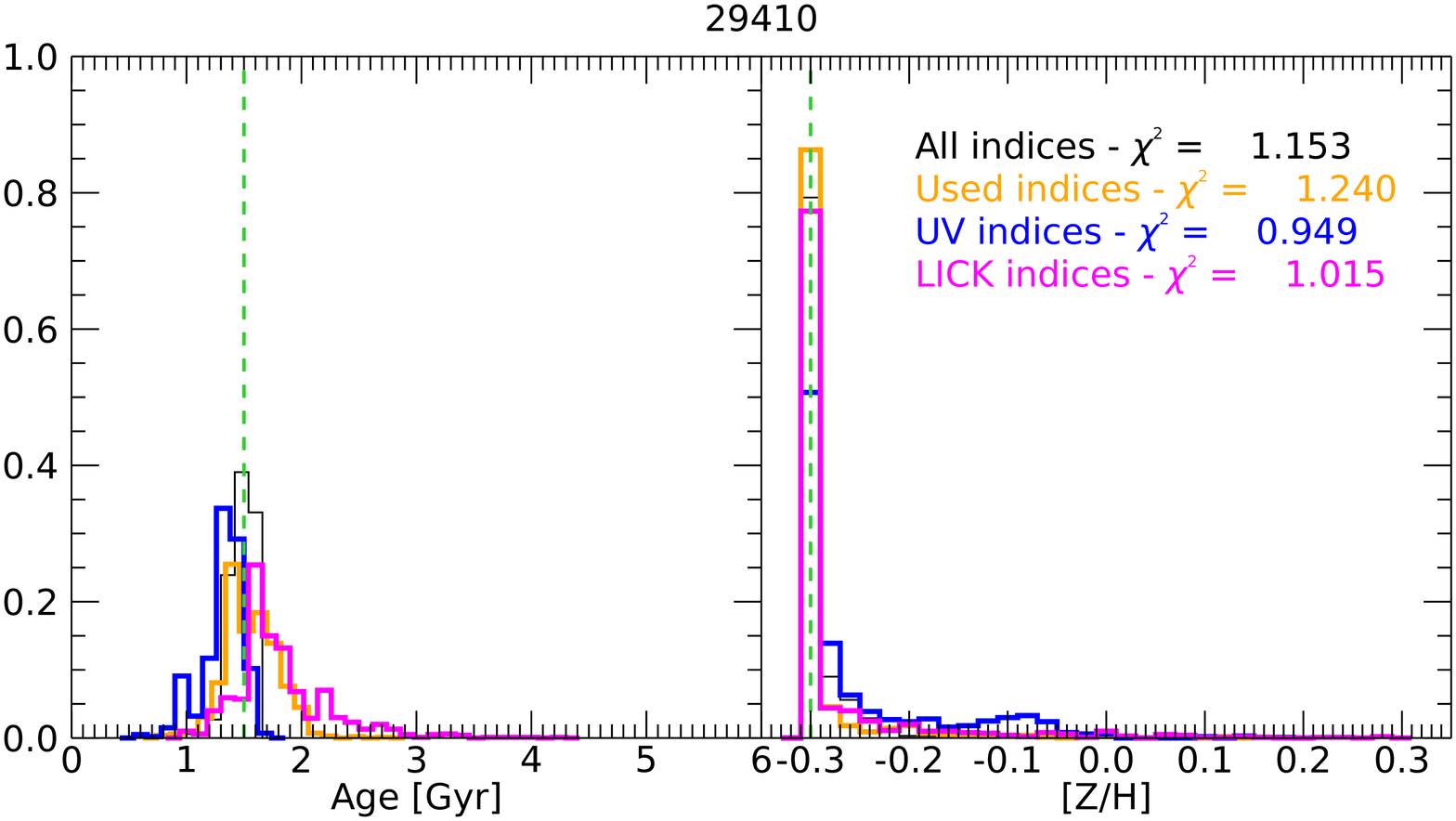}
\includegraphics[width=10cm]{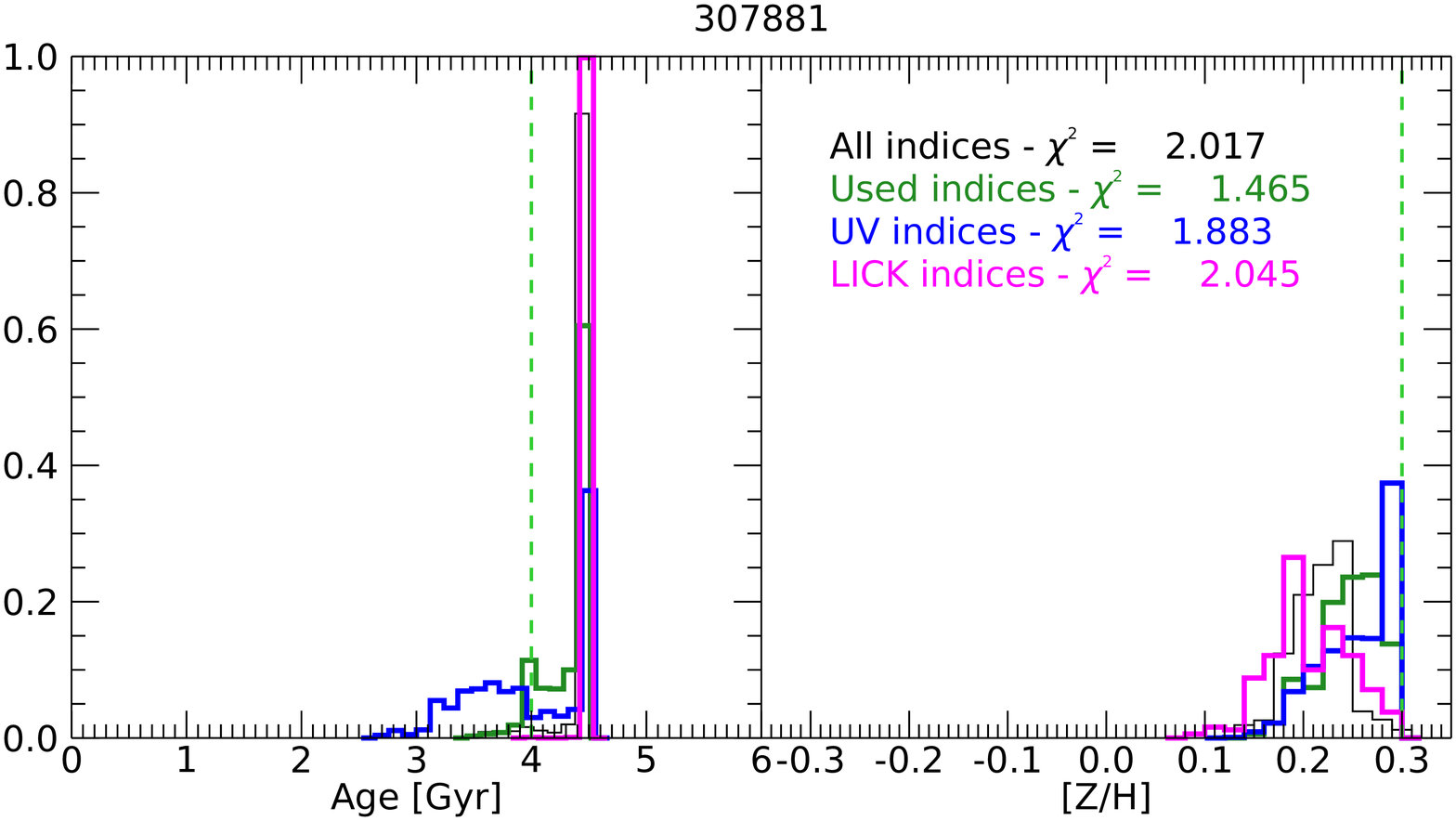}
\caption{\small{As in Figure \ref{fig_appA:set1} now considering UV indices (blue lines) and 
Lick indices (magenta).}}
\label{fig_appA:set2}
\end{centering}
\end{figure*}

The results are shown in Figure \ref{fig_appA:set1} and \ref{fig_appA:set2}. 

Differences among parameters obtained using different index subsets are negligible in all tests. The tests demonstrate that age and metallicity are retrieved with uncertainties similar to those obtained using real data, demonstrating that our analysis can be performed also on these low S/N spectra.

\
Regarding the comparison between UV and Lick indices (Figure \ref{fig_appA:set2}), it is interesting to note that age distributions generally peak at younger ages for UV indices and at older ages for Lick indices, such that the combination of them is centred at the true value. This test demonstrates the value of adding UV indices to the analysis of stellar populations.
\vspace{1cm}

For the second test we included in the analysis also the information on [$\alpha$/Fe] exploiting subsets of Lick indices in comparison with TMJ models as done in the real analysis. 
In this case, we constructed the mocks by perturbing 1000 times the values of the TMJ models assuming the error bars found from the previous tests. We choose to test a super-solar value of [$\alpha$/Fe]$=0.3$ for all the four mock spectra. The set of Lick indices used in each fit are those used for the real data and given in Table \ref{tab:allresultsalpha}.

\begin{figure*}
\begin{centering}
\includegraphics[width=13cm]{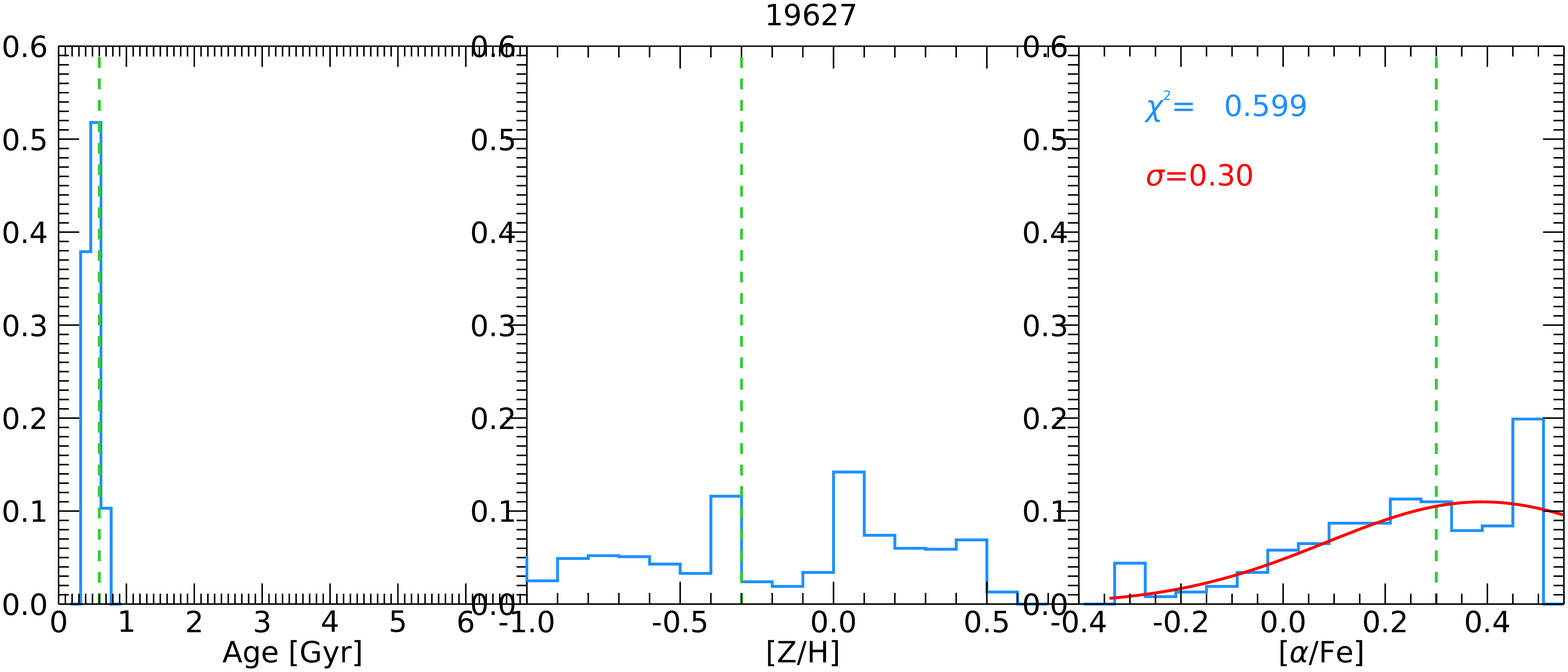}
\includegraphics[width=13cm]{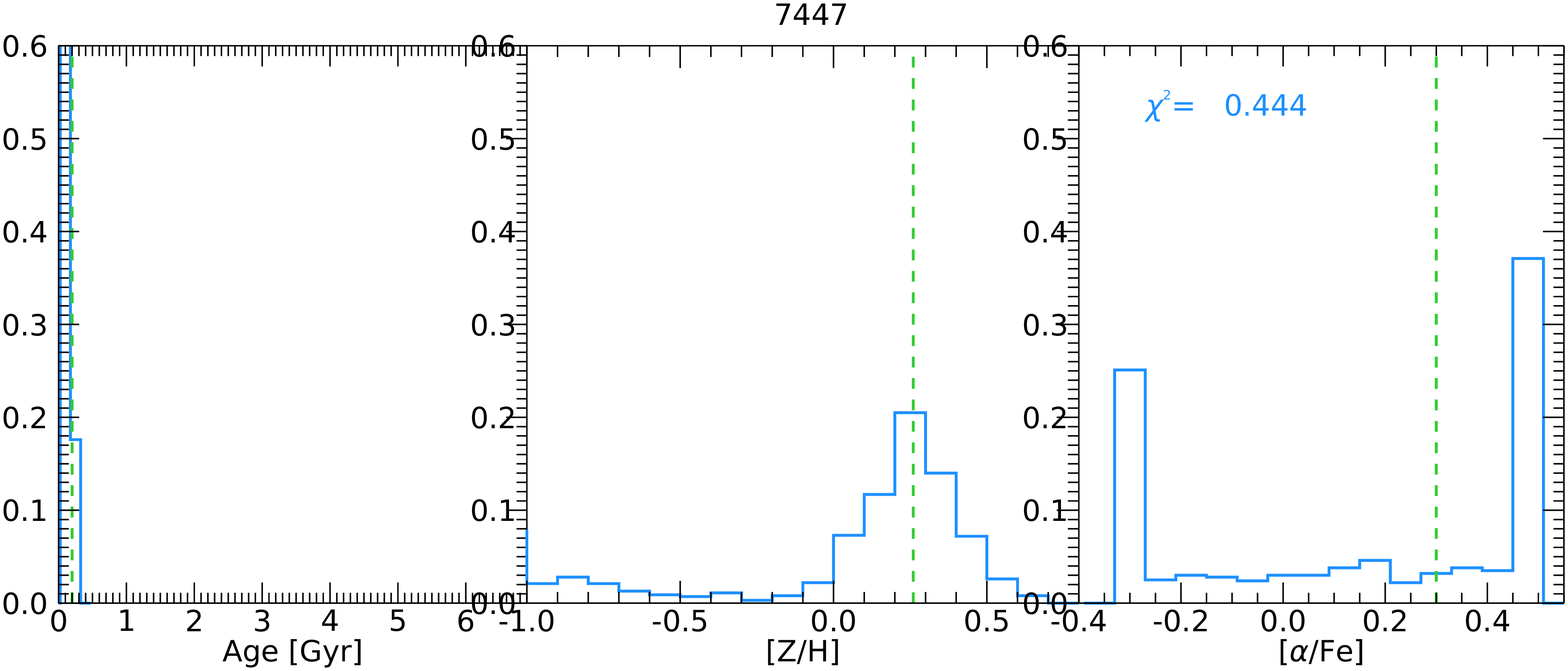}
\includegraphics[width=13cm]{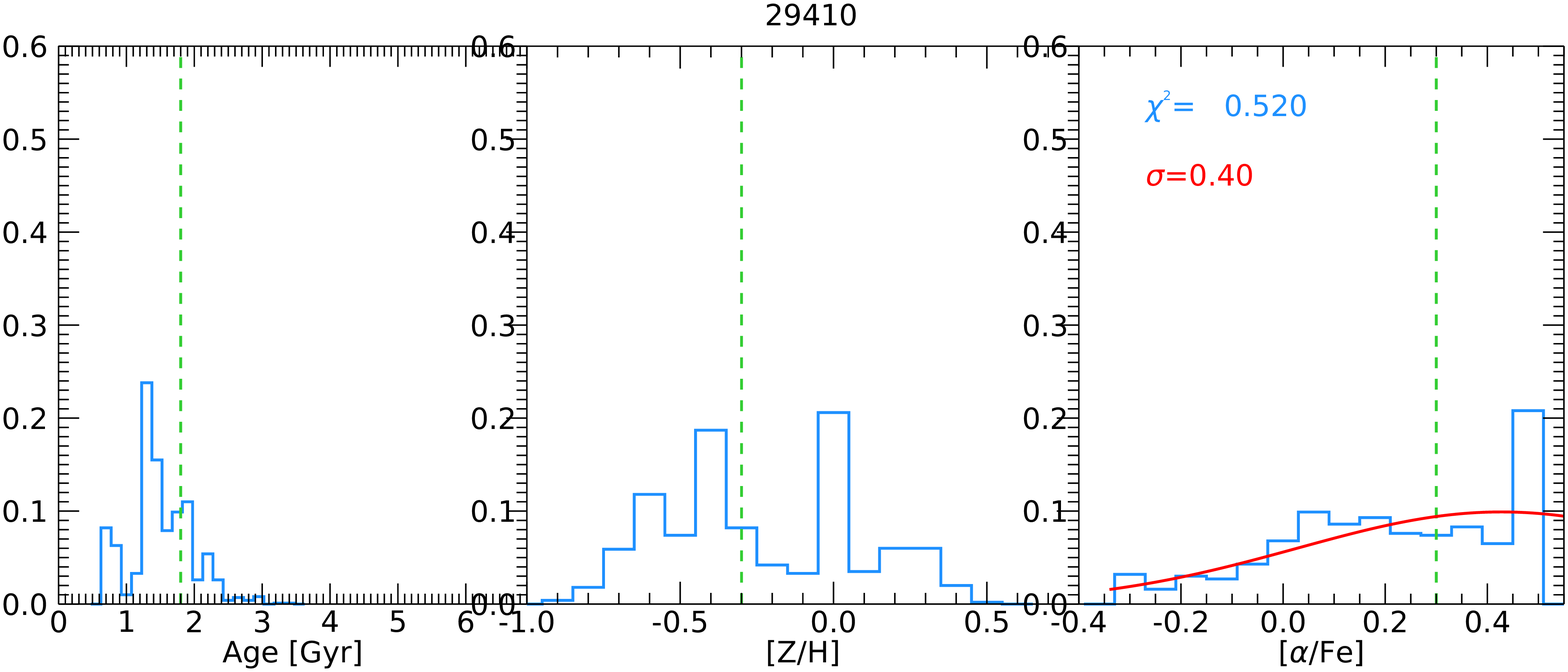}
\includegraphics[width=13cm]{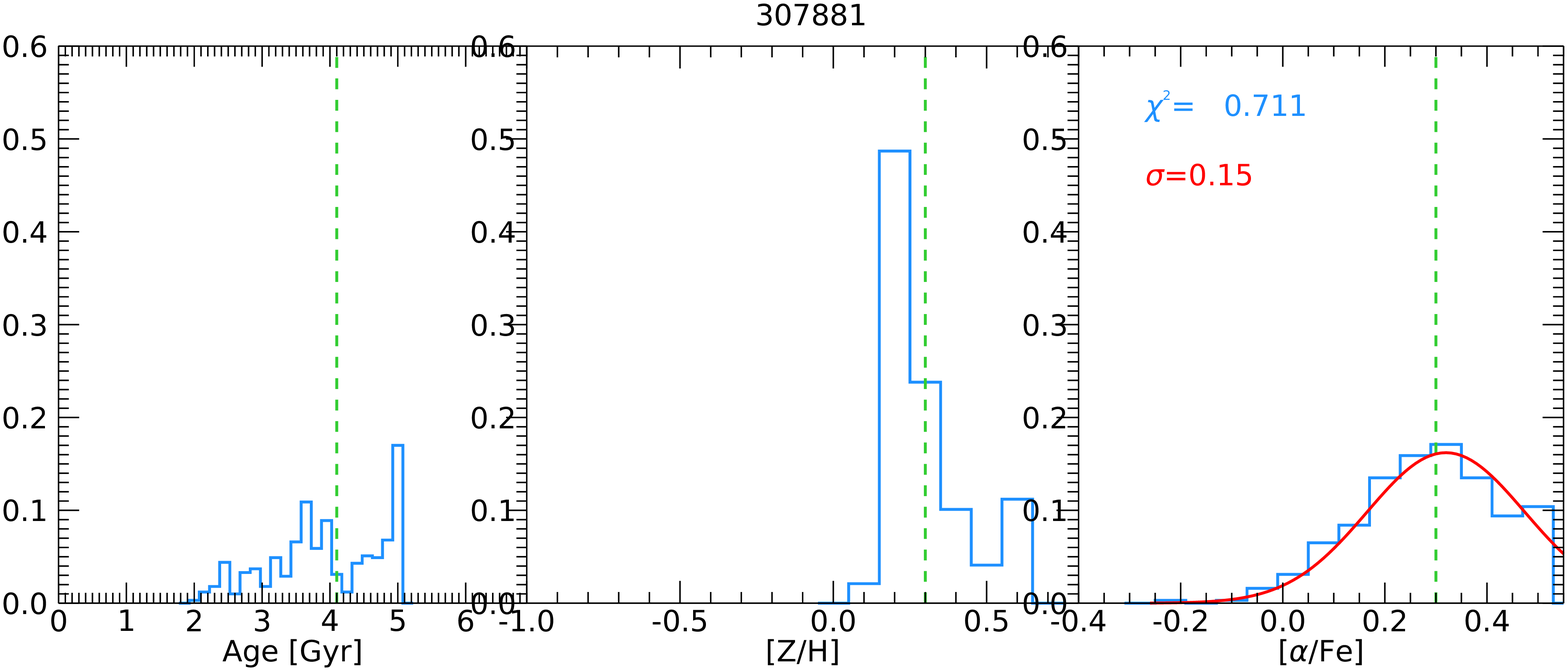}
\caption{\small{Mock spectra analysis: bestfit age, metallicity and [$\alpha$/Fe] distributions for different sets of Lick indices as used for real data analysis (see Table \ref{tab:allresultsalpha}). Vertical dashed green lines indicate the true input parameters
values. A Gaussian fit to the [$\alpha$/Fe] distribution is shown in red.}}
\label{fig_appA:lick}
\end{centering}
\end{figure*}

The results are shown in Fig. \ref{fig_appA:lick}. True values of age (left panels), metallicity (middle panels) and [$\alpha$/Fe] (right panel) are shown with vertical dashed lines in each panel. To better visualize the uncertainties on the derived [$\alpha$/Fe], the right panels include a Gaussian fit to the histogram (red line) and its correspondent $\sigma$ value.
From Fig. \ref{fig_appA:lick} it can be notice that only in case of the simulated 7447 object the [$\alpha$/Fe] value could not be constrained, exactly as we found when dealing with the real spectrum. This is expected since the available indices only include the Balmber lines of Hydrogen, that are mostly age-sensitive \citep{worthey94}.

These simulations demonstrate the feasibility of our analysis and ensure the results are robust.

\section{Comparisons of data and best-fits for indices.}
\label{appendixB}

\begin{figure*}
\begin{centering}
\includegraphics[width=9.5cm]{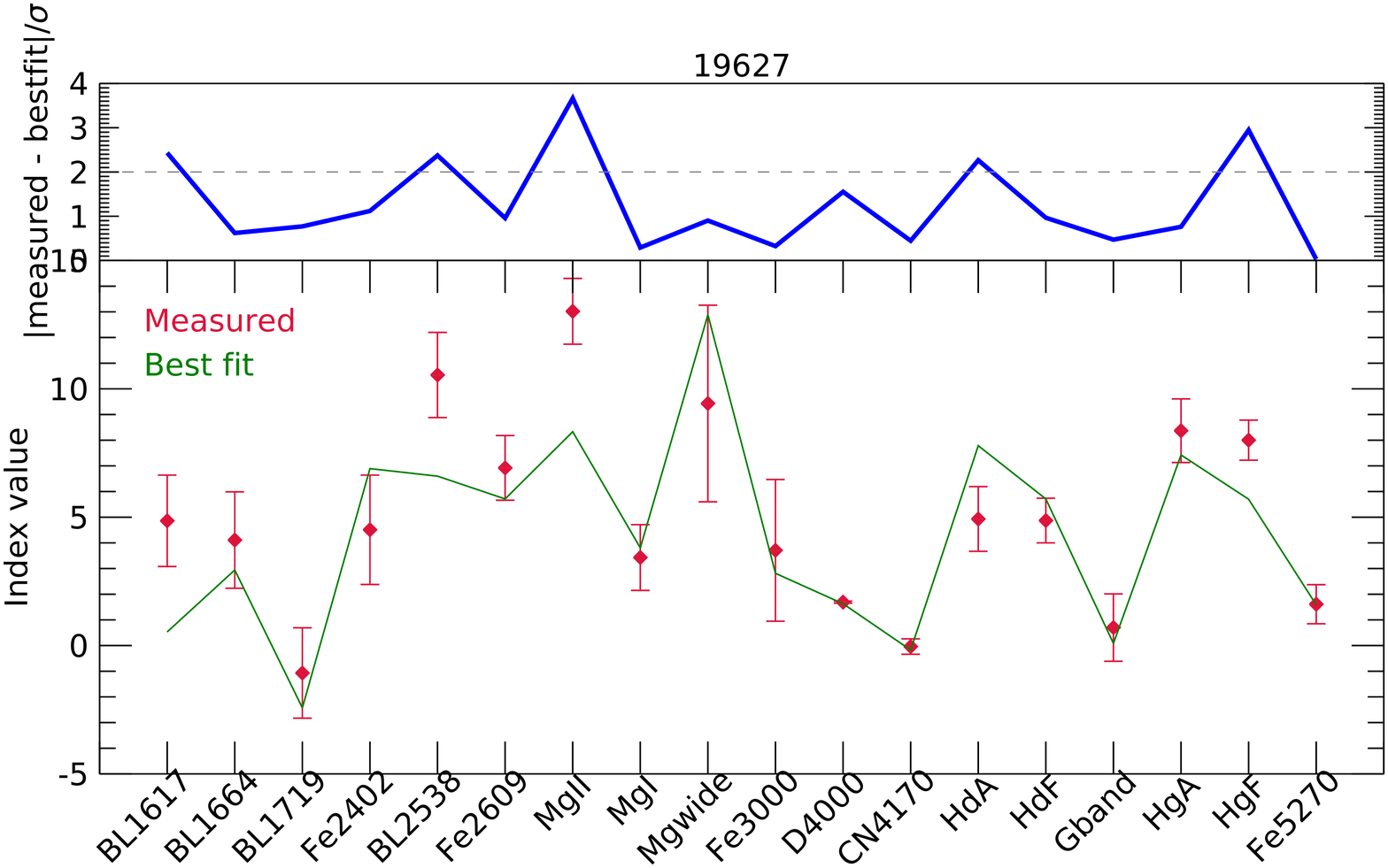}
\includegraphics[width=9.5cm]{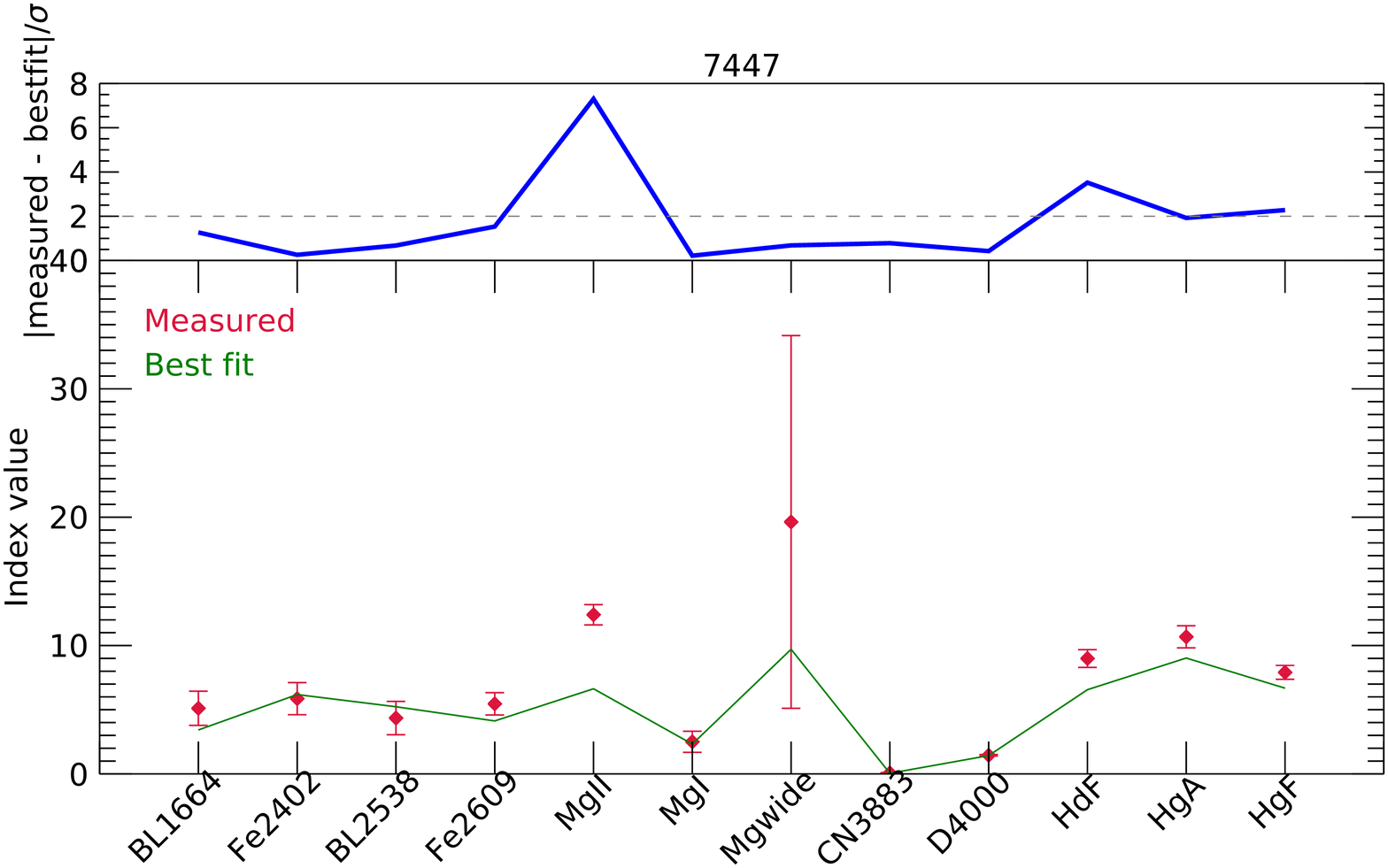}
\includegraphics[width=9.5cm]{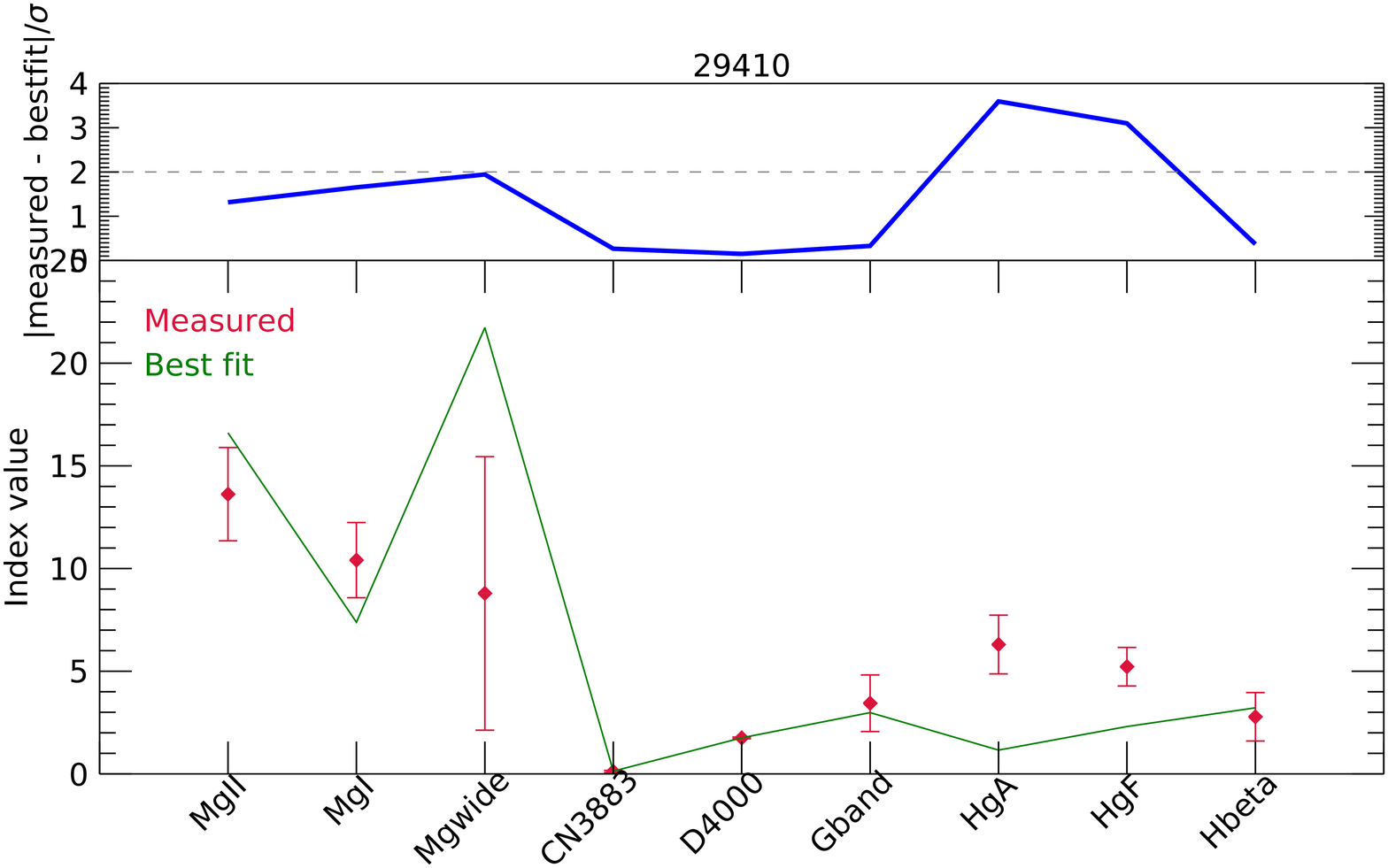}
\includegraphics[width=9.5cm]{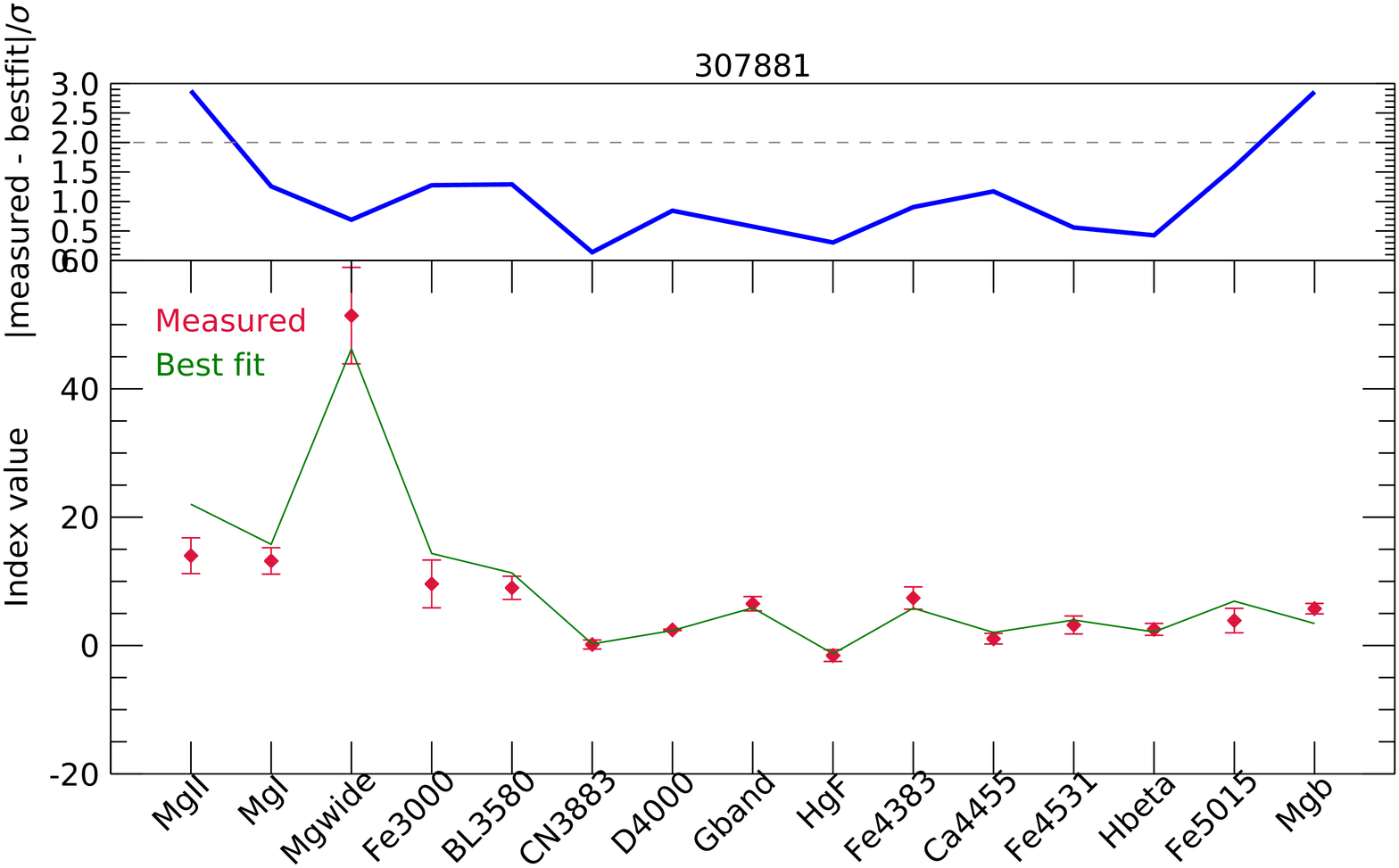}
\caption{\small{Comparison of best-fit model indices (green line) with measured values (red diamonds) for each galaxy. MS11-TMJ best-fit models are those reported in Table \ref{tab:allresults}. Upper panels (blue line) show the differences between data and models divided by the measured error, indicating how many standard deviations they differ. Horizontal grey line indicates the $2\sigma$ level.}}
\label{fig_appB:merged1}
\end{centering}
\end{figure*}

\begin{figure*}
\begin{centering}
\includegraphics[width=9.5cm]{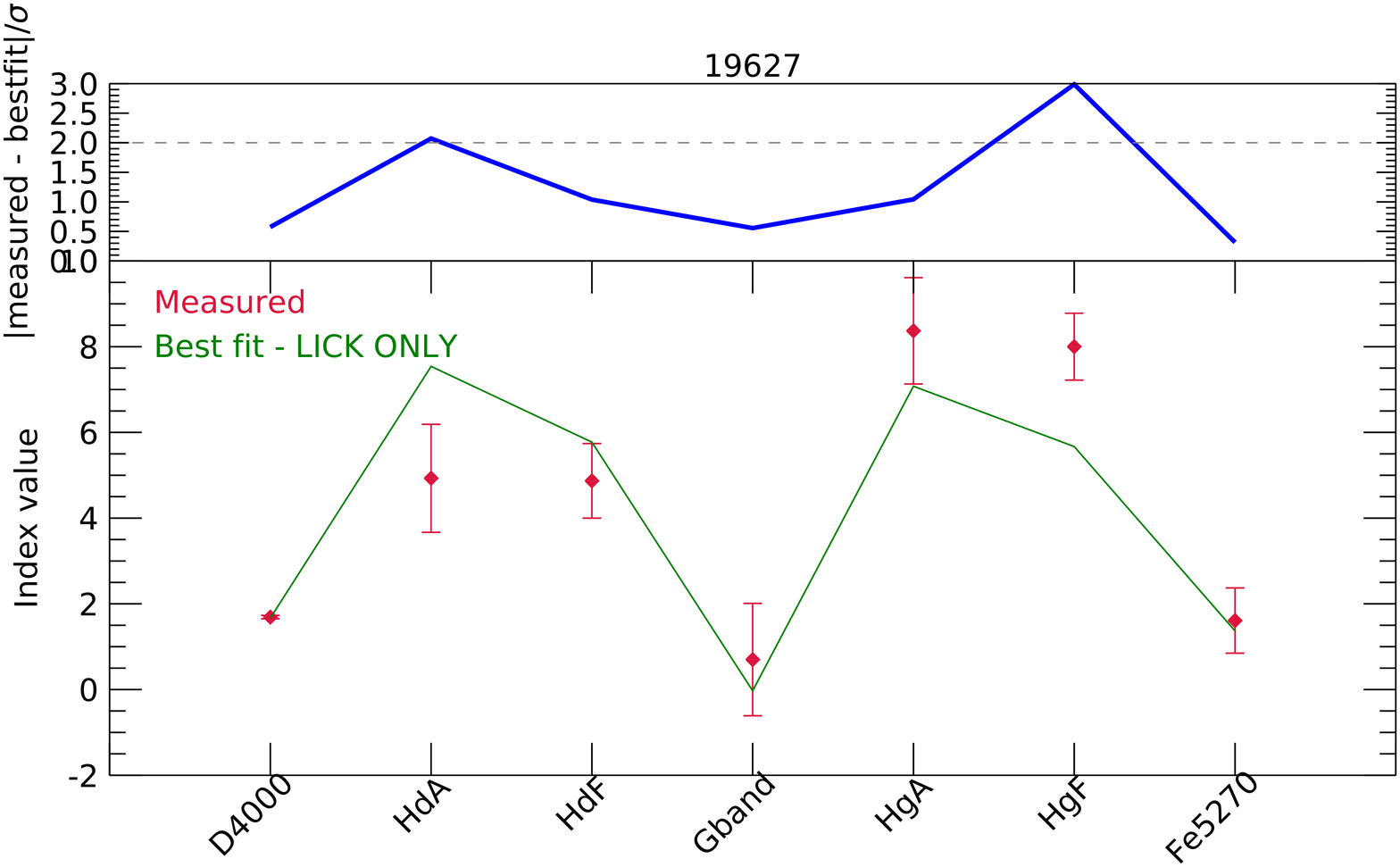}
\includegraphics[width=9.5cm]{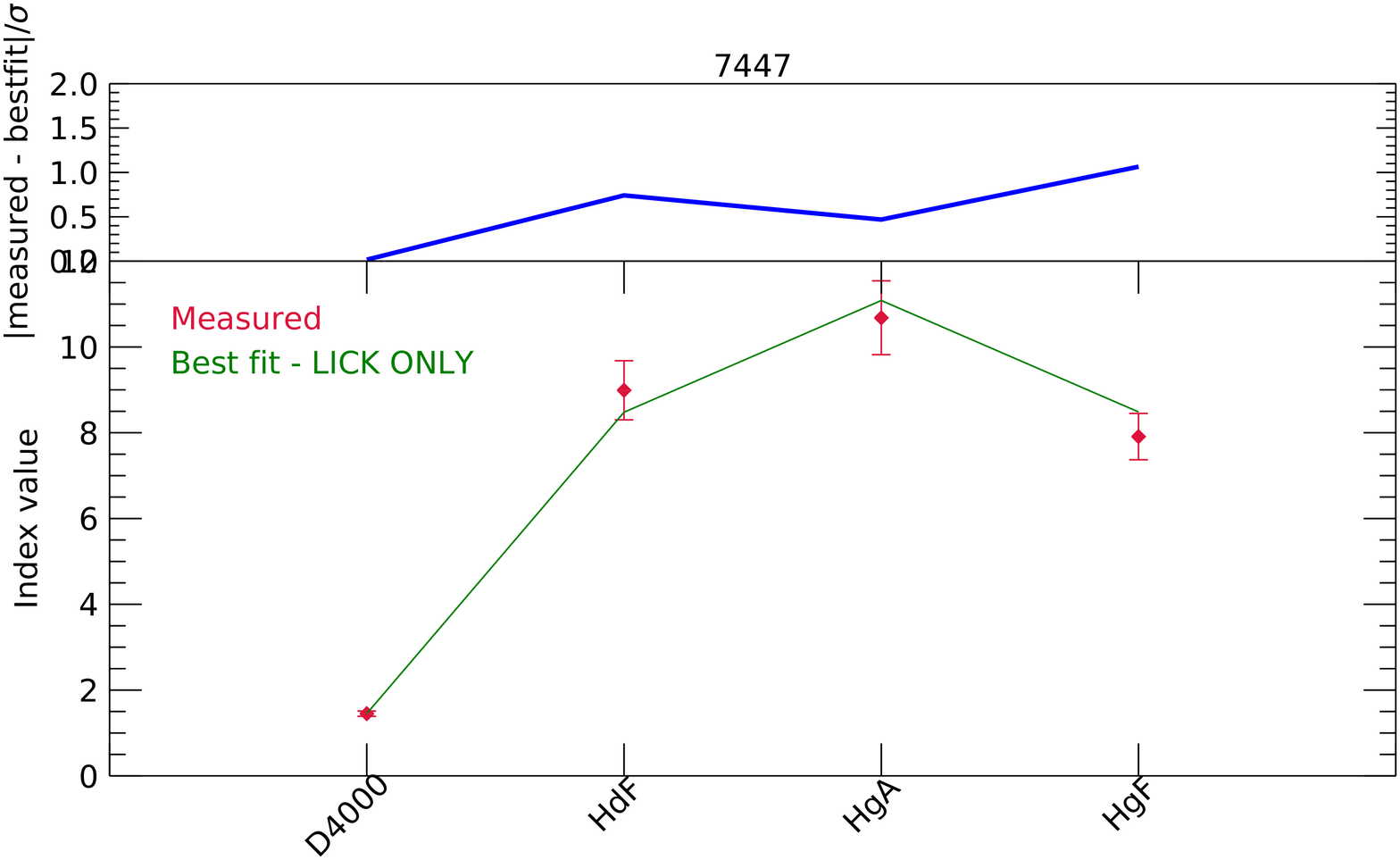}
\includegraphics[width=9.5cm]{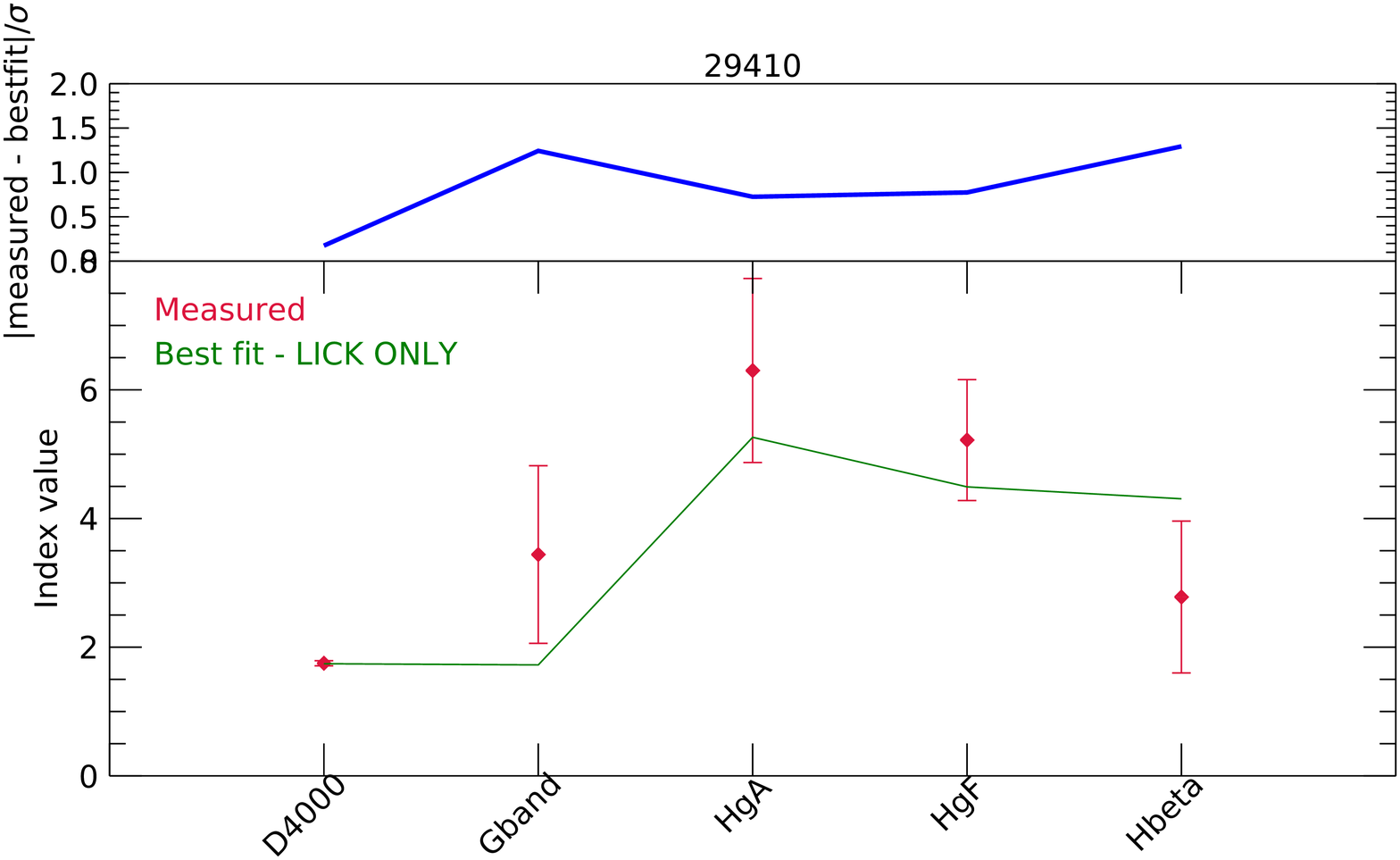}
\includegraphics[width=9.5cm]{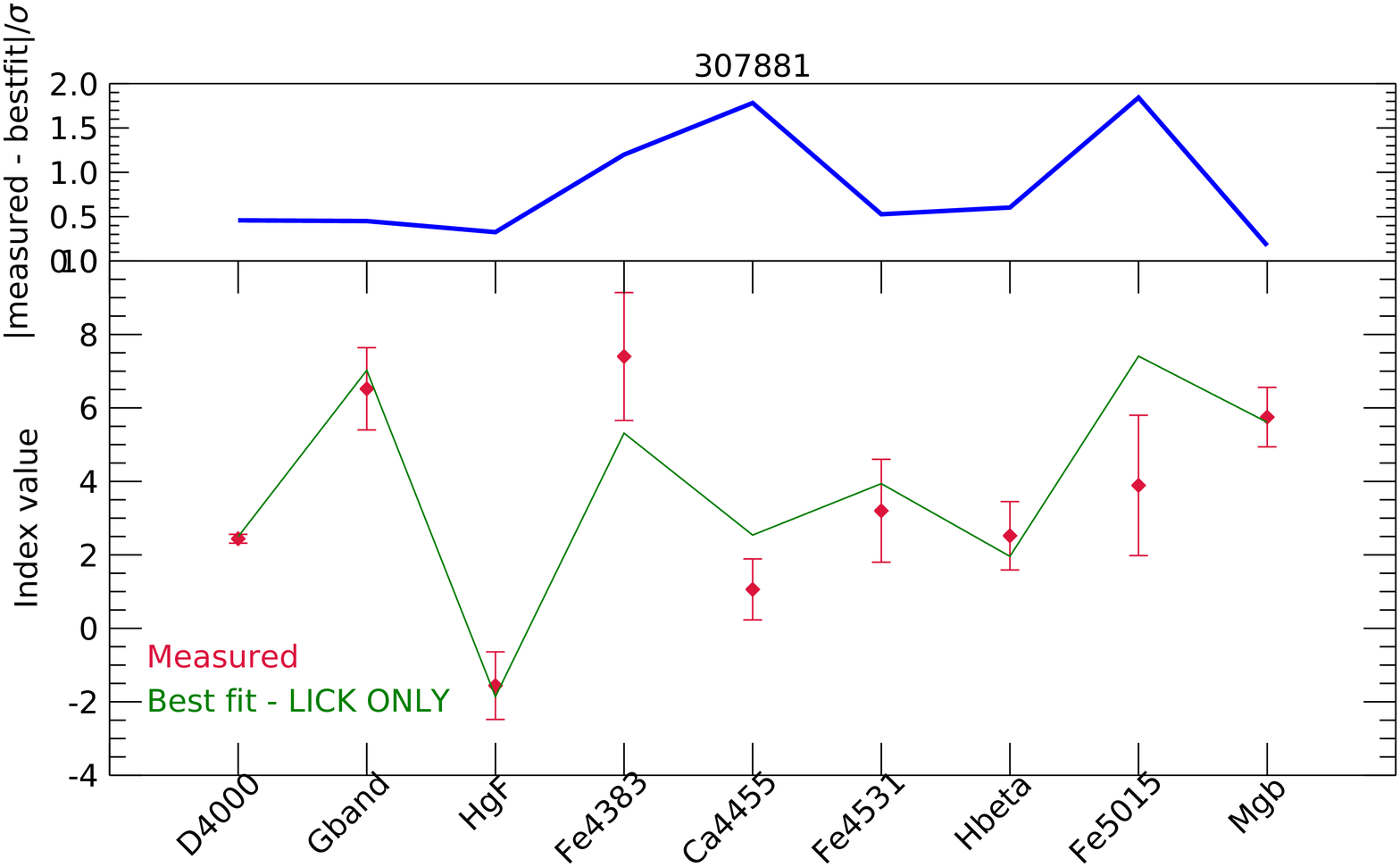}
\caption{\small{As in Figure~\ref{fig_appB:merged1} for the TMJ best-fit models reported in Table \ref{tab:allresultsalpha}.}}
\label{fig_appB:tmj1}
\end{centering}
\end{figure*}

Figure \ref{fig_appB:merged1} and \ref{fig_appB:tmj1} show the comparison of the measured indices for the four galaxies with the 
correspondent best-fit models. In particular, Figure \ref{fig_appB:merged1} refers to fits obtained with the MS11-TMJ merged models (Table \ref{tab:allresults}), while Figure \ref{fig_appB:tmj1} to fits obtained only using the TMJ models of Lick indices (Table \ref{tab:allresultsalpha}).

In each figure, the upper panel shows the values of the difference between data and best-fit model divided by the measured error, suggesting how many standard deviations data and model differ. It can be seen that generally most of the best-fit values agree with the data within $2 \sigma$.

\bsp	
\label{lastpage}
\end{document}